\documentclass[sigconf]{acmart}

\usepackage{lipsum}
\usepackage{subcaption}
\usepackage{macros}
\usepackage{makecell}
\usepackage{fancybox}
\usepackage{multirow}

\usepackage{microtype}
\usepackage[english]{babel}

\usepackage{csquotes}

\usepackage{wrapfig}

\usepackage{fancybox}
\usepackage{xcolor}

\usepackage{longtable}
\usepackage{colortbl}
\newcommand{\sig}[1]{\cellcolor{red!10}\textcolor{purple}{\textbf{#1}}}

\AtBeginDocument{%
  \providecommand\BibTeX{{%
    \normalfont B\kern-0.5em{\scshape i\kern-0.25em b}\kern-0.8em\TeX}}}

\definecolor[named]{ColorInstructor}{cmyk}{0,0.67,0.67,0.4}
\definecolor[named]{ColorStudent}{cmyk}{0,0.53,0.93,0}
\definecolor[named]{ColorPractitioner}{cmyk}{1,0.58,0,0.21}

\definecolor[named]{ColorGray}{cmyk}{0, 0, 0, 0.79}

\copyrightyear{2025}
\acmYear{2025}
\setcopyright{cc}
\setcctype{by}
\acmConference[UIST '25]{The 38th Annual ACM Symposium on User Interface Software and Technology}{September 28-October 1, 2025}{Busan, Republic of Korea}
\acmBooktitle{The 38th Annual ACM Symposium on User Interface Software and Technology (UIST '25), September 28-October 1, 2025, Busan, Republic of Korea}
\acmDOI{10.1145/3746059.3747772}
\acmISBN{979-8-4007-2037-6/2025/09}

\begin{document}

\title[\sys{}]{StoryEnsemble: Enabling Dynamic Exploration \& Iteration in\\the Design Process with AI and Forward-Backward Propagation}

\author{Sangho Suh}
\affiliation{%
  \institution{University of Toronto}
  \city{Toronto}
  \country{Canada}
}
\email{sangho@dgp.toronto.edu}

\author{Michael Lai}
\affiliation{%
  \institution{University of Toronto}
  \city{Toronto}
  \country{Canada}
}
\email{michaelkp.lai@utoronto.ca}

\author{Kevin Pu}
\affiliation{%
  \institution{University of Toronto}
  \city{Toronto}
  \country{Canada}
}
\email{jpu@dgp.toronto.edu}

\author{Steven Dow}
\affiliation{%
  \institution{University of California, San Diego}
  \city{La Jolla}
  \country{USA}
}
\email{spdow@ucsd.edu}

\author{Tovi Grossman}
\affiliation{%
  \institution{University of Toronto}
  \city{Toronto}
  \country{Canada}
}
\email{tovi@dgp.toronto.edu}

\renewcommand{\shortauthors}{Suh et al.}

\newcommand{\sys}[0]{StoryEnsemble}

\begin{abstract}
Design processes involve exploration, iteration, and movement across interconnected stages such as persona creation, problem framing, solution ideation, and prototyping. However, time and resource constraints often hinder designers from exploring broadly, collecting feedback, and revisiting earlier assumptions—making it difficult to uphold core design principles in practice. To better understand these challenges, we conducted a formative study with 15 participants—comprised of UX practitioners, students, and instructors. Based on the findings, we developed StoryEnsemble, a tool that integrates AI into a node-link interface and leverages forward and backward propagation to support dynamic exploration and iteration across the design process. A user study with 10 participants showed that StoryEnsemble enables rapid, multi-directional iteration and flexible navigation across design stages. This work advances our understanding of how AI can foster more iterative design practices by introducing novel interactions that make exploration and iteration more fluid, accessible, and engaging.
\end{abstract}

\begin{teaserfigure}
    \centering
    \includegraphics[trim=0cm 0cm 0cm 0cm, clip=true, width=0.9\textwidth]{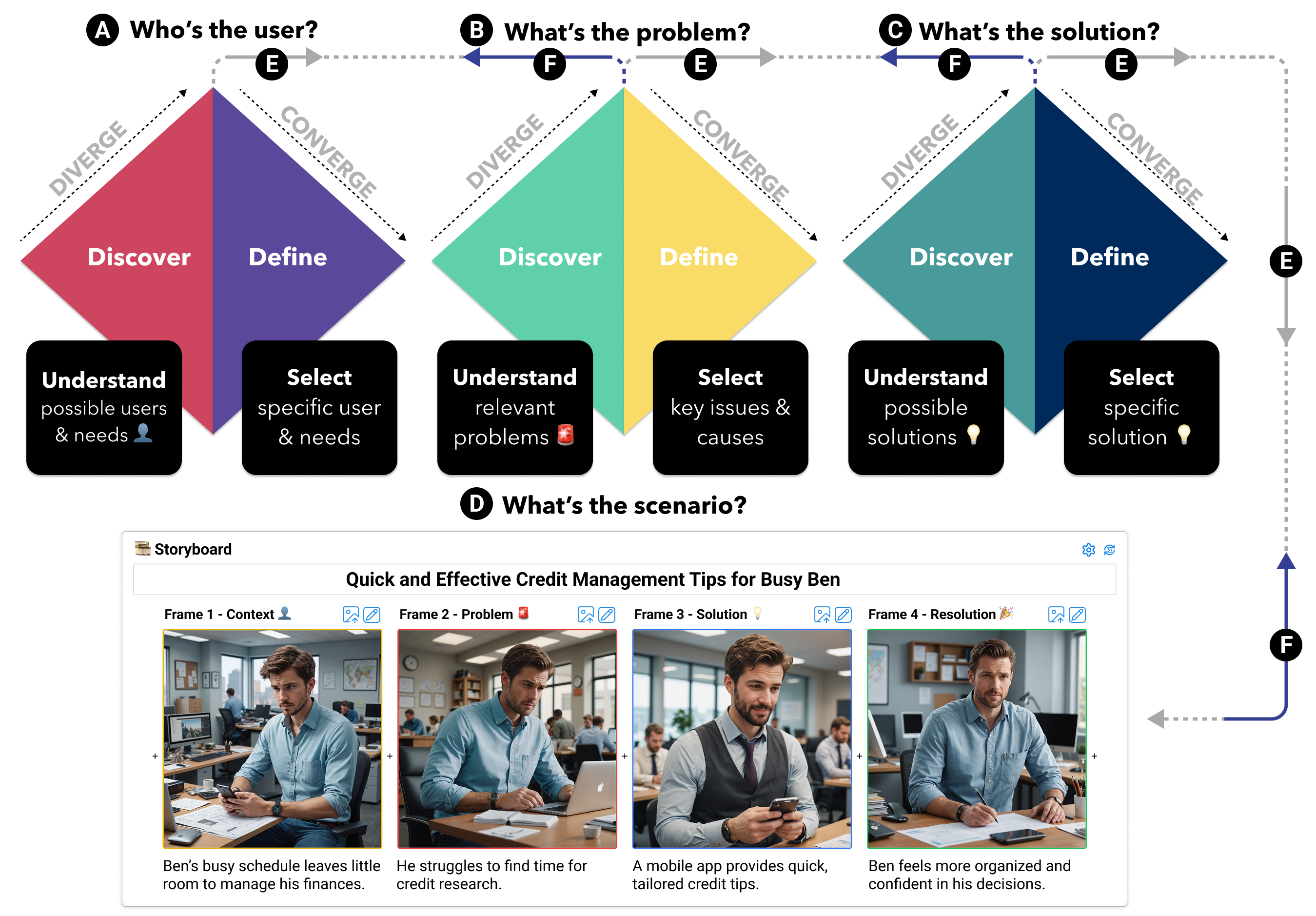}
    \caption{StoryEnsemble enables dynamic exploration and iteration in the design process by leveraging AI and supporting forward and backward propagation of changes across interconnected stages—(A) personas, (B) problems, (C) solutions, and (D) scenarios. Users can move fluidly between stages, revise earlier assumptions, and (E) propagate changes forward—enabling the kind of iteration and exploration that design frameworks promote but is often difficult in practice. They can also begin from later stages, such as solutions or scenarios, and (F) propagate changes backward to reframe upstream decisions.}
    \Description{Teaser figure}
    \label{fig:teaser}
\end{teaserfigure}



\maketitle

\section{Introduction}
\label{sec:intro}

\begin{displayquote}
    ``Few ideas work on the first try. Iteration is key to innovation.'' — Sebastian Thrun
\end{displayquote}

Design frameworks such as design thinking~\cite{archer1979whatever}, the Double Diamond model~\cite{doublediamond2023}, and scenario-based design~\cite{rosson2007scenario} provide structured approaches to addressing complex, open-ended problems. They have had an enormous influence for the past few decades, driving innovation across both academic and professional settings. Despite their diverse formats and implementations, these frameworks share core principles: iterative processes, divergent and convergent thinking, and user-centered storytelling~\cite{razzouk2012design, hehn2018design, micheli2019doing, wang2024research}.

Unfortunately, despite their value, these principles can be difficult to apply in practice. From the initial persona creation to the final high-fidelity design storyboard, time and resource constraints often limit individuals' and teams' ability to iterate, explore a broad range of ideas, or refine solutions through structured convergence~\cite{hartmann2008design}. Collecting meaningful feedback requires significant time and effort~\cite{dow2009efficacy, yen2024processgallery, jung2023toward, jane2024}, and when that feedback prompts revisions—such as reframing the problem or rethinking solutions—it can trigger a costly loop that requires revisiting earlier assumptions and reworking previous design artifacts such as personas, problem statements, scenarios, or storyboards. This overhead often deters individuals and teams from iterating~\cite{greenwood2019dissensus, tantiyaswasdikul2020design, xiang2024simuser}.

Moreover, because design stages are tightly interconnected, getting stuck in one stage—such as struggling to generate diverse ideas~\cite{wang2024roomdreaming} or define actionable problem statements—can stall the entire process. Without a clear problem framing or a range of initial directions, teams may struggle to develop meaningful prototypes or gather useful feedback. Similarly, at later stages, while artifacts like storyboards offer powerful ways to visualize user experiences and communicate design concepts, creating effective storyboards can require significant time and skills. This can result in oversimplified narratives that fail to capture nuanced interactions or encourage reflection and discussion~\cite{truong2006storyboarding}. When personas, problem statements, or scenarios remain underdeveloped, it becomes difficult to imagine concrete use cases or evaluate solutions in context. In practice, such bottlenecks often make it hard for teams to make meaningful progress and move through the design process effectively.

In light of such challenges, AI is emerging as a promising solution~\cite{schmidt2024simulating}. Recent work has demonstrated AI's capability to support various design tasks, including brainstorming~\cite{xu2024jamplate, suh2023sensecape, suh2024luminate}, persona generation~\cite{hamalainen2023evaluating, zhang2023personagen, shin2024understanding, schuller2024generating}, and visual content creation~\cite{antony2024id, wadinambiarachchi2024effects}. These tools can significantly reduce the labor required for key design activities, helping teams work within tight resource constraints. Notably, researchers have shown that large language models can create synthetic user personas comparable to those based on real user data~\cite{schuller2024generating}, potentially valuable for contexts involving vulnerable or hard-to-reach populations~\cite{jung2023toward, hamalainen2023evaluating}.

However, current AI-supported design tools still fall short of addressing the full complexity of design workflows, particularly the iterative nature of the process. Prior work has largely focused on isolated stages of the design process~\cite{schuller2024generating, zhang2024auto,hamalainen2023evaluating, shin2024understanding, bilgram2023accelerating}, overlooking the need to support the iterative and interconnected nature of design, where changes in one part of the process often require adjustments in others. As a result, the current landscape leaves designers with no tools that enable them to manage dependencies across stages and encourage reflection and iteration throughout the design process~\cite{do2001thinking, kirsh2010thinking, razzouk2012design}.

To address this gap and demonstrate how AI can facilitate key design principles across the design process, we developed StoryEnsemble, a system that supports widely shared principles across design methodologies—such as structured iteration~\cite{beckman2007innovation, dow2009efficacy, razzouk2012design, rose2022, Cao2025CompositionalSA}, divergent and convergent thinking~\cite{suh2024luminate}, and scenario-based design~\cite{rosson2007scenario, quesenbery2010storytelling, truong2006storyboarding}. StoryEnsemble enables dynamic exploration and flexible iteration by empowering users to explore diverse ideas, refine through structured feedback, propagate changes seamlessly across interconnected stages, and create easily editable storyboards. 
Our contributions include:

\begin{itemize}
\item StoryEnsemble,\footnote{\url{https://storyensemble-research.github.io/}} an interactive system that leverages a node-link interface and novel interaction techniques—such as forward and back propagation—to enable rapid exploration and flexible iteration. 
\item A formative study identifying challenges faced by students, instructors, and practitioners in applying iterative, exploratory, and feedback-driven design principles in practice.
\item A system evaluation study that advances our understanding of how AI can support workflows grounded in core design principles—such as iteration, divergence-convergence, and scenario construction—particularly in design education and practice.
\end{itemize}

In the remainder of the paper, we review related work on design frameworks and AI design tools, then present our formative study identifying key challenges that students, instructors, and practitioners face when applying design principles. We then describe StoryEnsemble and its features, followed by an evaluation with 10 participants that demonstrates how it enhances exploration, iteration, and scenario development. We conclude with implications for future AI-assisted design systems, as well as for other domains involving interconnected, iterative workflows.

\section{Related Work}
\label{sec:related-work}

We first examine common principles across design methodologies, then explore how generative AI is transforming creative workflows, and review existing AI design tools.

\subsection{Principles Across Design Methodologies and Practical Challenges}

The landscape of design methodologies is diverse, with frameworks varying significantly in their structure, terminology, and specific approaches~\cite{gama2023developers, micheli2019doing}. Research has documented well over a hundred distinct design approaches across various domains~\cite{hehn2018design}. From design thinking and the Double Diamond model to human-centered design, participatory design, service design, and scenario-based approaches, designers have numerous options for characterizing and structuring their design processes. This diversity reflects the fact that no single approach is universally optimal and applicable. Organizations and practitioners frequently adapt frameworks to their specific contexts, leading to continuous evolution and customization of design processes~\cite{kleinsmann2017capturing, gama2023developers}.

Despite this variety, comparative analyses have shown that the underlying principles transcend specific methodologies~\cite{hehn2018design}. These shared principles include iterative refinement, alternating phases of divergent and convergent thinking, human-centered design, and user-centered storytelling~\cite{quesenbery2010storytelling}. While different frameworks visualize and structure these principles distinctly—some making the divergent-convergent pattern explicit (e.g., Double Diamond~\cite{doublediamond2023}), others emphasizing empathy and iteration (e.g., design thinking~\cite{ideo}) or scenario-based design—the core approaches to problem-solving remain consistent across methodologies. Many frameworks that appear linear in representation are fundamentally iterative in practice, incorporating feedback loops that enable practitioners to revisit earlier stages as new insights emerge~\cite{ideo}.

However, a notable gap exists between theoretical frameworks and their practical application~\cite{stolterman2008nature, goodman2011understanding, arteaga2024across, pruitt2003personas, wang2024roomdreaming, rogers2004new}. Despite their theoretical importance, these principles often prove challenging to implement due to time constraints, resource limitations, and the labor-intensive nature of key design tasks~\cite{kleinsmann2017capturing}. For instance, while design thinking emphasizes the importance of multiple iterations to refine ideas~\cite{rose2022}, time constraints frequently limit iteration cycles to one or a few. Even in semester-long courses, the design cycle may only be completed once, or iteration may be intentionally minimized to avoid major changes to a project.

Practitioners across educational and professional settings face several common challenges. Novice designers often gravitate toward familiar problems, frequently addressing the same types of design challenges and getting fixated~\cite{maceli2024incorporating}. Teams under deadline pressure tend to implement theoretically non-linear processes in a more linear, constrained manner~\cite{greenwood2019dissensus}. Additionally, collecting and incorporating meaningful feedback becomes difficult within tight project timelines~\cite{dow2009efficacy, schrage1999serious}, further widening the gap between ideal iterative refinement and practical implementation.

Our work builds on decades of research on design by contributing a system that addresses a longstanding challenge in design practice: enabling easy, seamless iteration across interdependent stages. By demonstrating how to lower the overhead of iteration, our work aims to bridge the gap between theoretical ideals and the realities of design processes.

\subsection{Generative AI for Creative Processes}

The recent advances in AI opened up new interaction possibilities and are radically transforming creative workflows across various fields~\cite{chung2023, choi2024, son2024, almeda2024prompting}, ranging from creative writing~\cite{lee2022coauthor, kim2023metaphorian, masson2025textoshop} to design~\cite{wang2024roomdreaming, masson2024directgpt} and content creation~\cite{wang2024lave, cao2025compositional}. For example, in TableBrush, users can generate stories based on a narrative arc sketched by the user~\cite{chung2022talebrush}. Liu et al. explored how text-to-image AI can be incorporated into the 3D design workflows, showing that text-to-image models can assist with producing reference images, preventing design fixation, and inspiring design considerations~\cite{liu20233dall}. These advancements highlight the transformative potential of AI in supporting more dynamic and flexible creative processes.

\begin{figure*}[htb!]
    \centering
    \includegraphics[width=\textwidth]{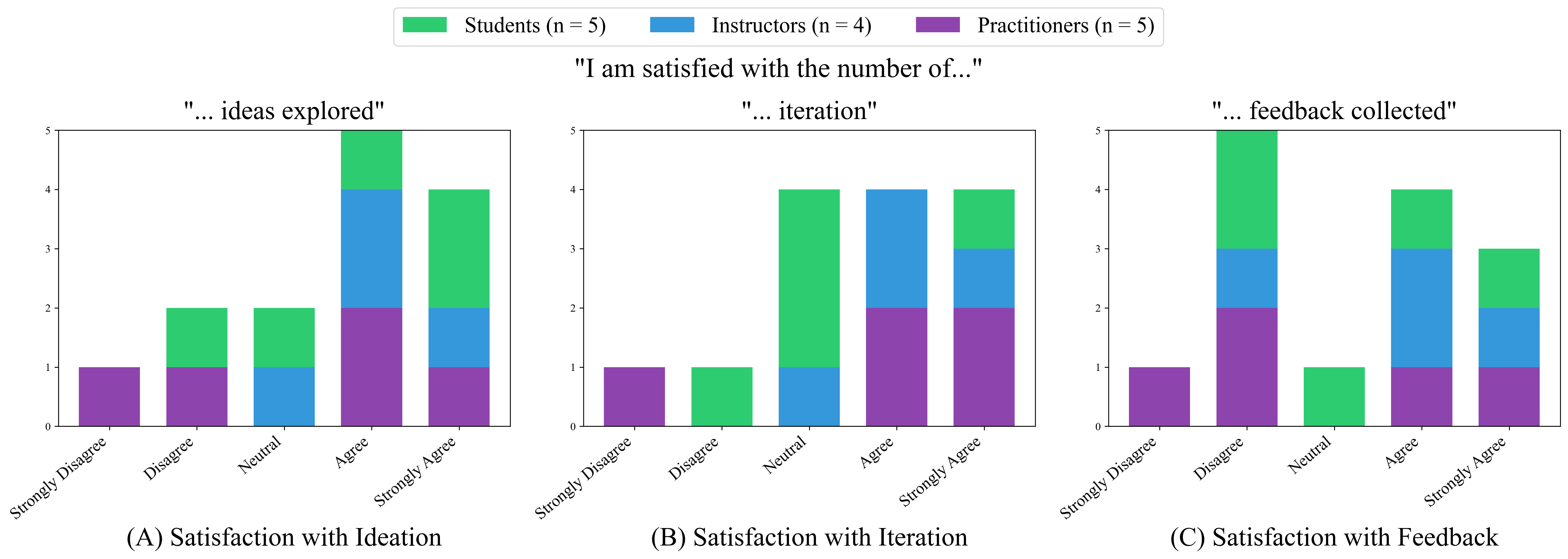}
    \caption{Results from a formative study with \raisebox{0.3ex}{\colorbox[HTML]{2ecc71}{\hspace{1em}\rule{0pt}{0.5ex}}} students, \raisebox{0.3ex}{\colorbox[HTML]{3498DB}{\hspace{1em}\rule{0pt}{0.5ex}}} instructors, and \raisebox{0.3ex}{\colorbox[HTML]{8E44AD}{\hspace{1em}\rule{0pt}{0.5ex}}} practitioners and their satisfaction with ideation, iteration, and feedback collection.}
    \label{fig:formative-satisfaciton}
\end{figure*}

AI is increasingly being integrated into the creative process to expand idea exploration. In Luminate, Suh et al. explored a novel interaction paradigm where the system generates a range of ideas designed to cover the design space instead of producing a single response to a prompt~\cite{suh2024luminate}. Similarly, Wang et al. contributed RoomDreaming, a system that enables interior designers to co-design with AI~\cite{wang2024roomdreaming}. While both systems focus on supporting designers in exploring diverse ideas and iterating on their designs, RoomDreaming does not address the interconnected nature of stages in design thinking, where changes in one stage, such as problem definition, often require adjustments in subsequent stages like persona development or solution generation.

Generative AI is also being explored for its potential to automate feedback and enhance the iterative process. For instance, Duan et al. developed a Figma plugin where LLMs generate feedback on UI mockups, identifying subtle design errors and improving textual elements~\cite{duan2024generating}. Similarly, Xiang et al.'s SimUser system automates usability feedback, producing insights that closely mirror those generated by human users~\cite{xiang2024simuser}. Assuming feedback will always be accessible with the advances in AI, E et al. explored the tradeoffs of providing access to feedback
throughout a design task versus only providing feedback after~\cite{jane2024}. These systems demonstrate how AI can facilitate rapid iteration by providing timely, automated feedback, allowing designers to refine their work more efficiently. 

Our work builds on these efforts by utilizing AI-generated feedback to facilitate transitions across various design processes, from persona creation to design artifact creation, demonstrating the new role that feedback can play in the AI-powered framework for iterative design and creative exploration. While the traditional design process has used feedback primarily for evaluation, our approach highlights its potential as a generative tool—one that can be invoked on demand and propagate changes forward and backward in the design process, interconnecting different design stages to stimulate new directions and support flexible iteration throughout the design process.

\subsection{AI Tools for Design}

Generative AI has become increasingly important in supporting individual stages within the design process, such as generating persona~\cite{schuller2024generating, zhang2024auto, choi2024proxona} and prototyping~\cite{hamalainen2023evaluating, bilgram2023accelerating, shin2024understanding, salminen2024deus, benharrak2024writer}. He et al. proposed Cinemassist, a system that supports creative exploration of potential cinematic compositions for users creating 3D animations. For each keyframe in the animation, users are presented with different scene options and frame-level options~\cite{he2024interactive}. 
Wang et al. developed AIdeation to support search and combination of reference images for visual concept designers \cite{Wang2025AIdeationDA}. 
The UIDEC tool scaffolded the design brainstorming process by emphasizing design under constraints and using AI to generate examples \cite{Shokrizadeh2025DancingWC}.
LLMs have also been used to generate user research data, providing designers with fast, scalable tools for modeling user behaviors and needs~\cite{salminen2024deus, schuller2024generating, Hung2024SimTubeGS}. 
Other tools focus on AI-assisted novel artifact generation, such as sketching \cite{Lin2025InkspireSD}, motion comics \cite{chen2025dancingboard}, or manipulable design tokens \cite{Shi2025BrickifyEE}.
These advancements suggest that AI can reduce the manual effort involved in various design stages, allowing designers to focus on more complex creative tasks.

While many AI tools have focused on supporting tasks within specific stages of the design process, limited attention has been paid to augmenting the overall process and bridging fragmentation between stages. Our work contributes to the growing body of research on AI-assisted design by addressing this gap, demonstrating how AI can help manage interdependencies and support iteration across related artifacts in a cohesive manner.

\section{Formative Study: Understanding Challenges, Opportunities, and Practices with Survey \& Interview}
\label{sec:formative-study}

To understand how generative AI can enable new design workflows, we recruited 15 participants for two formative studies involving a survey and interviews.

\textbf{Survey.} The survey study aimed to identify challenges related to exploration, iteration, and feedback collection, along with potential opportunities and concerns regarding AI integration. 
While we intended to keep the scope open to various design processes, we referenced \textit{design thinking} in questions about design practices to provide a familiar reference point and facilitate discussion around common activities such as problem framing, ideation, and iteration.

\textbf{Interview.} The interview study focused on understanding the practices surrounding storyboarding for feedback collection among UX designers. We concentrated our interviews on only UX professionals, since industry practices around storyboarding—compared to academic settings—are less well-documented and offer opportunities for new insights, such as how storyboards are used in communicating with stakeholders.

\subsection{Procedure \& Participants for \\Two Formative Studies}

For the survey study, we recruited 8 participants (gender: 4M, 4F; age: M = 30, SD = 7.7, range = [24, 46]; referred to as F1-8).
Out of the eight participants, five had design learning experience as students (through HCI courses at universities or UX bootcamp), four had experience teaching design process as instructors (M = 4.3 years of teaching experience, SD = 5.3, range: [1, 12]), and five had experience as design practitioners (M = 4.6 years of design practice experience, SD = 3.2, range: [2, 10]). Participants completed an online survey where they described challenges they faced during the design process, particularly with respect to exploration, iteration, and feedback collection. They were then asked how these challenges might be addressed and whether AI could support these processes. Each participant received a \$15 gift card.

In the interview study, we conducted 60-minute semi-structured interviews with 7 professional UX designers (4M, 3F; F8-15) who had an average of 7.1 years of experience (SD = 4.7, range: [2, 15]) in UX design and storyboarding. On average, they had received feedback on their UX scenarios (in text) 12.6 times (SD: 7.9; range: [3, 25]) and storyboard 9.6 times (SD: 9.3; range: [3, 30]). Participants discussed the types of storyboards they used—text-only scenarios, visual storyboards, or photos—and how they collected feedback from stakeholders. They also described the kinds of feedback they typically sought. Each participant received a \$25 gift card.

We provided participants with a definition and overview of design thinking principles during recruitment and again at the beginning of both studies to ensure a shared understanding of the terminology and process.

\begin{figure*}[h!]
    \centering
    \includegraphics[alt={Storyboards created by study participants with AI assistance.}, width=\linewidth]{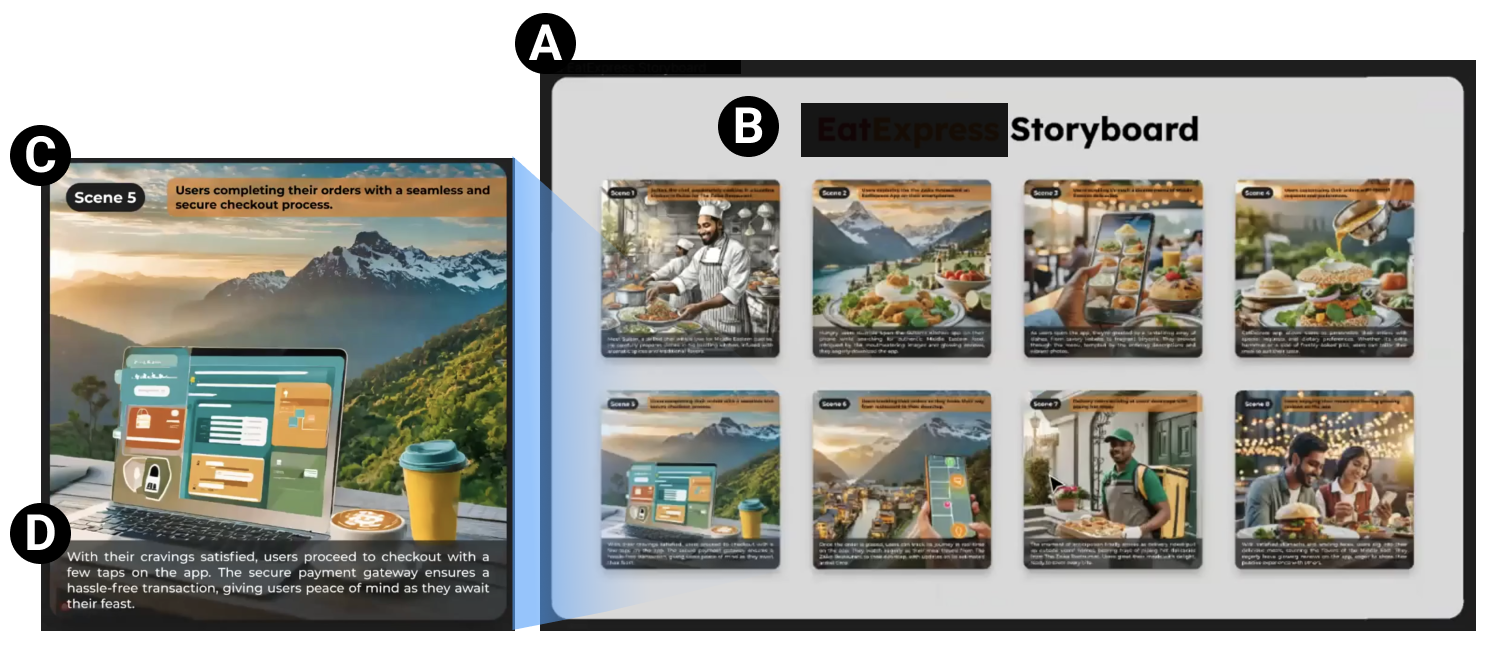}
    \caption{A screenshot of (A) F15's storyboard, with the (B) title (redacted) at the top. (C) is a frame featuring an image that shows the interface with (D) caption. This, along with general storyboard visuals~\cite{truong2006storyboarding}, informed our storyboard design (Fig.~\ref{fig:storyboard}).}
    \label{fig:storyboard-reference}
\end{figure*}

\subsection{Identified Needs (Ns) from Survey with Students, Instructors, and Practitioners}

\subsubsection{Challenges in Applying Design Thinking in Projects}
\label{formative-problem-in-dt}

\begin{center}
    \setlength{\fboxsep}{4pt}%
    \fbox{%
    \begin{minipage}{7.8cm}
    \begin{Bitemize}
        \item[] \textbf{N1} Support time-consuming tasks in design thinking
        \item[] \textbf{N2} Support ideation at any design phase
        \item[] \textbf{N3} Support translation of user needs to problem statement
    \end{Bitemize}
    \end{minipage}}
\end{center}

F1 (UX designer) shared that \textbf{time and budget constraints} often limited the full application of design thinking. F2 (practitioner) and F6 (student) noted that \textbf{ideation} was difficult, with initial ideas often basic and better ones emerging later in the process. Integrating diverse user perspectives was also a challenge, sometimes leading to overlooked groups (F3, student). F4 (practitioner) and F5 (student) found \textbf{translating user needs and challenges into actionable problem statement} challenging. 
Finally, F7 (instructor \& UX consultant), who has been teaching HCI and UX courses for 12 years, touched on the \textbf{theory-practice gap}, explaining that the widely taught five-stage design thinking model oversimplifies the process, saying: 

\blockquote{``The process is described as linear, but in reality, it's not that linear. Students should understand that not every problem will require all the steps and that things need to be customized on a per-problem basis.''}

To address these challenges, participants suggested adapting design thinking according to the stakeholders' needs, time constraints, and budget (F1) and using collaboration or AI for ideation (F2). Other solutions included getting help from team members to refine problem statements (F4), building systems to guide problem definition, and using AI to generate feedback (F5).

\subsubsection{Challenges in Transitioning Between Design Stages}

\begin{center}
    \setlength{\fboxsep}{4pt}%
    \fbox{%
    \begin{minipage}{6.5cm}
    \begin{Bitemize}
        \item[] \textbf{N4} Support divergent and convergent thinking
        \item[] \textbf{N5} Support synthesis and translation of ideas
        \item[] \textbf{N6} Support returning to previous design phases
    \end{Bitemize}
    \end{minipage}}
\end{center}

F5 (student) stressed the value of the \textbf{double diamond process, which encourages both divergent and convergent thinking}. F6 (student) pointed out the importance of \textbf{synthesizing ideas}, noting that while generating ideas is commonly highlighted, the ability to \textbf{distill and translate those ideas} is often overlooked yet crucial. F7 (instructor) mentioned the \textbf{flexibility to revisit previous phases of the design process}, stressing that it is okay and it should be encouraged to return to earlier stages if new insights emerge.

\subsubsection{Challenges in Receiving Feedback and Iterating Design}
\label{formative-experience}

\begin{center}
    \setlength{\fboxsep}{4pt}%
    \fbox{%
    \begin{minipage}{8.2cm}
    \begin{Bitemize}
        \item[] \textbf{N7} Support more frequent design iteration via feedback
        \item[] \textbf{N8} Facilitate efficient feedback collection at any design phase
    \end{Bitemize}
    \end{minipage}}
\end{center}

With respect to ideation, student participants (F3, F5, F8) indicated that the number of ideas they explored depended on the course requirement and instructor. F3 said: ``The \textbf{exploration was driven by instructor feedback and collaborative discussions with my peers}, which encouraged me to think beyond the obvious.'' F2 and F3 (students) shared that while they ``felt that the number of ideas explored could have been greater,''  ``\textbf{time constraint}'' is always a problem.

Regarding iteration, except for F6, every student participants (F3, F5, F7-8) mentioned they iterated at least once, while varying in the number of iteration. For example, F2 had a chance to iterate only once during a workshop, while F8 did not iterate ``during the workshop.'' However, they all agreed on the feedback's role in iteration. F8 said, ``teacher feedback come in as a good `reason' for team members to agree on going back.'' F6, who said there was no iteration in her class, attributed this to the absence of feedback, saying ``classes don't offer enough time to engage in feedback.'' F2 and F3 both mentioned they iterated based on feedback from ``instructor and peers.'' Students (F3, F5-6, F8) not satisfied with the number of iterations in the course noted that ``the course's pace ... pushed [them] to move forward without enough refinement'' (F3). To address this problem, they suggested giving students ``\textbf{more chances to [get] feedback}'' (F6) and teach the importance of feedback and iteration to team members who ``do not want to go back and iterate and consider [iteration] very inefficient'' (F8).

While all instructor participants (F4, F6-7) strongly agreed that \textit{it is important to return to previous stage of the design process and iterate}, its \textbf{time-consuming and laborious nature} made it challenging. F7 said that even for ``12-week courses, [he finds] it hard for students to go back to previous stages.'' He added that the most he asks the students to do is to iterate \textit{within} specific stages---``especially at the stages of lo-fi, mid-fi, and evaluation.'' 
F6 believed that ``learning how to iterate is a skill that requires time.''

Practitioner participants (F1, F6-8) also \textbf{acknowledged the importance of iteration}. However, the number of times they iterate varied depending on the project schedule (F7), client (F1), and feedback (F5). F1 and F5 noted that they ``return to previous stages'' when there are new insights from formative studies, users' feedback, or stakeholders' requests to add or remove some features. 

The practices and experiences surrounding collecting feedback varied. Some (F3, F8) reported collecting feedback more frequently than others (F2, F5-6). F3 collected feedback at every stage, F5 collected feedback only on specific stages such as solution or testing stages. 
As shown in Fig.~\ref{fig:formative-satisfaciton}, practitioner participants (F1, F5, F7-8) were also generally \textbf{dissatisfied with the number of feedback they collect}. 
Several participants (F3, F5-6) expressed dissatisfaction with the current status quo and suggested potential solutions. F5 suggested using generative AI to generate ``feedback... so that a more frequent feedback process can be easily realised.'' F3 suggested integrating ``regular feedback checkpoints into the course structure, ensuring that feedback is shared consistently at each stage of the design process.''

\subsubsection{Challenges with Using AI for Design Thinking}
\label{formative-ai-for-dt}

\begin{center}
    \setlength{\fboxsep}{4pt}%
    \fbox{%
    \begin{minipage}{8cm}
    \begin{Bitemize}
        \item[] \textbf{N9} Reduce manual effort in requesting AI feedback
        \item[] \textbf{N10} Synergize human and AI capabilities for co-creation
    \end{Bitemize}
    \end{minipage}}
\end{center}

Many participants (5 of 8) had experience using AI. The most common use case was for generating ``a wide range of ideas'' (F3). Several participants (F2, F3, F8) shared that they used it to explore ``diverse perspectives [to] spark new thoughts and directions on a problem'' and ``get the divergent thinking going.'' 
In addition to ideation, F4 used it to create storyboard and ``analyze feedback to identify common themes.'' All participants supported the idea of integrating AI in their design process. F1 felt the current practice is ``still very manual'' and that ``there has to be a way to \textbf{make it more automatic}.'' However, F3 and F4 also emphasized that it has to be a \textbf{co-creative process}, saying ``AI should complement human creativity and judgment, not replace it. The best outcomes often come from a \textbf{synergy between human insight and AI's capabilities}'' (F4). Several participants (F5, F6, F7) mentioned using AI to generate ``feedback'' (F5) ``at every stage of the design'' (F7) to help ``break free from design fixation'' (F8). F7 shared that it can perhaps ``help with the iteration by suggesting more ideas/solutions.'' F7 went on to suggest that given that ``the constrains of 10-weeks course projects (while students take multiple courses) and the timelines in industry often mean that \textbf{steps have to be skipped} and you can \textbf{only do one full cycle} (without much opportunities to go back),''  ``we could \textbf{\textit{sub-contract} some aspects to AI} [to] free up more time for designers to address these challenges.''

\subsection{Identified Needs (Ns) from Interview with UX Designers on Storyboarding Practices}

\subsubsection{Storyboard is essential with many benefits but the costly nature of producing it prevents its use.} All the UX designers acknowledged that the costly nature of storyboarding prevented them from using them as much as they would like considering the \textbf{many benefits compared to alternatives} such as text-based scenarios. F12 said: ``What's awesome about storyboards is that it just makes ideas easier to discuss and gives you a fuller picture. A higher level view.'' However, she stated that she \textbf{uses storyboard sparingly}, saying ``I've used storyboards when I've had a greater amount of time which is unfortunate, or when I really needed help explaining a beginning to end concept.'' Several participants (F12, F13) working as freelancers also mentioned that the \textbf{clients who hire them avoid requesting storyboarding} as that would require them to pay more for the time spent on creating storyboards, highlighting the costly nature of storyboarding even in terms of finance. 

\begin{center}
    \setlength{\fboxsep}{4pt}%
    \fbox{%
    \begin{minipage}{8cm}
    \begin{Bitemize}
        \item[] \textbf{N11} Support image consistency across frames in storyboard
    \end{Bitemize}
    \end{minipage}}
\end{center}

\subsubsection{Ensuring image consistency in storyboard is challenging when AI is used to generate images for storyboard.} During the interview, F15 showed storyboard (see Fig.~\ref{fig:storyboard-reference}) he created using ChatGPT's image generation capability. While AI eased the process of creating storyboard for efficient communication with his clients, he noted the challenges present in his current workflow such as \textbf{maintaining consistency across image frames in the storyboard}. He explained that he creates each frame separately, which led to inconsistency in the generated images. As shown in Fig.~\ref{fig:storyboard-reference}, images varied in style and tone. For example, some images were in drawing style (1st frame), while others were photo realistic.

\section{\sys{}}
\label{sec:system}

We first present the design goals for \sys{}, grounded in needs (\textbf{Ns}) identified through formative studies, a literature review, and pilot studies. We then provide an overview of the interface and its features, followed by an example workflow.

\subsection{Design Goals (DGs)}

The DGs are organized into three themes: (1) \textit{core innovations}, (2) \textit{creative support}, and (3) and \textit{usability across diverse contexts}.

\vspace{0.1em}
\noindent\underline{\textit{I. Core innovations}} \\[0.3em]
\textbf{DG1. Enable non-linear engagement (N7).} 
Allow users to start at or engage with any design stage, with automatic updates ensuring consistency across related artifacts, supporting the iterative workflows emphasized in many design methodologies.

\noindent\textbf{DG2. Enable on-demand feedback (N8, N9).} 
Offer real-time feedback on design artifacts and streamline its incorporation, encouraging users to reflect and continuously refine ideas with tight feedback loops.

\vspace{0.1em}
\noindent\underline{\textit{II. Creative support}} \\[0.3em]
\textbf{DG3. Promote diverse idea exploration (N2, N5, N6).}
Encourage broader ideation while avoiding AI-induced homogenization~\cite{anderson2024homogenization}. Provide tools for rapid generation and variation to help users overcome fixation.

\noindent\textbf{DG4. Support divergent and convergent thinking (N4).}
Enable fluid movement between generating possibilities (\textsc{Discover}) and narrowing options (\textsc{Define}) based on insights and feedback, mirroring the structure of the Double Diamond model.

\vspace{0.1em}
\noindent\underline{\textit{III. Usability across diverse contexts}} \\[0.3em]
\noindent\textbf{DG5. Simplify artifact generation using AI (N1, N3, N11).} Reduce effort needed to create personas, problem statements, solution ideas, and storyboards using AI.

\noindent\textbf{DG6. Support flexible AI engagement (N8, N9, N10).}
Accommodate three levels of AI involvement to match varying design contexts and user preferences: (1) \textit{no AI involvement}, where users manually create and refine all design artifacts; (2) \textit{partial AI assistance}, where users provide key information and AI fill gaps; and (3) \textit{full AI generation} for exploration, iteration, and feedback. 

\noindent\textbf{DG7. Ensure adaptability to diverse contexts (N10).} Design the system to be simple yet robust, so that it can scale across a range of design methodologies and implementation settings—from short workshops to extended courses—and support users with varying levels of expertise.

\begin{figure}
    \centering
    \includegraphics[alt={\sys{} interface. There are (A) `Start brainstorming' and `Add empty node' buttons for adding new ideas. There are also 5 rows of cards, in order: (B) a single context card that reads "Urban sustainability", (C) three persona cards which include an image and name, (D) three problem cards which include an image and title, (E) three solution cards which include an image and title, and (F) a card which depicts a design storyboard with four frames. There are (G) edges that connect personas to problems, problems to solutions, and solutions to the storyboard.},width=0.5\textwidth]{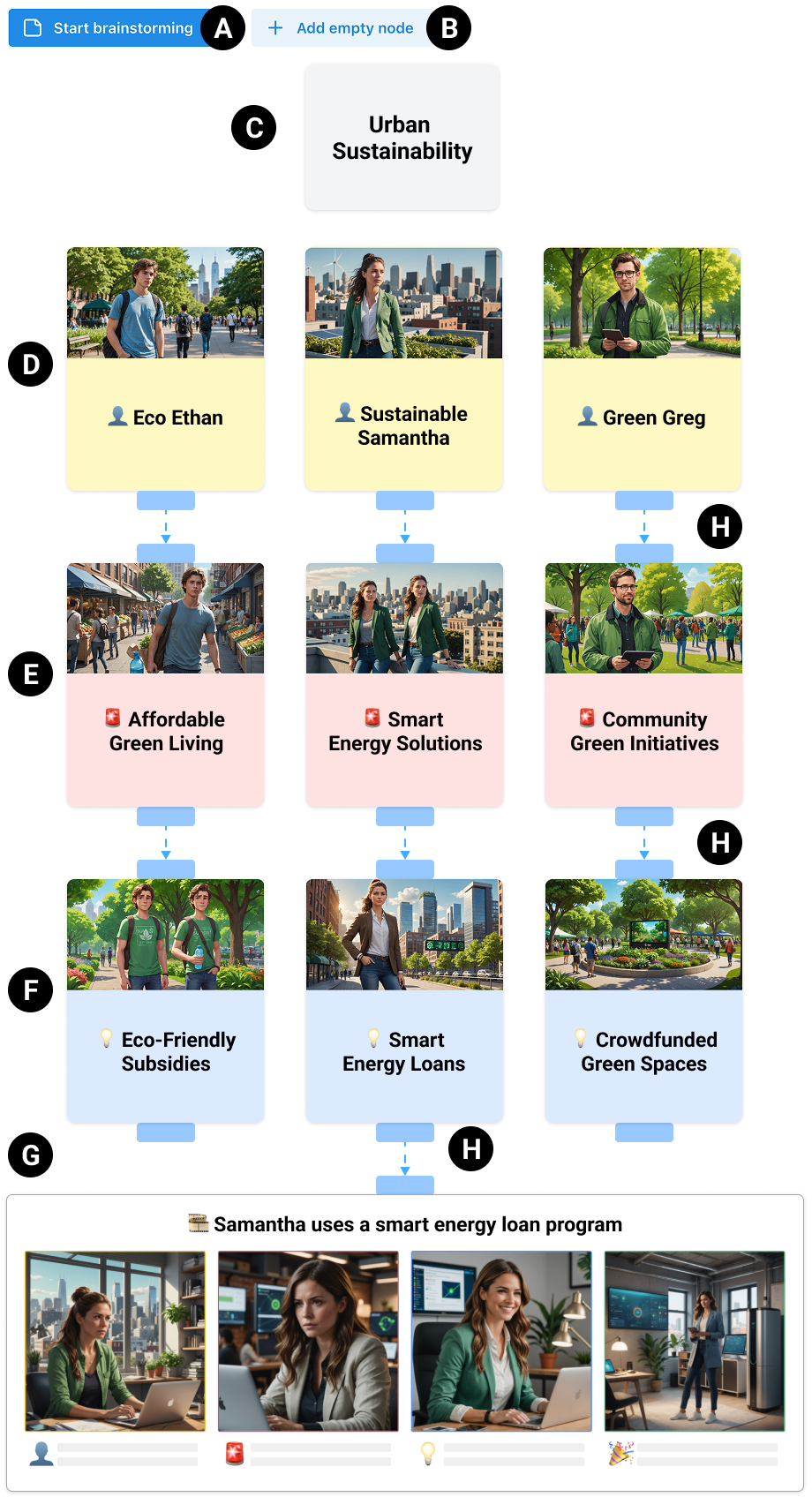}
    \caption{\textsc{\sys{} Interface}: (A) opens up a panel (Fig.~\ref{fig:start-brainstorming}) to add context and generate chains of ideas on a specific theme, such as \textsf{Urban Sustainability}, (B) allows users to add empty nodes and edit node content manually (Fig~\ref{fig:manual}) or with AI (Fig.~\ref{fig:revise-using-ai}). The system features different nodes types: (C) context node (e.g., \,\raisebox{-2.9pt}{\includegraphics[scale=0.25]{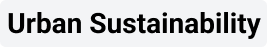}}) for grouping ideas, (D) persona nodes, (E) problem nodes, (F) solution nodes, and (G) a storyboard node. (H) Edges (\,\raisebox{-2.7pt}{\includegraphics[scale=0.23]{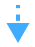}}\,) map the relationships and dependencies between ideas, showing how ideas influence each other throughout the design process. The figure shows outputs generated using the\,\raisebox{-2.7pt}{\includegraphics[scale=0.23]{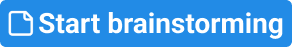}}\,feature, detailed in Section~\ref{sec:features}.}
    \label{fig:interface}
\end{figure}

\begin{figure}[!htb]
    \includegraphics[alt={A card representing a persona named "Eco Emily." The card includes an image of "Eco Emily" at the farmers' market with a bicycle. The card also includes text attributes which describe the persona's location, bio, needs, challenges, and a description. A toolbar below has buttons for editing the persona or creating other related cards.}, width=0.4\textwidth]{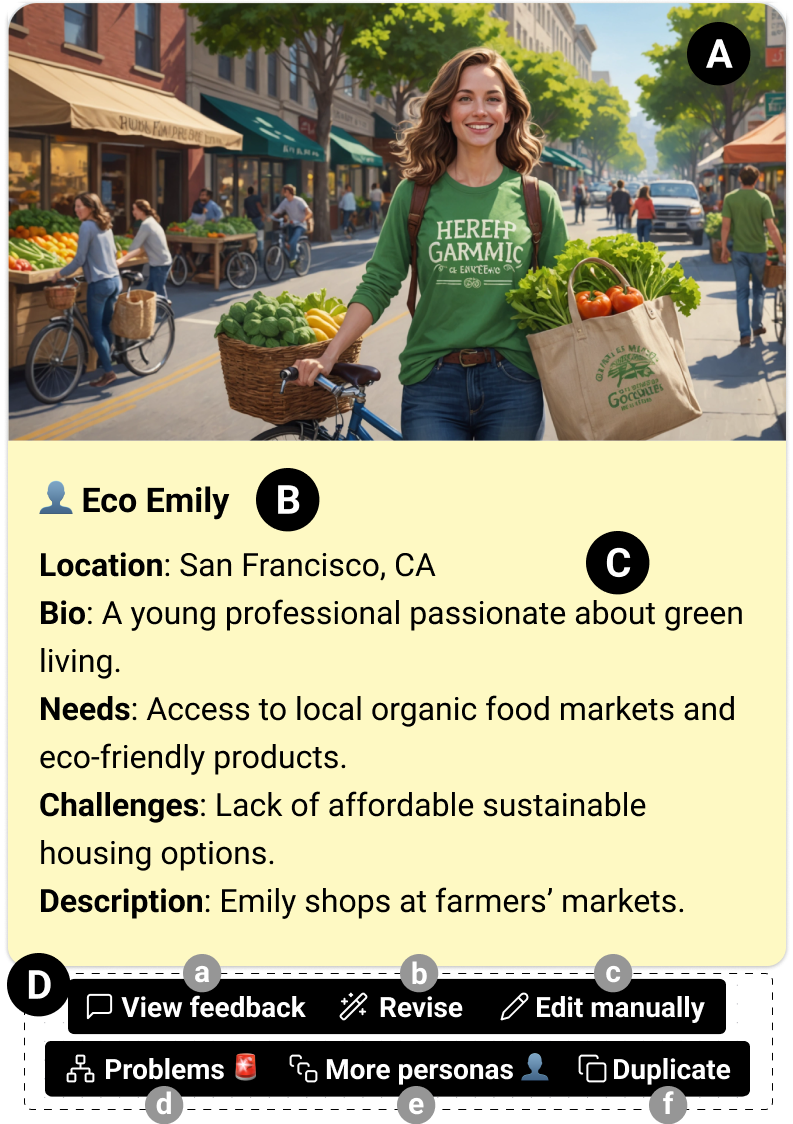} 
    \caption{\,\raisebox{-0.5pt}{\includegraphics[scale=0.11]{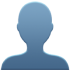}}\,\textsc{Persona node}: An (A) illustrative image, (B) name and (C) attributes which represent a persona. Each node includes (D) a contextual toolbar that appears when the node is selected. Toolbar options include: (a) view AI-generated feedback (Fig.~\ref{fig:view-feedback}), (b) revise using AI (Fig.~\ref{fig:revise-using-ai}), (c) edit manually (Fig.~\ref{fig:manual}), (d) generate problem node, (e) generate more personas (Fig.~\ref{fig:more-nodes}), and (f) duplicate the node.}
    \label{fig:individual-node}
\end{figure}

\subsection{Interface: Basic Components}

Here, we describe the basic components of \sys{}, which structure the design process into modular nodes that users can create, edit, and iterate on—either manually, with AI assistance, or through a hybrid approach. As shown in Fig.~\ref{fig:interface}, \sys{} offers four node types (Fig.~\ref{fig:interface}D-G), each corresponding to one of the first four \textsc{phases} (\textsc{Empathize}-\textsc{Define}-\textsc{Ideate}-\textsc{Prototype}) of the 5-stage design thinking process defined by Stanford d.school~\cite{ideo}: 

\noindent\textbf{\,\raisebox{-2pt}{\includegraphics[scale=0.15]{figures/persona-icon.png}}\, Persona Node (Fig.~\ref{fig:interface}D \& \ref{fig:individual-node})} [\textsc{Empathize}]: Helps users define target users through attributes like \textsf{name}, \textsf{location}, \textsf{needs}, and \textsf{challenges}. These can be filled in manually, completely generated using AI, or partially completed by users with AI assisting to fill in remaining fields. This flexibility supports cases where designers already have established personas (e.g., from user interviews or research) that they want to incorporate directly, or where they have specific characteristics in mind but need help expanding the persona with additional details.

\noindent\textbf{\,\raisebox{-2pt}{\includegraphics[scale=0.15]{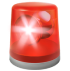}}\, Problem Node (Fig.~\ref{fig:interface}E)} [\textsc{Define}]: Allows users to define key challenges, with fields such as \textsf{context}, \textsf{stakeholders}, and \textsf{objectives}. Users can input their own insights, leverage AI to generate complete problem statements, or take a hybrid approach where they specify certain aspects and use AI to develop or refine others—providing flexibility for cases where designers have identified specific issues but need help articulating them fully.

\noindent\textbf{\,\raisebox{-2pt}{\includegraphics[scale=0.15]{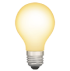}}\, Solution Node (Fig.~\ref{fig:interface}F)} [\textsc{Ideate}]: Enables users to propose solutions by outlining attributes like \textsf{problems addressed}, \textsf{key features}, and \textsf{benefits}. As with the other nodes, users can manually create solutions, generate them entirely with AI, or combine their own ideas with AI-generated content—allowing designers to start with a specific solution concept and use AI to elaborate on its details and implementation.

\begin{figure*}
    \centering
    \includegraphics[alt={Interface displaying a design storyboard. This storyboard includes (A) a title and four storyboard frames. Each frame has (B) a type (Context, Problem, Solution, and Resolution respectively), an image, and a (C) caption. There is (D) an frame edit button and (E) a plus button between each frame to add new frames. In the top right there are (F) settings to change the image style and to regenerate the images.},width=\textwidth]{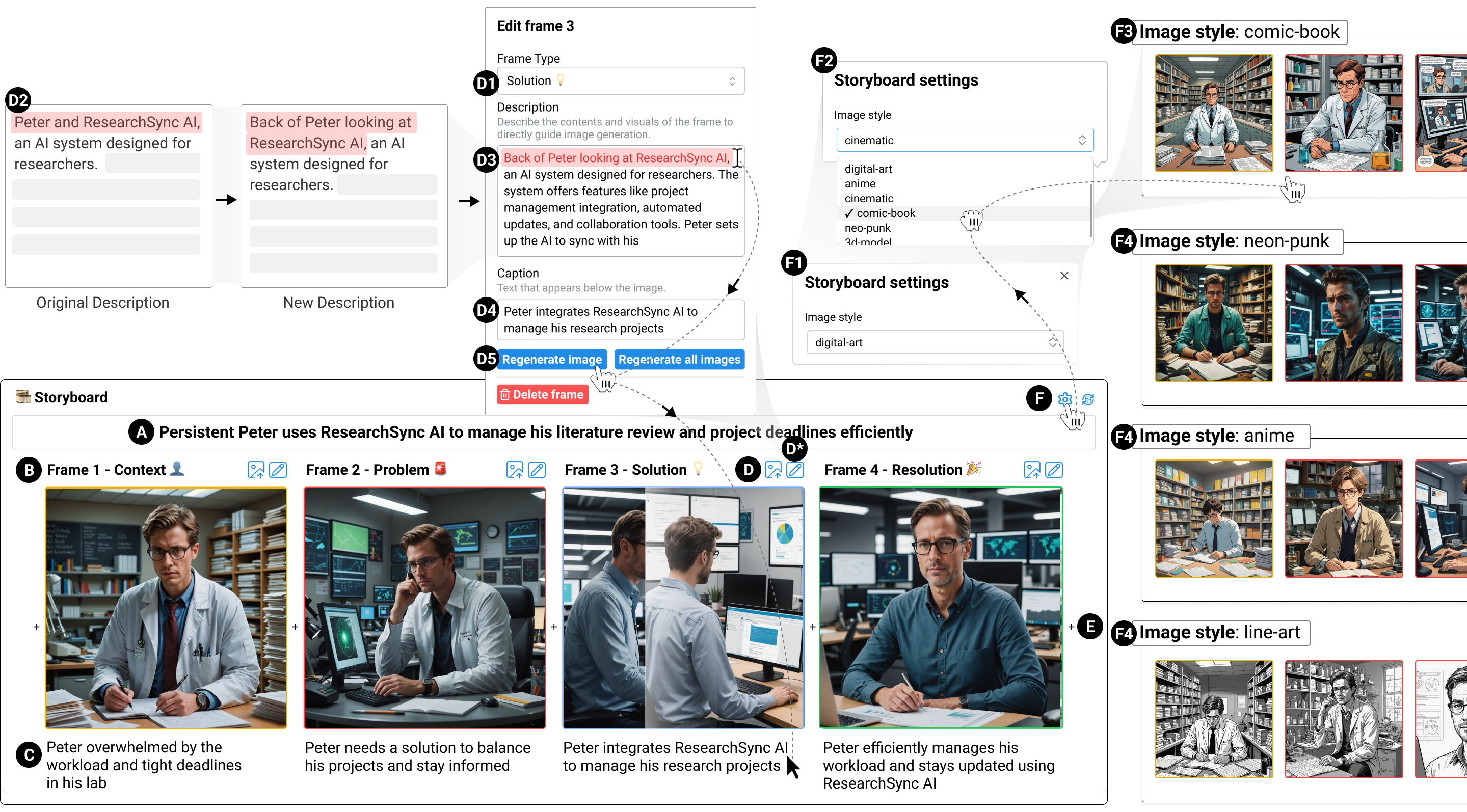}
    \caption{\textsc{Storyboard}: Each storyboard has (A) a title and (by default) four frames. Each frame represents a (B) specific type (\textsc{Context}\,\raisebox{-0.5pt}{\includegraphics[scale=0.11]{figures/persona-icon.png}}\,, \textsc{Problem}\,\raisebox{-0.5pt}{\includegraphics[scale=0.11]{figures/problem-icon.png}}\,, \textsc{Solution}\,\raisebox{-0.5pt}{\includegraphics[scale=0.11]{figures/solution-icon.png}}\,, \textsc{Resolution}\,\raisebox{-0.5pt}{\includegraphics[scale=0.11]{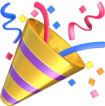}}\,) and comes with an image and a (C) caption. Users can (D) upload an image of choice to the frame or (D*) edit it: they can update the (D1) frame type, (D2$\rightarrow$D3) description used for guiding image generation, caption text (D4), and then (D5) regenerate image with new description (`\textsf{Back of Peter looking at ResearchSync AI}'). Users can also (E) add new frames to extend the narrative. The (F) storyboard settings panel enables customization of visual styles: users can choose an image style (F1$\rightarrow$F2), such as (F3) \code{comic-book} or (F4) other options like \code{neon-punk}, \code{anime}, and \code{line-art}, and regenerate all images in the selected style. The pipeline for generating the storyboard is detailed in Fig.~\ref{fig:pipeline-building-storyboard}.}
    \label{fig:storyboard}
\end{figure*}

\noindent\textbf{\,\raisebox{-2pt}{\includegraphics[scale=0.15]{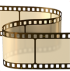}}\, Storyboard Node (Fig.~\ref{fig:interface}G, \ref{fig:storyboard})} [\textsc{Prototype}]: Enables users to create visual narratives for their design concepts, either by generating frames based on earlier stages using AI or by manually authoring them for full creative control. Fields include \textsf{frame type}, \textsf{image}, and \textsf{caption}, supporting rapid exploration and feedback collection. While prototypes can range from low-fidelity (as in this static storyboard) to high-fidelity (e.g., interactive prototypes in Figma), \sys{} focuses on low-fidelity storyboards to facilitate quick idea exploration and early feedback, aligning with the goal of providing something ``tangible enough to receive feedback in the short-term''~\cite{tham2022keywords}.

\begin{figure*}
    \centering
    \includegraphics[alt={This figure illustrates a feedback-driven system for updating design artifacts. The user selects a problem statement, `Low Financial Literacy in Young Adults' and clicks the `View feedback' button to generate a list of feedback questions, such as `What current financial literacy initiatives are in place?' and `Are there specific financial topics that young adults struggle with?' The user clicks a button to incorporate the second feedback question and inputs a response `Credit management and debt accumulation.' An updated card displays the revised title `Financial Literacy on Credit Management in Young Adults,' which integrates the feedback.},width=\textwidth]{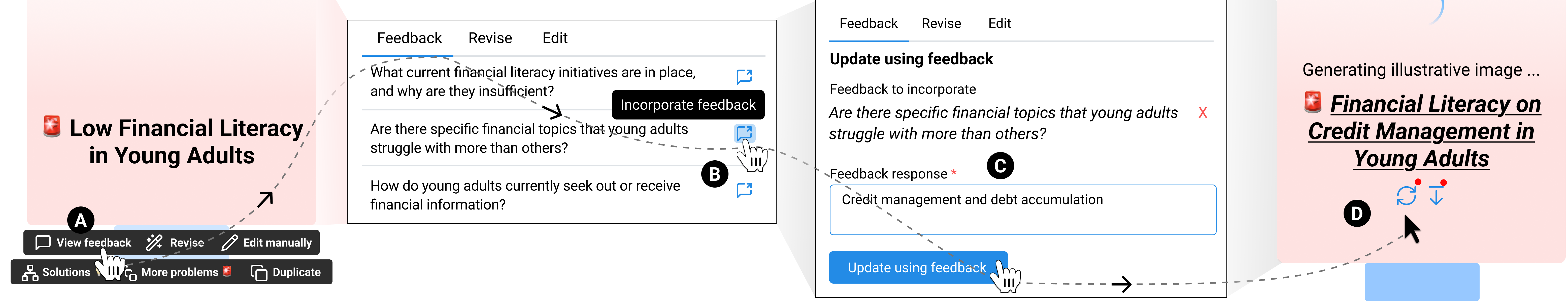}
    \caption{\textsc{View Feedback}: (A) Users can generate feedback for each node (\,\raisebox{-2.7pt}{\includegraphics[scale=0.23]{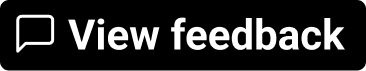}}\,) and (B) review a list of AI-generated feedback to select feedback to incorporate. (C) Users provide a response to a feedback question to (D) update the node accordingly.}
    \label{fig:view-feedback}
\end{figure*}

No specific node represents the \textsc{Test} stage, but feedback mechanisms are built into each node type to support iterative refinement (Fig.~\ref{fig:view-feedback}), which we elaborate in Sec~\ref{sec:view-feedback}.

\textit{Node Elements.} Each node features illustrative images (Fig.~\ref{fig:individual-node}A) to enhance engagement and relatability---helping facilitate empathy with personas, and making problem and solution nodes more visually compelling and easier to connect with (DG5). To maintain consistency for nodes chained together, as shown in Fig.~\ref{fig:interface}, nodes such as persona or solution nodes that are generated after their prior nodes use character description generated from prior nodes when generating illustrative images (See `\textsf{Generate visual character descriptions}' in Table~\ref{table:prompts-images} for prompt and Fig.~\ref{fig:pipeline-building-storyboard} for pipeline). 

\textit{Connection Mechanisms.} \sys{} enables users to establish dependencies between different stages of the design process through connection handles. Persona nodes have a single connection handle at the bottom, allowing them to serve as starting points that influence subsequent stages. Problem and solution nodes feature connection handles at both top and bottom, enabling them to receive input from earlier stages and propagate changes to later ones. Storyboard nodes, positioned at the end of the process, have only a top connection handle to receive input from earlier design artifacts. These handles, shown in Fig.~\ref{fig:interface}H, allow users to establish dependencies between different stages—for example, linking a persona to a problem node (\,\raisebox{-1pt}{\includegraphics[scale=0.11]{figures/persona-icon.png}}\,\,\raisebox{-2pt}{\includegraphics[scale=0.2]{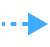}}\,\,\raisebox{-1pt}{\includegraphics[scale=0.11]{figures/problem-icon.png}}\,).

These connections serve dual purposes: they visualize relationships between design artifacts and function as pathways for propagating changes. When a node is updated, the system identifies connected nodes that might be affected and offers users the option to cascade these changes forward or backward through the dependency network. This interconnected structure ensures that modifications in one stage can be propagated throughout the design process, supporting the iterative nature of design thinking.

\begin{figure}
    \centering
    \includegraphics[alt={`Start brainstorming' form to provide designer input for AI-generated design ideas. At the top, there's a dropdown to select the `Design stage' with `Storyboard' selected. Below, a text field asks users to describe a design context. Additional optional fields include `Persona ideas,' `Problem ideas,' `Solution ideas,' and `Storyboard description.' A "Generate ideas" button is at the bottom to start the process.}, width=0.47\textwidth]{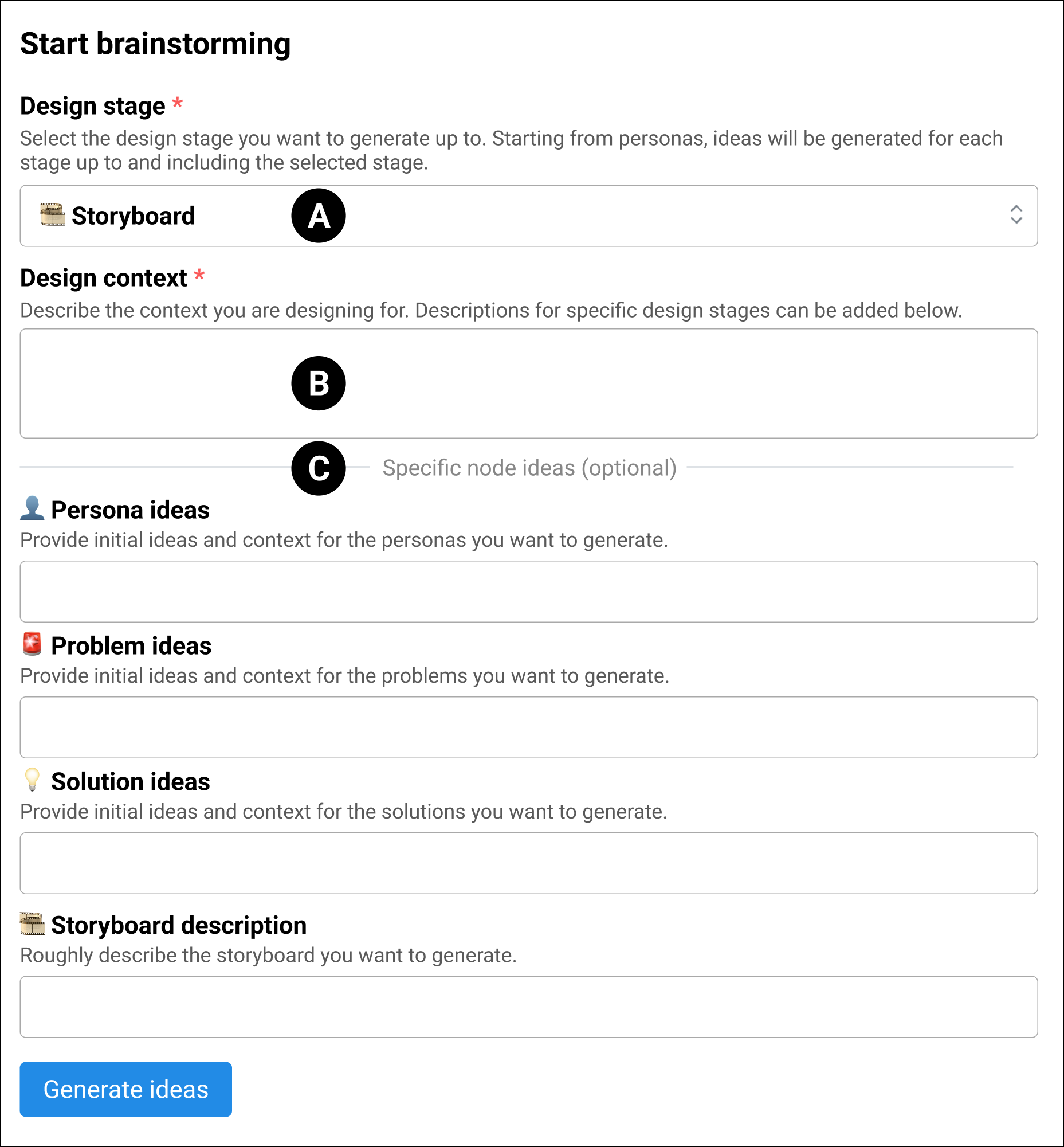}
    \caption{\textsc{Start Brainstorming}: A panel that opens when users select the\,\raisebox{-2.7pt}{\includegraphics[scale=0.23]{figures/start-brainstorming-btn.png}}\,button (Fig.~\ref{fig:interface}A). It allows users to generate and explore multiple design artifacts. Users can (A) specify the design stage (personas, problems, solutions, storyboards) they want to generate up to, (B) provide a design context text prompt, and (C) add optional context or text guidance for each stage. For example, users can add optional context like `\textsf{PhD students who publish at CHI, IUI, or UIST}' to steer the kinds of personas that get generated.}
    \label{fig:start-brainstorming}
\end{figure}

\begin{figure*}
    \centering
    \includegraphics[alt={`Generate More Ideas' feature. At the top, the (A) selected idea is `Urban Gardening Workshops.' Below, users can input new ideas in a (B) text box labeled `Solution ideas,' which contains examples like `Community Cleanup Drives' and `Sustainable Living Clinics.' (C) AI-generated suggestions are displayed as tags under the input box. At the bottom, (D) four new ideas are displayed as visual cards: `Community Clean-Up Drives,' `Sustainable Living Clinics,' `Recycling Awareness Programs,' and `Eco-Friendly Product Demos.'},width=\textwidth]{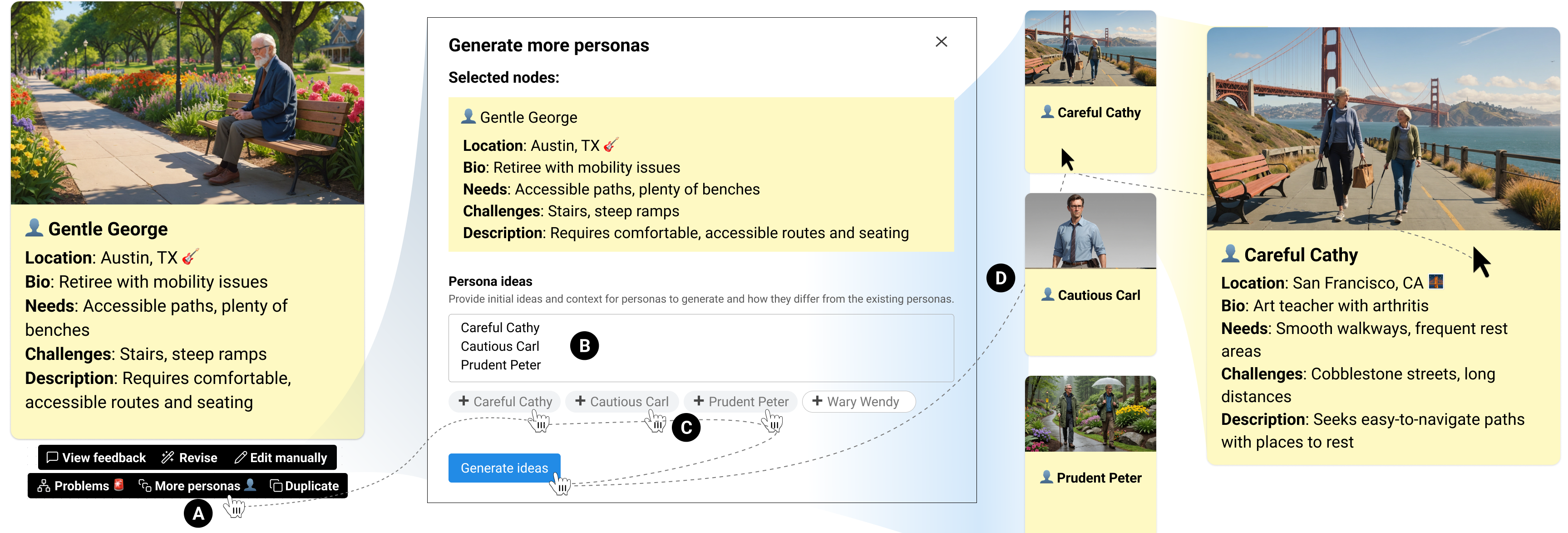}
    \caption{\textsc{Generate More}: Users can build on (A) a selected idea by (B) entering their own suggestions in natural language or (C) utilizing AI-generated recommendations. For example, a user could type a prompt like `\textsf{people who live in the same state (e.g., Austin, TX) as \textcolor{ACMOrange}{Gentle George}}' or `\textsf{...in different states (e.g., California)}' to steer the generation. The system then generates (D) additional ideas (e.g., \textcolor{ACMOrange}{\code{Careful Cathy}} from San Francisco) to support further exploration.}
    \label{fig:more-nodes}
\end{figure*}

\begin{figure}
    \centering
    \includegraphics[alt={A diagram illustrating the `Revise with AI' feature with four labeled sections. Section A displays the original node content, Section B shows a user typing a natural language prompt, Section C presents AI-generated instructions for revising the content, and Section D highlights the updated node content with changes underlined.},width=0.48\textwidth]{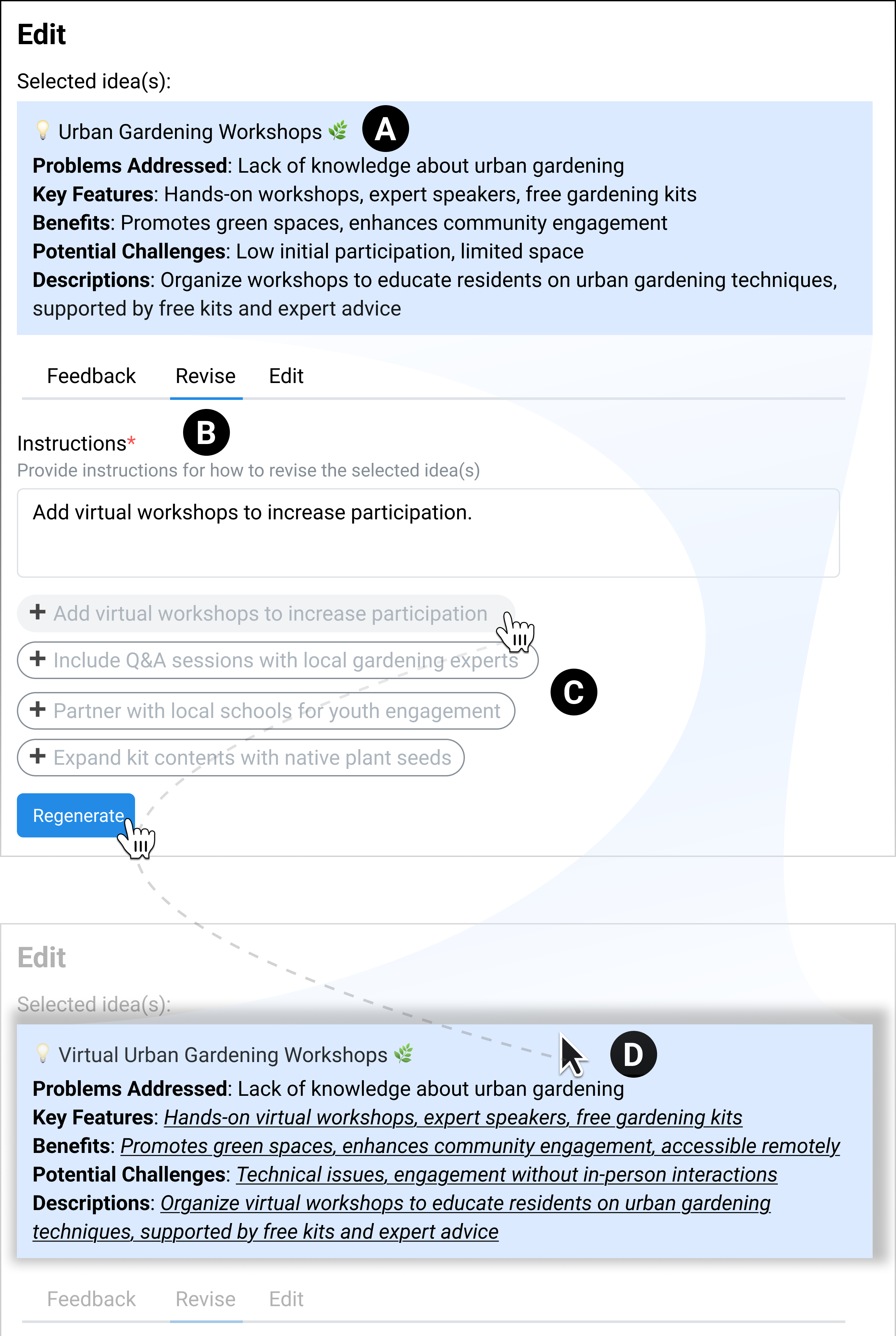}
    \caption{\textsc{Revise with AI}: Users can easily update (A) idea content by providing (B) natural language prompts (e.g., `\textsf{Add virtual workshops to increase participation}'). (C) AI suggests instructions for iteration (e.g., `\code{Partner with local schools for youth engagement}'), and (D) revised content is automatically \underline{underlined} for clarity.}
    \label{fig:revise-using-ai}
\end{figure}

\begin{figure}
    \centering
    \includegraphics[alt={Interface that allows users to manually edit a persona named `Eco Emily' with fields for location, bio, needs, challenges, and description, and a `Save' button at the bottom.}, width=0.48\textwidth]{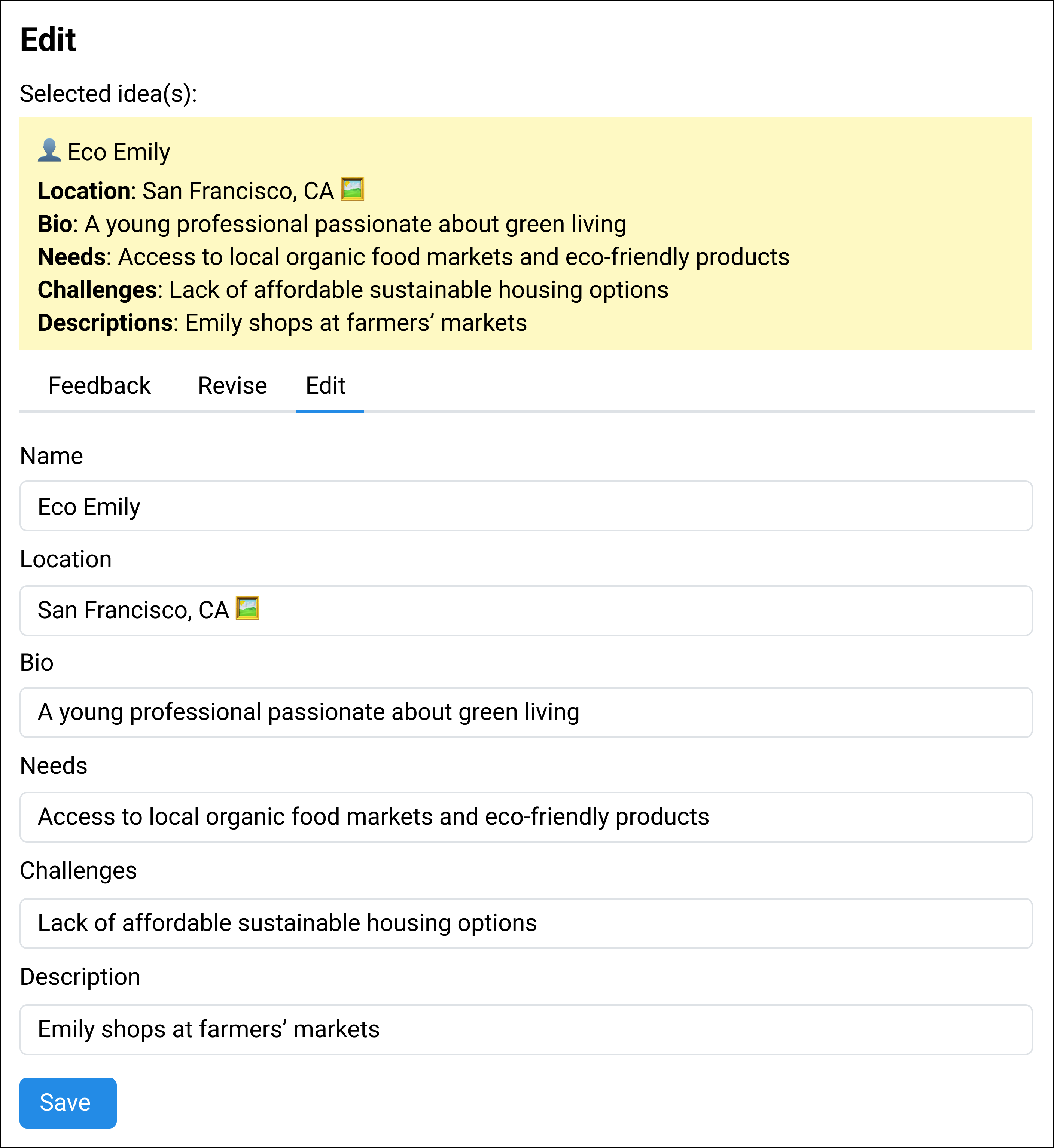}
    \caption{\textsc{Designer input}: Users can manually edit the contents of a node using an interface with text fields by selecting the\,\raisebox{-2.7pt}{\includegraphics[scale=0.23]{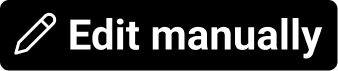}}\,button on a persona, problem, or solution node. This supports flexible engagement with AI, allowing users to rely fully, partially, or not at all on AI (DG6).}
    \label{fig:manual}
\end{figure}

\subsection{Features for Supporting Rapid Exploration and Flexible Iteration}
\label{sec:features}

We added features in \sys{} designed to support rapid exploration and flexible iteration (\textbf{DG3}). This design decision is grounded in the principle that ``design is often an interplay between ideation and iteration''~\cite{greever2015articulating, rose2022}. In \sys{}, users can iterate within and across design stages (\textbf{DG1}). 
Below, we describe the supporting features and their use cases in Sec~\ref{sec:example-scenario}.

\textit{Start Brainstorming (Fig.~\ref{fig:interface}A,~\ref{fig:start-brainstorming}; \textbf{DG5, 3}).} In the early stages of design, users often struggle to get started due to lack of knowledge~\cite{dorner1999approaching, suh2024luminate}. To assist with this, \sys{} offers a brainstorming feature where users can start from scratch or build upon initial ideas. The system leverages AI to generate diverse ideas based on minimal input, helping users overcome creative blocks and encouraging exploration of different possibilities. This feature specifically supports the divergent thinking phase of the first diamond (\textsc{Discover}) in the Double Diamond model, allowing users to broadly explore the persona-problem-solution space by generating multiple persona, problem, and solution ideas. It addresses common challenges in design practice, such as the difficulty of generating diverse ideas at the start of a project, when team members may be fixated on a few familiar personas, problems, or solutions.

\textit{Add Empty Node (Fig.~\ref{fig:interface}B; \textbf{DG3, 1}}). Tools need flexibility and support diverse contexts and user needs. This feature allows users to add any empty node and manually fill the contents in---using the edit panel shown in Fig.~\ref{fig:manual}---rather than relying on AI-generated content. This feature is designed to account for situations where users already have a persona, problem statement, solution, or storyboard design, such as those informed by prior research or user interviews. By supporting manual input, it upholds the principle of human-centered design, which emphasizes grounding decisions in engagement with end users. In practice, StoryEnsemble users can complete the persona and problem nodes based on actual user data, then leverage AI—as needed or as appropriate in a given context—to explore potential solution ideas and generate storyboards to communicate their design concepts.
    
\textit{Generate More (Fig.~\ref{fig:more-nodes}; \textbf{DG5,3}).} One of the key strengths of generative AI is its ability to rapidly produce new ideas~\cite{suh2024luminate}. To ease the ideation process, \sys{} features a \textsc{Generate More} option that allows users to seamlessly generate additional ideas or variations of existing ones, as shown in Fig.~\ref{fig:more-nodes}. This feature ensures that users are not limited by their initial ideas and can explore beyond them. Users can use this feature for all node types except for storyboard. It helps users efficiently explore the persona-problem-solution spaces and discover ideas they might have difficulty thinking about before converging on a specific direction, as prescribed in the \textsc{Develop} phase of the Double Diamond model (Fig.~\ref{fig:teaser}). This sustains divergent thinking beyond initial ideation, encouraging users to push beyond their perspectives.

\textit{Revise with AI (Fig.~\ref{fig:revise-using-ai}; \textbf{DG1}).} This feature allows users to provide any instruction as prompt to update the content of any node. Users can open the panel (Fig.~\ref{fig:revise-using-ai}) by selecting any node and clicking the\,\raisebox{-2.7pt}{\includegraphics[scale=0.23]{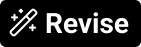}}\,button, as shown in Fig.~\ref{fig:individual-node}D. \sys{} intelligently supports partial completion workflows: when a user manually adds data to certain fields but leaves others empty, the Revise with AI feature automatically recognizes these gaps. Upon switching to the Revise tab in the popup, the system adds `\textsf{Fill in missing values}' to the instruction suggestions. This allows users to quickly complete their partially defined nodes without having to explicitly request or formulate this common task. For example, if a designer has specific demographic information about a persona from research but lacks insights about their challenges, they can simply select this auto-generated instruction to have AI suggest appropriate challenges based on the existing persona attributes.

\textit{Generate Next Node (Fig.~\ref{fig:individual-node}d)}. After selecting any node (except storyboard), users can use the Generate Next Node option from the contextual toolbar to create several downstream nodes. For instance, selecting it from a persona node—\,\raisebox{-2.7pt}{\includegraphics[scale=0.23]{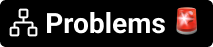}}\,as in Fig.~\ref{fig:individual-node}d—allows the user to generate related problem nodes, while selecting it from a problem node can yield multiple solution ideas. This feature enables users to extend their ideas efficiently. By allowing generation from any stage in the process, this feature encourages rapid exploration.
    
\textit{View Feedback (Fig.~\ref{fig:view-feedback}, Iterating \underline{within} Design Stage; \textbf{DG2}).}\label{sec:view-feedback} \sys{} enables users to gather AI-generated feedback for refining individual nodes, such as personas, problem statements, solutions, and storyboards. When feedback is incorporated and the content of a node is updated, the system recognizes the potential impact on other connected nodes. For example, revising a problem statement may necessitate changes to the personas or solutions linked to that problem. The system provides a visual cue (\,\raisebox{-2pt}{\includegraphics[scale=0.25]{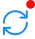}}\,), signaling users that updates in one node could affect related nodes. Users are made aware of these dependencies through icons that appear on the connected nodes, allowing them to decide whether to update the content in those nodes. This ensures that feedback-driven changes flow naturally within a design stage while maintaining consistency across the artifacts. This feedback mechanism supports the transition toward convergent thinking phases (\textsf{Define} and \textsf{Deliver} in the Double Diamond model), enabling users to evaluate and narrow down options based on critical assessment of their ideas. The AI-generated feedback provides diverse perspectives that might otherwise require extensive user research or stakeholder consultation, making the convergent phases more thorough and evidence-based.

\begin{figure*}
    \centering
    \begin{subfigure}[t]{0.48\textwidth}
    \includegraphics[alt={}, trim=0cm 0cm 0cm 0cm, clip=true, width=1\columnwidth]{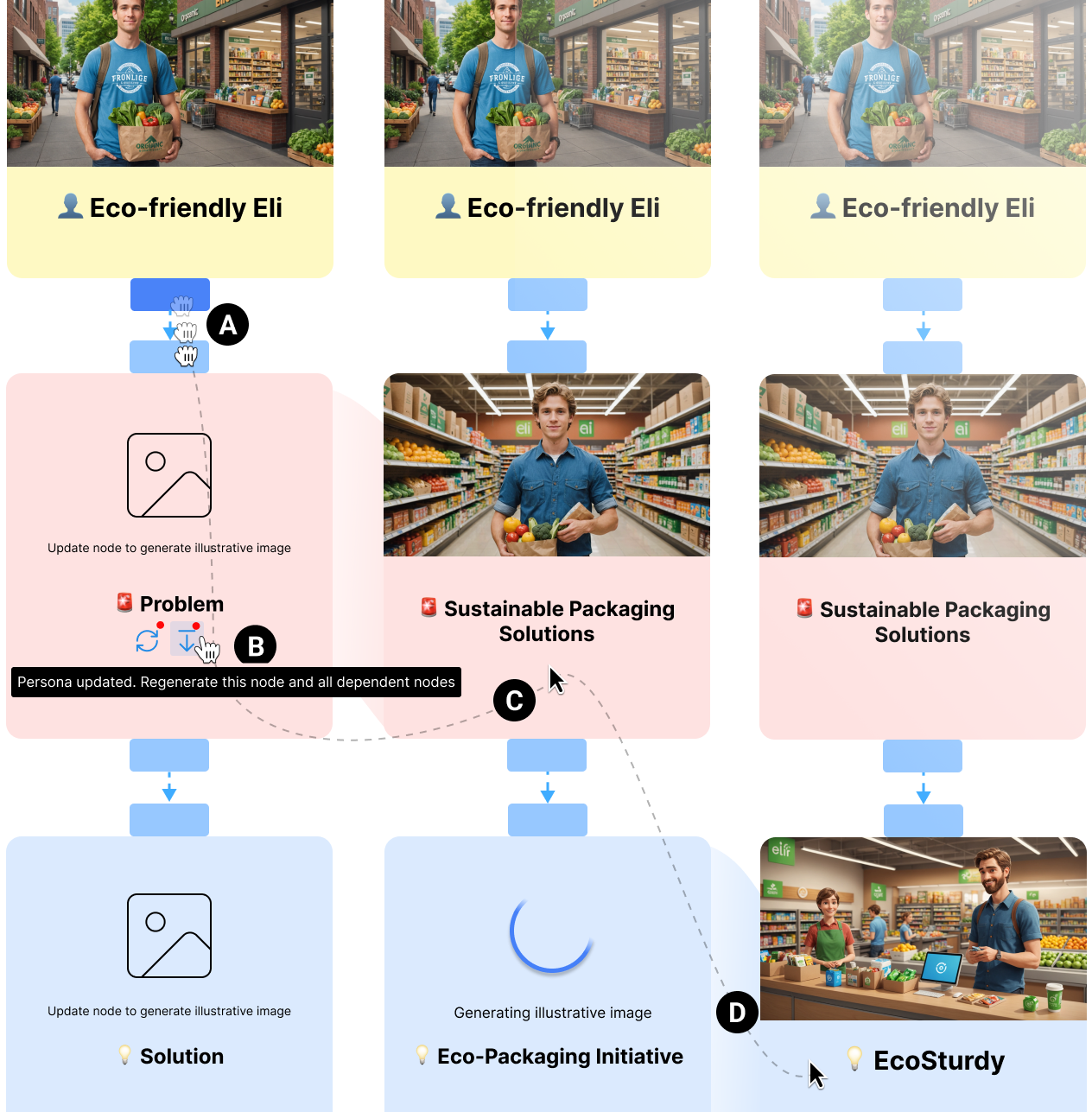}
    \caption{Forward propagation: Users can (A) connect from a upstream node to downstream node (e.g., persona$\rightarrow$problem), and then (B) click forward prop icon (\,\raisebox{-2pt}{\includegraphics[scale=0.25]{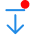}}\,) to trigger (C$\rightarrow$D) updates to later stages.}
    \label{fig:forward-propagation}
    \end{subfigure}
    \hfill
    \centering
    \begin{subfigure}[t]{0.48\textwidth}
        \includegraphics[alt={}, trim=0cm 0cm 0cm 0cm, clip=true, width=\textwidth]{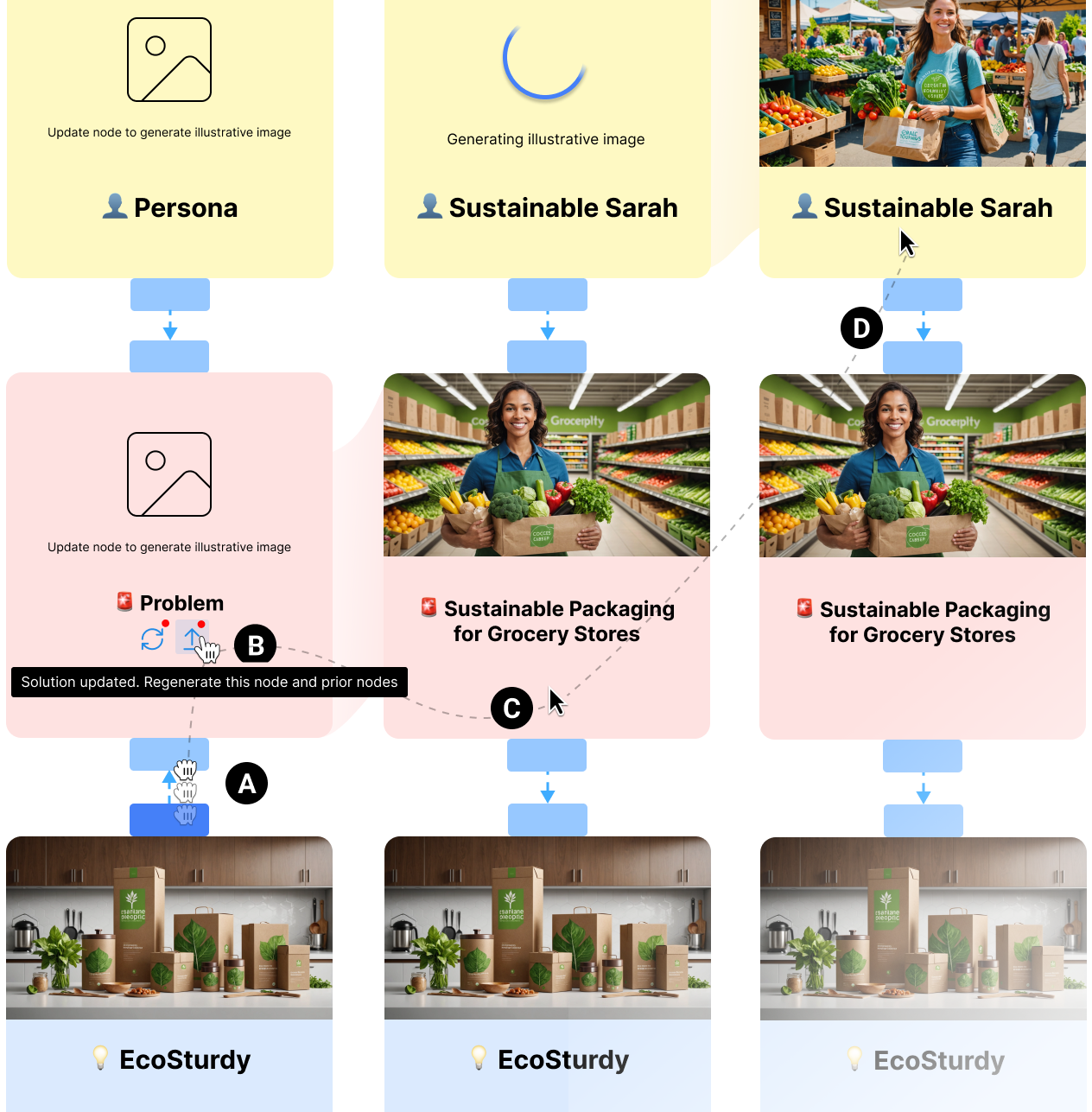}
        \caption{Back propagation: Users can (A) connect from a downstream node to upstream node (e.g., solution$\rightarrow$problem), and then (B) click back prop icon (\,\raisebox{-2pt}{\includegraphics[scale=0.25]{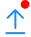}}\,) to trigger (C$\rightarrow$D) updates to earlier stages.}
        \label{fig:backpropagation}
    \end{subfigure}
    \caption{StoryEnsemble supports forward and backward propagation. Propagation can be triggered either by user-initiated handle connections between nodes (shown above) or by making changes to an existing node. Users can freely connect any node—upstream or downstream—to another. This enables users to fill in missing content, maintain coherence across stages, and iterate fluidly without being constrained by a fixed direction or sequence.}
    \Description[short description]{}
    \label{fig:propagation-mechanisms}
\end{figure*}

\textit{Cascading Changes (Fig.~\ref{fig:propagation-mechanisms}, Iterating \underline{Across} Design Stages; \textbf{DG3, 1}).} In addition to supporting feedback-driven iterations within a stage, \sys{} facilitates the propagation of changes across different stages of the design process. When a node is updated—via manual editing (Fig.~\ref{fig:manual}), AI-assisted revision (Fig.~\ref{fig:revise-using-ai}), feedback incorporation (Fig.~\ref{fig:view-feedback}), or creating new connections between nodes (Fig.~\ref{fig:propagation-mechanisms})—the system identifies how these changes may influence other nodes connected to that stage. For example, modifying a persona may prompt updates to the problem statement and solution nodes that are dependent on it.

Whenever a node is updated via any of the four interactions, \sys{} displays `\textit{update}' (\,\raisebox{-2pt}{\includegraphics[scale=0.25]{figures/icon_update.png}}\,) and `propagate change' (forward propagation: \,\raisebox{-2pt}{\includegraphics[scale=0.25]{figures/icon_forwardprop.png}}\,; backward propagation: \,\raisebox{-2pt}{\includegraphics[scale=0.25]{figures/icon_backprop.png}}\,) icons that appear on affected nodes, as shown in Fig.~\ref{fig:propagation-mechanisms}. These visual indicators alert users that a change in one area of the design may require adjustments in related stages. Unlike in the \textsc{View Feedback} feature, where users can refine individual nodes within a stage, cascading changes allow for holistic updates across multiple stages of the design process. Users can choose to propagate changes \textit{all the way to the last connected node} (\,\raisebox{-2pt}{\includegraphics[scale=0.25]{figures/icon_forwardprop.png}}\,) or \textit{to the first node in the chain} (\,\raisebox{-2pt}{\includegraphics[scale=0.25]{figures/icon_backprop.png}}\,) or manually \textit{update the connected nodes one by one} (\,\raisebox{-2pt}{\includegraphics[scale=0.25]{figures/icon_update.png}}\,), ensuring flexibility and control over the iterative process. By highlighting potential interdependencies between design stages, \sys{} reduces the manual effort required to maintain consistency, allowing users to focus on creative exploration and iteration without losing track of critical connections between artifacts.

    \textit{Multi-Node Selection and Batch Operation (Fig. \ref{fig:multi-node-selection}; \textbf{DG5, DG4, DG1}).} 
    \sys{} enables users to select multiple nodes simultaneously and perform operations on them as a batch. For example, users can select multiple persona nodes and generate corresponding problem nodes that take all selected personas as context, allowing for efficient exploration of problem spaces across different user types. Similarly, multiple problem nodes can be selected to generate diverse solutions addressing various challenges. This capability extends to feedback as well, where users can request feedback on multiple nodes at once to evaluate them comparatively. By supporting batch operations, \sys{} streamlines the workflow when exploring diverse design directions, enabling users to quickly branch out and consider multiple pathways through the design space without having to perform repetitive individual actions.

    \textit{Semantic Zoom (Zoomed Out - Fig.~\ref{fig:interface}; In - Fig.~\ref{fig:individual-node} | \textbf{DG4}).} As users generate multiple personas, problems, solutions, and storyboards, the interface can become complex and potentially overwhelming. To address this challenge, \sys{} implements semantic zoom functionality that adapts content visibility based on zoom level. When zoomed out, nodes display only their titles—as in Fig.~\ref{fig:interface}—while still allowing access to full content via hover, which reveals a detailed popup window, eliminating the need to zoom in. This supports rapid exploration of the design space by allowing users to generate and compare many ideas without visual clutter. By providing an overview while preserving access to details, \sys{} enables users to maintain awareness of their full idea landscape while focusing on specific areas of interest. This aligns with both divergent thinking principles—supporting the generation and navigation of many possibilities—and convergent thinking phases where users need to compare and evaluate options efficiently across the design space.

    \begin{figure*}
    \label{fig:multi-node-selection}
    \centering
    \begin{subfigure}[t]{0.32\textwidth}
        \includegraphics[width=\linewidth]{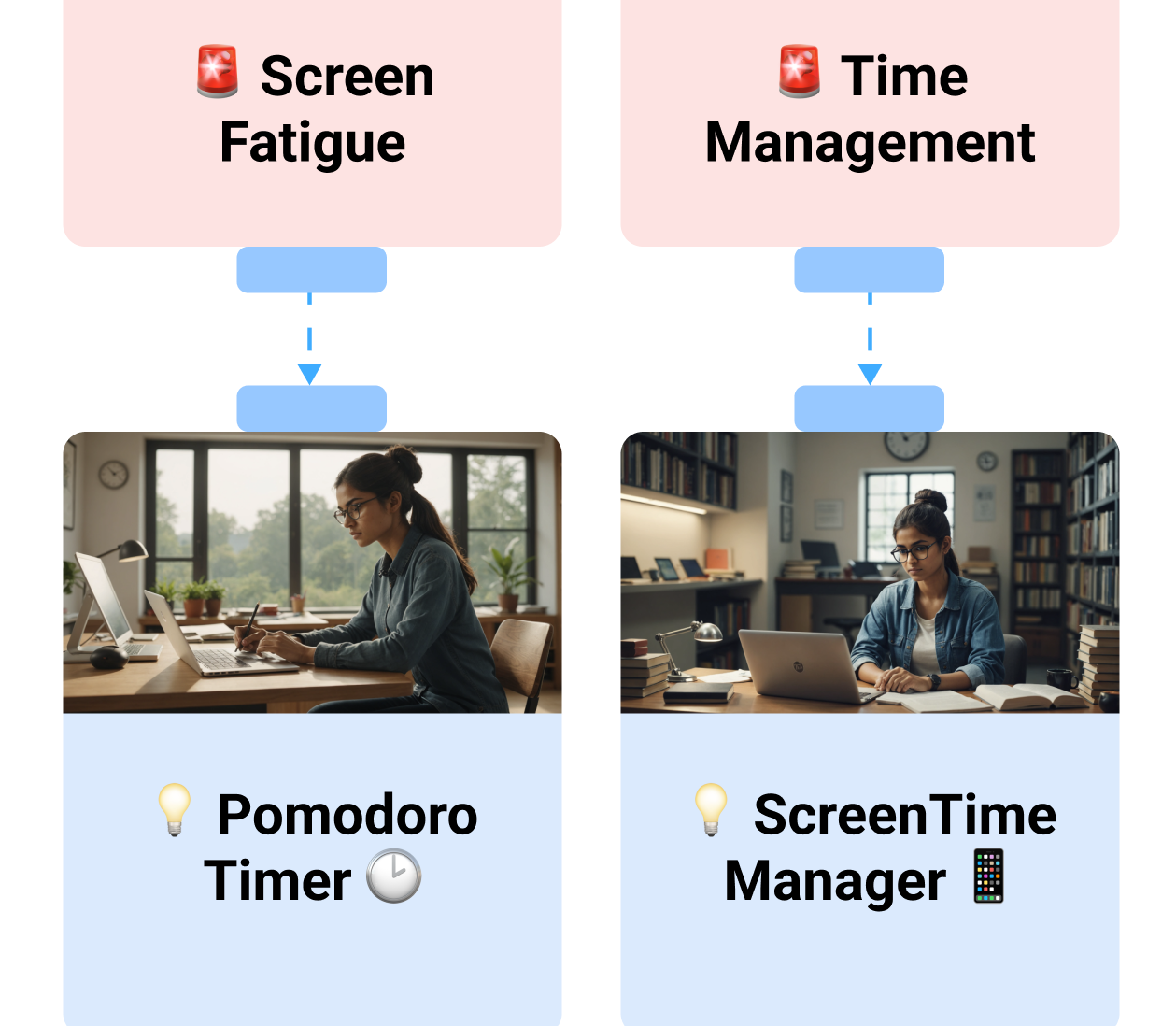}
        \caption{One-to-One: Using the \textsc{Start Brainstorming} feature results in multiple one-to-one pairs (e.g., Fig.~\ref{fig:interface}).}
        \label{fig:one-to-one}
    \end{subfigure}
    \hfill
    \begin{subfigure}[t]{0.32\textwidth}
        \includegraphics[width=\linewidth]{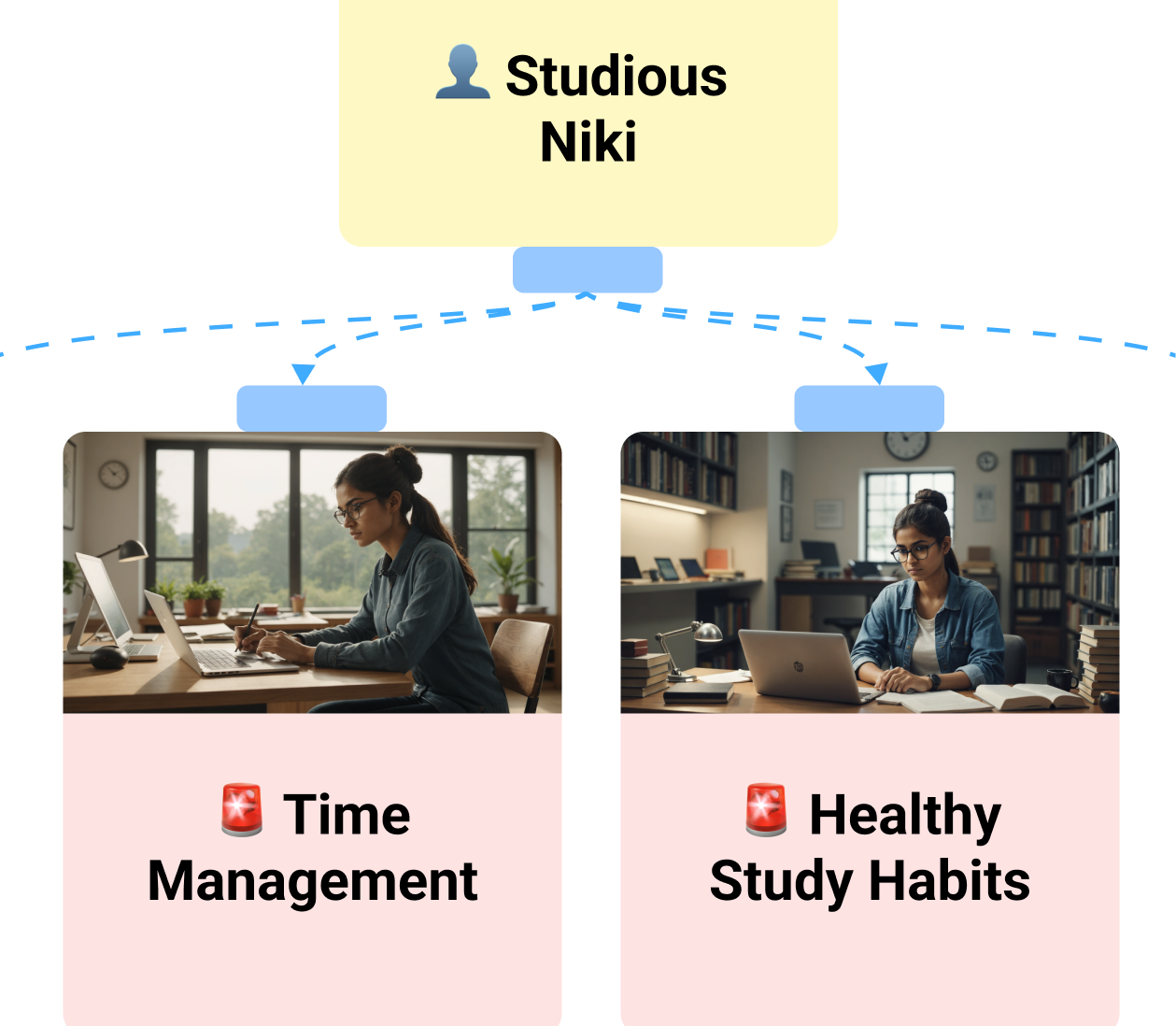}
        \caption{One-to-Many: Using toolbar's \textsc{Generate Next Node} feature (e.g., Fig.~\ref{fig:individual-node}d) from a single node creates multiple downstream nodes.}
        \label{fig:one-to-many}
    \end{subfigure}
    \hfill
    \begin{subfigure}[t]{0.32\textwidth}
        \includegraphics[width=\linewidth]{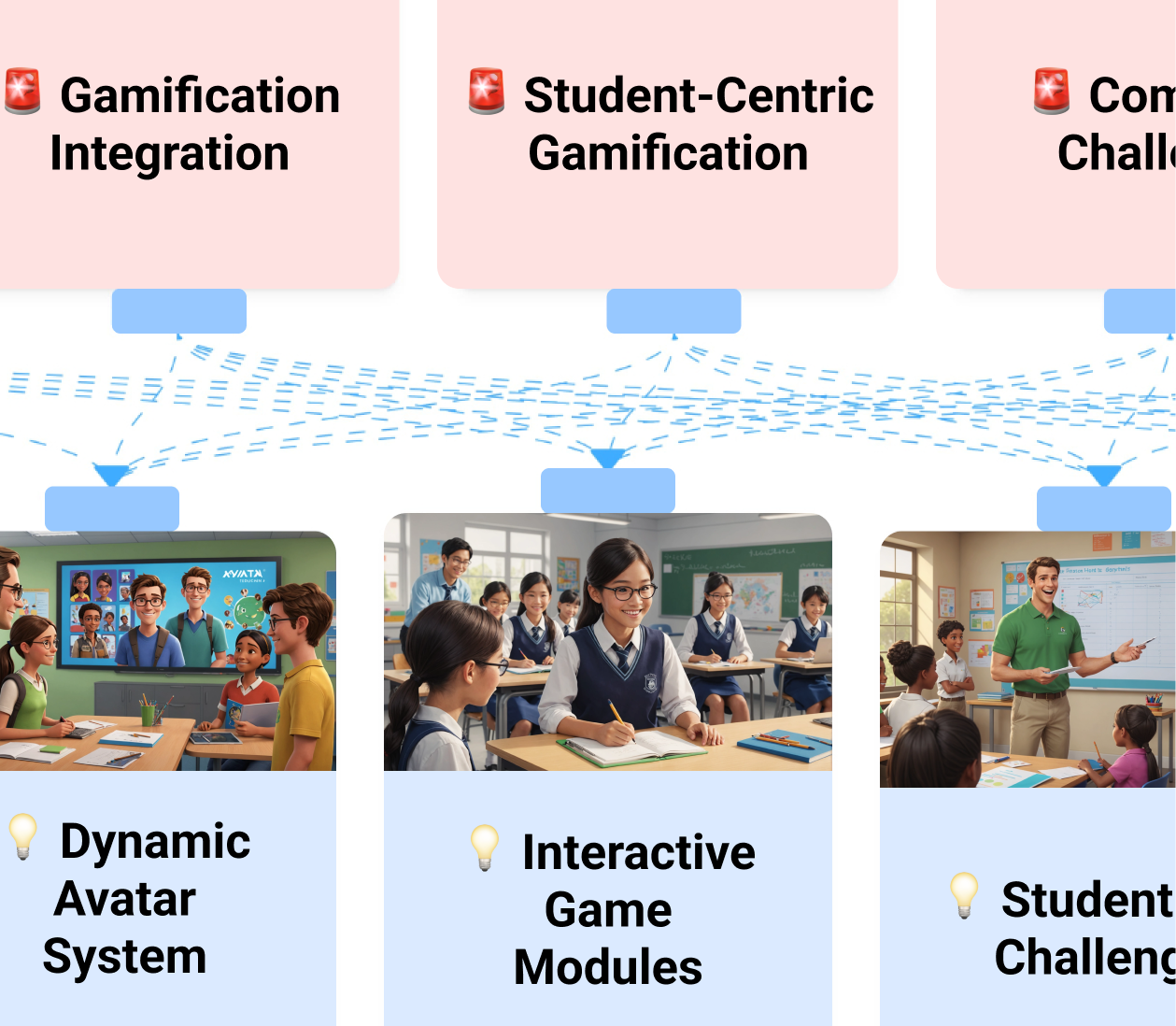}
        \caption{Many-to-Many: Selecting multiple nodes and using toolbar's \textsc{Generate Next Node} feature yields multiple downstream nodes, each generated using the selected nodes.}
        \label{fig:many-to-many}
    \end{subfigure}
    \caption{Three node-to-node patterns that showcase different workflows and exploration strategies.}
    \Description[short description]{Three panels showing different multi-node selection patterns and corresponding batch operations in a node-based design tool.}
    \label{fig:multi-node-selection}
\end{figure*}

\subsection{Example Scenario: Designing a Sustainable Packaging Solution}
\label{sec:example-scenario}

Alex, an eco-conscious freelance designer, is using \sys{} to create a \textsf{sustainable packaging solution for grocery stores}. As Alex progresses through the design process, he utilizes both divergent and convergent thinking tasks, guided by the system’s features. 
Alex's journey through this design process demonstrates how StoryEnsemble supports the complete design cycle while following the Double Diamond model's structure. Throughout this scenario, Alex moves through both diamonds of the process: first engaging in divergent thinking to explore multiple personas (\textsc{Discover}) before converging on \,\raisebox{-1pt}{\includegraphics[scale=0.11]{figures/persona-icon.png}}\,\textcolor{ACMOrange}{\textsf{Sustainability-Seeking Sara}} (\textsc{Define}), then diverging again to consider various solutions (\textsc{Develop}) before converging on \,\raisebox{-1pt}{\includegraphics[scale=0.11]{figures/solution-icon.png}}\,\textcolor{ACMDarkBlue}{\textsf{Plant-based Packaging}} (\textsc{Deliver}). StoryEnsemble supports this natural flow between expansive exploration and focused refinement, demonstrating how generative AI can operationalize core design principles that are often difficult to implement fully in practice due to time and resource constraints.

\textbf{\textsc{Empathize} Stage (\,\raisebox{-1pt}{\includegraphics[scale=0.11]{figures/persona-icon.png}}\,Persona Node) - Divergent Task}.
Alex begins by brainstorming types of users who would benefit from sustainable packaging. Using the AI-generated persona feature, he explores a range of possibilities and settles on \,\raisebox{-1pt}{\includegraphics[scale=0.11]{figures/persona-icon.png}}\,\textcolor{ACMOrange}{\textsf{Sustainability-Seeking Sara}}, a young professional who cares about the environment but struggles to find eco-friendly packaging at her local grocery store. The system helps Alex define \textcolor{ACMOrange}{\textsf{Sara}}’s attributes, such as her environmental concerns, grocery shopping habits, and desire for sustainable products. This divergent phase encourages Alex to explore various user profiles before selecting one that resonates with the problem space.

\textbf{\textsc{Define} Stage (\,\raisebox{-1pt}{\includegraphics[scale=0.11]{figures/problem-icon.png}}\,Problem Node) - Convergent Task}.
Alex needs to focus on clarifying the core problem that his solution will address. Using the problem node, Alex enters insights based on \textcolor{ACMOrange}{\textsf{Sara}}'s persona, allowing the AI to generate potential problem statements. The chosen problem is: \,\raisebox{-1pt}{\includegraphics[scale=0.11]{figures/problem-icon.png}}\,\textcolor{ACMRed}{\textsf{Limited availability of sustainable packaging options in grocery stores}}. The \textsc{View Feedback} feature prompts Alex to refine the problem further by considering suggestions, such as including the economic challenges of transitioning to sustainable packaging. This phase represents a convergent task, as Alex narrows down and refines the problem into a clear, actionable statement.

\textbf{\textsc{Ideate} Stage (\,\raisebox{-1pt}{\includegraphics[scale=0.11]{figures/solution-icon.png}}\,Solution Node) - Divergent Task}.
In the ideation phase, Alex tries to explore as many potential solutions as possible by using the \textsc{Generate More} feature.  To guide solutions towards environmentally friendly packaging options, Alex inputs `\textsf{biodegradable packaging}' as a guiding prompt. StoryEnsemble returns several AI-generated alternatives, including\,\raisebox{-1pt}{\includegraphics[scale=0.11]{figures/solution-icon.png}}\,\textsf{\textcolor{ACMDarkBlue}{Reusable Packaging Programs}} and\,\raisebox{-1pt}{\includegraphics[scale=0.11]{figures/solution-icon.png}}\,\textsf{\textcolor{ACMDarkBlue}{Plant-based Packaging}}. This stage is all about divergent thinking, as Alex explores multiple avenues and evaluates which solutions are most viable. After reviewing the options, Alex decides to focus on\,\raisebox{-1pt}{\includegraphics[scale=0.11]{figures/solution-icon.png}}\,\textsf{\textcolor{ACMDarkBlue}{Plant-based Packaging}} and begins fleshing out the key features and benefits of this solution.

\textbf{\textsc{Prototype} Stage (\,\raisebox{-1pt}{\includegraphics[scale=0.11]{figures/storyboard-icon.png}}\,Storyboard Node) - Convergent Task}.
Now in the prototype stage, Alex shifts to a convergent task by focusing on turning his solution into a tangible concept. Using the storyboard node, Alex creates a visual narrative to illustrate \textcolor{ACMOrange}{\textsf{Sara}}’s journey—from struggling to find sustainable packaging, discovering the plant-based packaging solution, and adopting it for her groceries. Alex uses \textsc{Revise with AI} to tweak the storyboard frames, ensuring the visuals and captions align with the design goals. This low-fidelity prototype allows Alex to quickly generate a storyboard that communicates the solution’s value and gathers early feedback.

\textbf{Iteration and Feedback Collection – Divergent and Convergent Tasks}.
Throughout the process, Alex engages in both divergent and convergent tasks. The \textsc{Forward Propagation} feature helps Alex ensure that updates to \textcolor{ACMOrange}{\textsf{Sara}}’s persona (e.g., adding her involvement in a community recycling program) are reflected in other stages like the problem statement and solution nodes. The\,\raisebox{-2.7pt}{\includegraphics[scale=0.23]{figures/view-feedback-icon.png}}\,feature allows Alex to gather diverse feedback at each stage, a divergent task, but then Alex converges by incorporating that feedback into a refined version of the solution. The Cascading Changes feature ensures consistency across the nodes, making it easy for Alex to iterate without losing coherence.

\textbf{Final Stage: Conclusion and Refinement – Convergent Task}.
In the final phase, Alex reviews the project holistically. By using Cascading Changes, Alex ensures that the connections between the persona, problem, and solution are solid. This final convergence solidifies Alex's design, making sure that the artifacts work together to support a cohesive, user-centered solution.

\subsection{Implementation Details}

\sys{} is built as a web application using the React framework. The interactive canvas is implemented with ReactFlow, an open-source library for building node-based interfaces. For AI-powered features, the system integrates OpenAI's API for text generation and Stability for image generation. Each node tracks its dependency changes, and the prompts used in the system—including those for regenerating node content—are listed in Appendix~\ref{sec:pipeline-n-prompts}.

\section{User Evaluation}
\label{sec:evaluation}

To assess the perceived usefulness of \sys{}, we conducted a user study. Specifically, we aimed to understand: (1) how \sys{} supports workflows inspired by core design principles such as iteration, divergence and convergence, and scenario-ased reasoning, (2) the usefulness of \sys{}'s features, and (3) potential applications of \sys{}, especially for learning, teaching, and using the iterative, exploratory design process. 

\textit{Study Setup \& Rationale.} Given the novel aspects of \sys{}, we opted for an exploratory study, a method increasingly used to evaluate AI systems offering novel experiences~\cite{suh2023sensecape, brade2023promptify, jiang2023graphologue, suh2024luminate}. Like recent systems, \sys{} introduces a variety of features (as shown in Sec~\ref{sec:system}), including innovative ones like backward propagation (Fig.~\ref{fig:backpropagation}). This made an exploratory approach well-suited for examining the system’s broader potential and the perceived utility of each feature. Another key motivation was to understand how \sys{} could support diverse user groups—students, instructors, and practitioners—across a variety of design methodologies and workflows. The open-ended nature of this inquiry reinforced the need for an exploratory study; a more narrowly defined, hypothesis-driven approach (requiring a baseline) would have limited our analysis and prevented us from a broader examination of the system's flexibility, usefulness, and overall potential. While a hypothesis-driven approach is equally valuable, we see it as more appropriate as a next step—to build on the findings from this exploratory work to form hypotheses and evaluate them in a more controlled setting.

\subsection{Procedure}

\begin{figure}[htb!]
    \centering
    \includegraphics[alt={Study procedure}, width=0.48\textwidth]{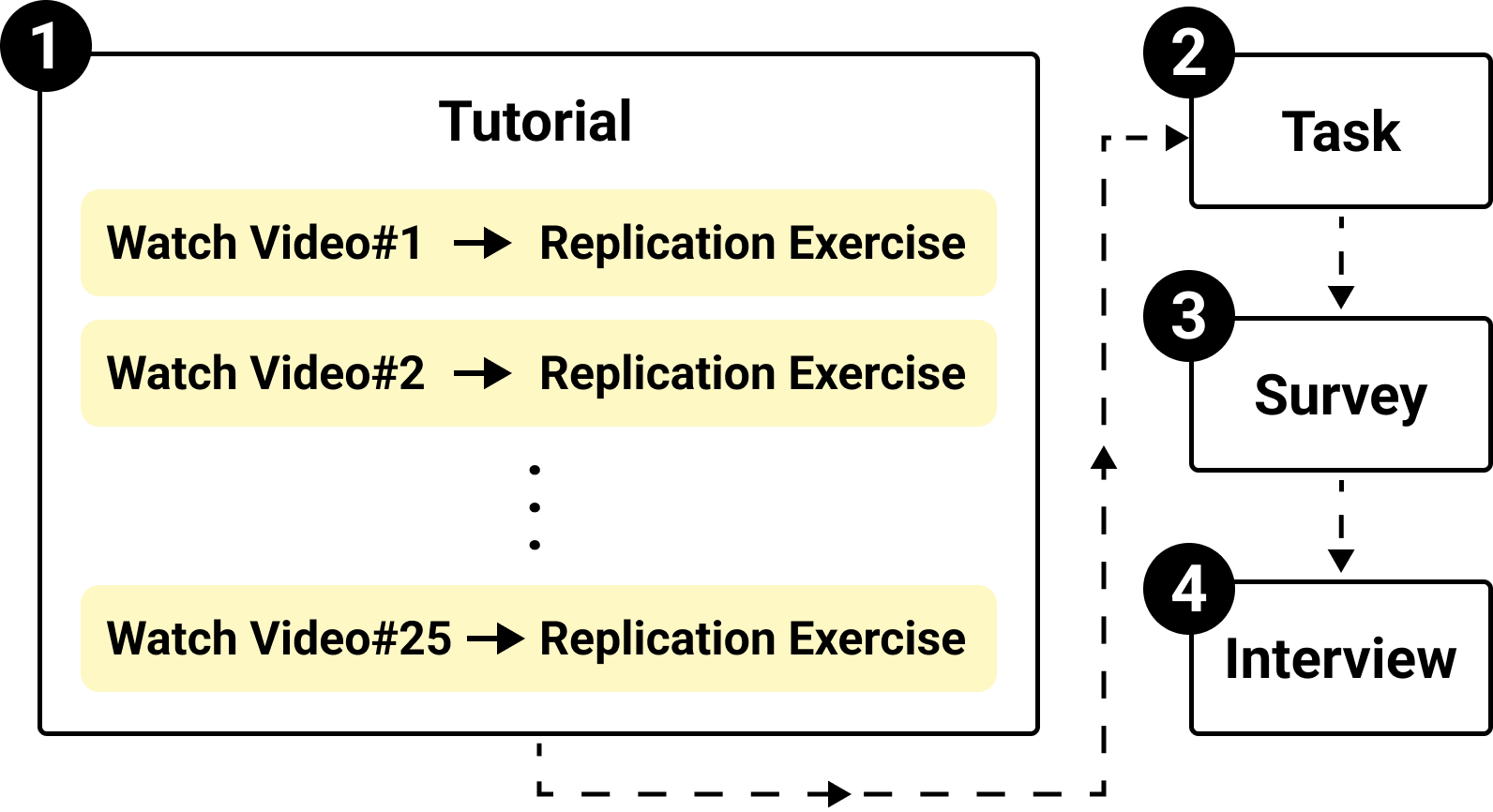}
    \caption{User study procedure}
    \label{fig:study-procedure}
\end{figure}

The 90- to 120-minute evaluation study (Fig.~\ref{fig:study-procedure}) was conducted in four main phases: tutorial, task, survey, and interview. Each phase was designed to guide participants through the process and gather insights on their experiences and interactions.

\textit{Tutorial (40-60 min).} The study began with a self-paced tutorial session to familiarize participants with the system and its features. For consistency across participants, they were asked to watch a set of pre-recorded tutorial videos (20+ videos totaling 37.5 min) demonstrating how to use \sys{} for various tasks, such as idea generation and iteration within the design process. Participants were instructed to replicate the actions shown in the videos.

\textit{Task (20-30 min).} After the tutorial, participants were asked to complete a task using the system while engaging in the think-aloud. The task involved selecting a topic of their choice, with a list of pre-selected topics (Table~\ref{table:task-topics}) provided for reference, though they were also free to choose a topic outside of the list. We instructed participants to iterate on their designs at least once, in order to gather insights into their experience using the system’s iteration features and understand how and why they chose to use them. This session lasted approximately 20–30 minutes, depending on the complexity of the task and participant engagement.

\textit{Survey (15 min).} Once participants completed the task, they were given a brief survey to gather quantitative feedback on their experience. The survey included questions from the Creativity Support Index~\cite{cherry2014quantifying} (CSI) and System Usability Scale~\cite{bangor2009determining} (SUS) questionnaires, as well as questions about the perceived effectiveness of individual features. To provide participants with a familiar frame of reference, we referred to the design process as design thinking in some questions, similar to our approach in the formative study.

\textit{Interview (15 min).} Finally, participants took part in a semi-structured interview to provide more in-depth qualitative feedback. The interview focused on participants’ perceptions of the system’s strengths and weaknesses, the ease of use of the AI tools, and how effectively the system supported their design process.  As with the survey, we occasionally referred to \textit{design thinking} when probing participants’ process, in order to facilitate discussion and align with a commonly recognized design framework. Participants were also asked about any challenges they encountered during the task and their thoughts on how the system could be improved.

\subsection{Participants}
We recruited 10 participants (age: M = 34.6, SD = 10.9, range = [24, 60]; gender: 5F, 5M) who represented the perspectives of three user groups: 4 students', 5 instructors' (M = 8 years of experience, SD = 8.1, range: [1, 20]), and 6 practitioners' (M = 5.2 years of experience, SD = 2.6, range: [3, 10]). Throughout the paper, as shown in Table~\ref{tab:participant_info}, we refer to participants as P1–P10, using subscripts to indicate their current profession: student (stud), instructor (inst), or practitioner (prac). For example, P4\textsubscript{stud} or P10\textsubscript{inst}.

\begin{table}[h]
\centering
\begin{tabular}{llcl}
\toprule
\textbf{} & \makecell[l]{\textbf{Profession}} & \textbf{Age} & \makecell[l]{\textbf{Years of Experience}} \\
\midrule
\multirow{1}{*}{P1} & Practitioner & 
\multirow{1}{*}{24} & 3 years \\

\midrule

\multirow{1}{*}{P2} & Practitioner & \multirow{1}{*}{32} & 5 years \\

\midrule

\multirow{1}{*}{P3} & Student & \multirow{1}{*}{26} & 1 year \\

\midrule

\multirow{3}{*}{P4} & & \multirow{3}{*}{29} & 2 years \\
& Student & & + 1 year (Instructor) \\
& & & + 3 years (Practitioner) \\

\midrule

\multirow{1}{*}{P5} & Practitioner & \multirow{1}{*}{31} & 5 years \\

\midrule
\multirow{1}{*}{P6} & Practitioner & \multirow{1}{*}{37} & 10 years \\

\midrule
\multirow{2}{*}{P7} & \multirow{2}{*}{Student} & \multirow{2}{*}{28} & 1 year \\
& & & + 1 year (Instructor) \\

\midrule
\multirow{2}{*}{P8} & \multirow{2}{*}{Instructor} & \multirow{2}{*}{46} & 12 years \\
& & & + 5 years (Practitioner) \\

\midrule
\multirow{1}{*}{P9} & Instructor & \multirow{1}{*}{60} & 20 years \\

\midrule
\multirow{2}{*}{P10} & \multirow{2}{*}{Instructor} & \multirow{2}{*}{33} & 6 years \\
& & & + 1 year (Student)\\
\bottomrule
\end{tabular}
\caption{Participant demographics for user evaluation}
\label{tab:participant_info}
\end{table}
\subsection{Measures}

We describes measures used to examine (1) how \sys{} supports exploration and iteration and (2) the usefulness of \sys{}'s features. 

\subsubsection{Creativity Support \& Usability.}
To assess whether \sys{} provides effective creativity support and is usable, we administered two standard questionnaires for measuring them: the Creativity Support Index (CSI) and the System Usability Scale (SUS) surveys. 

\subsubsection{Exploration Measures} 

To assess how participants explored the design space while using \sys{}, we defined the following as exploration measures. 

\begin{enumerate}
    \item \textit{Number of{ persona, problem, solution, storyboard }nodes.} This measure is informed by prior work in information retrieval and HCI~\cite{zhang2020users, urgo2022learning, suh2023sensecape}, where the breadth of information exploration is often assessed by counting concept nodes or domain-specific terms created in mind maps or knowledge structures. In our case, this refers to the number of nodes that participants created during their task, corresponding to major stages of the design process: empathizing with the user, defining the problem, ideating solutions, and creating storyboards. By counting persona, problem, solution, and storyboard nodes, we can gauge how thoroughly participants engaged with each stage. Higher counts may indicate more extensive exploration and ideation, reflecting deeper engagement with the system's core tasks.
    \item \textit{Usage frequency of features designed for exploration.}
    To measure the extent to which participants used features designed for exploration, we tracked the frequency of interactions with specific system tools, such as the \textsc{Start Brainstorming} for generating ideas and \textsc{Generate More} option for expanding on ideas from selected idea. These metrics help us understand how often participants relied on \sys{}’s generative AI capabilities to enhance their exploration of ideas, as well as their willingness to try different features for broadening the solution space.
\end{enumerate}

\begin{table*}[t]
\centering
\caption{Creativity Support Index (CSI) Results (N=10). The highest value is shown in \textbf{bold}, and the second highest in \underline{underline}. Following the guidelines by the authors of CSI~\cite{cherry2014quantifying}, the \textit{Collaboration} factor was included in the paired-factor comparison questions despite it being out of scope in this study as the score reflects its perceived importance to users and could thereby be useful for future reference.}
\begin{tabular}{lcc}
\hline
Factor & Av.g. Score (SD) & Avg. Factor Count \\
\hline
Enjoyment (``\textit{I enjoyed using the tool}'') & \textbf{18} (2.1) & 1.8 \\
Exploration (``\textit{It allowed me to track different ideas}'') & \underline{17.8} (2.5) & \textbf{5.1} \\
Expressiveness (``\textit{It allowed me to be very expressive}'') & 15.6 (4.6) & \underline{3.9} \\
Immersion (``\textit{The tool was engaging}'') & 15.3 (4.4) & 1.3 \\
Results Worth Effort (``\textit{I was satisfied with what I got}'') & 17.6 (2.5) & \textbf{5.1} \\
Collaboration (``\textit{I could collaborate with others}'') & \sig{N/A} & 2.8 \\
\hline
Overall CSI Score & 72.0 (15.7) & \\
\hline
\label{tab:csi}
\end{tabular}
\end{table*}

\subsubsection{Iteration Measures} 

According to \cite{rose2022}, \textit{an iteration} can be defined as ``a new version of a design or changes to design.'' Additionally, we measure the number of times users utilize the propagate changes feature. The measures are as follows. 

\begin{enumerate}
    \item 
    \textit{\# of Individual Node Edits.}
    This measure tracks how often participants edited a node (e.g., persona, problem, solution) and used the Update function (\,\raisebox{-2pt}{\includegraphics[scale=0.25]{figures/icon_update.png}}\,) to apply changes to a single connected node. By monitoring the number of edits and updates that propagate to only one related node, we can assess how participants iterated within specific parts of the design process while keeping the scope of changes localized. This can indicate whether participants preferred to make targeted refinements to adjacent nodes rather than making broader, system-wide adjustments across multiple connected nodes.
    \item \textit{\# of Forward Propagate Edits.} This measure counts the number of times participants edited a node and chose to \textsc{Forward Propagate} (\,\raisebox{-2pt}{\includegraphics[scale=0.25]{figures/icon_forwardprop.png}}\,) changes to subsequent nodes in the workflow (e.g., updating a problem node and propagating changes to connected solution nodes). The number of nodes updated by forward propagations was also tracked. By tracking forward propagation, we can assess how frequently participants recognized that changes in one stage should affect subsequent stages, a hallmark of effective iterative process.
    \item \textit{\# of Back Propagate Edits.} This measure captures how often participants edited a node and chose to \textsc{Back Propagate} (\,\raisebox{-2pt}{\includegraphics[scale=0.25]{figures/icon_backprop.png}}\,) changes to preceding nodes (e.g., modifying a solution node and propagating changes back to the problem node). The number of nodes updated by backward propagations was also tracked. This metric helps us understand how participants handled iterative feedback loops, ensuring that updates made at later stages of the design process were reflected in earlier stages. It highlights the system's ability to support non-linear design processes where iteration occurs across multiple stages.
\end{enumerate}

\subsubsection{Perceived Utility Measures} To evaluate the perceived utility of \sys{}'s features, we analyzed responses from the post-study survey and interviews. This included participants' agreement ratings (1: Strongly Disagree, 5: Strongly Agree) to statements like, `\textsf{Being able to propagate changes down ↓ to dependent nodes is useful for iterating on ideas}' and questions during interview to elicit further details on their assessment of the features. 

\section{Results}
\label{sec:results}

We present an analysis of survey responses, interviews, and participant interactions with \sys{},\footnote{Due to time constraints, one participant (P6) was unable to complete the task, reducing the system usage analysis to N=9. All other analyses are based on N=10.} detailing their assessment of the system, how they utilized its features, the perceived value of each, and the diverse ways \sys{} can enhance learning, teaching, and the practice of exploratory, iterative design process.

\textit{Creativity Support \& Usability.} Participants rated \sys{} as providing high creativity support~\cite{cherry2014quantifying}. As shown in Table~\ref{tab:csi}, the participants found \sys{} to provide the most support in terms of \textsc{Enjoyment} and \textsc{Exploration}. On the other hand, the CSI factor count revealed that they regard `\textsf{getting satisfactory results}' (\textsc{Results Worth Effort}) and `\textsf{being able to explore diverse ideas}' (\textsc{Exploration}) as the important factors when using \sys{} for iterative design process, followed by `\textsf{being able to be expressive}' (\textsc{Expressiveness}). System Usability Scale (M = 84.8, SD = 10.8), coupled with self-reported responses, showed that participants found \sys{} extremely usable (\textit{Excellent} according to \cite{bangor2009determining}) and passing the cutoff score (68) for production by a large margin. P10\textsubscript{inst} echoed this, saying it is ``intuitive'' and that he can see ``students or designers learning this in a few minutes.''

\begin{figure*}[htb!]
    \centering
    \begin{subfigure}[t]{\textwidth}
        \includegraphics[alt={Stacked bar charts showing how participants rated the usefulness of features: "Rapidly generating ideas using AI," "AI-generated suggestions," "AI-generated feedback," "Propagating changes down to dependent nodes," "Propagating changes up to earlier nodes" and "Creating more nodes based on a selected node" for exploring, iterating, evaluating and visualizing ideas. The majority of participants responded with "4 - Agree" or "5 - Strongly Agree" on most measures.}, trim=0cm 0cm 0cm 0cm, clip=true, width=\textwidth]{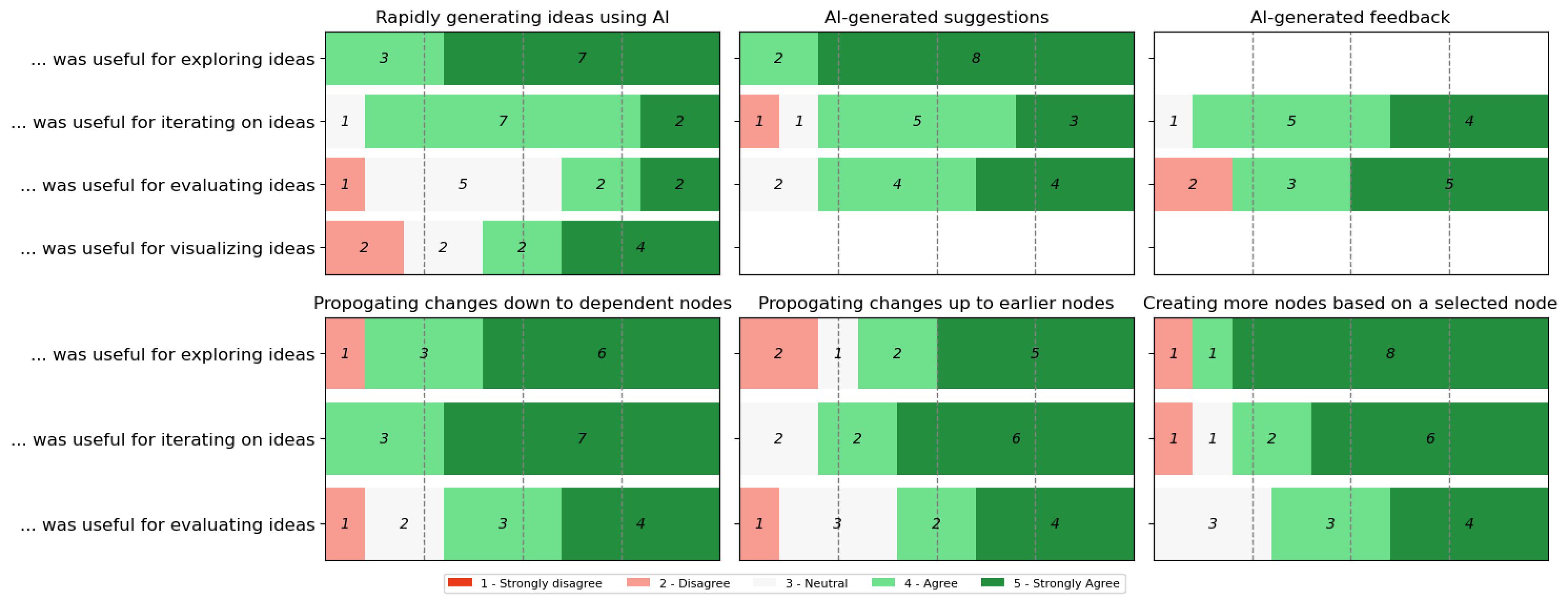}
        \caption{Perceived usefulness of each feature for ideation, iteration, evaluation, and visualization of ideas.}
        \label{fig:likert-features}
    \end{subfigure}
    \begin{subfigure}[t]{0.9\textwidth}
        \includegraphics[alt={Stacked bar charts showing how participants rated system artifacts: "AI-generated ideas," "AI-generated suggestions" and "AI-generated feedback" for usefulness, relevance and accuracy. The majority of participants responded with "4 - Agree" or "5 - Strongly Agree" on most measures.}, trim=0cm 0cm 0cm 0cm, clip=true, width=\textwidth]{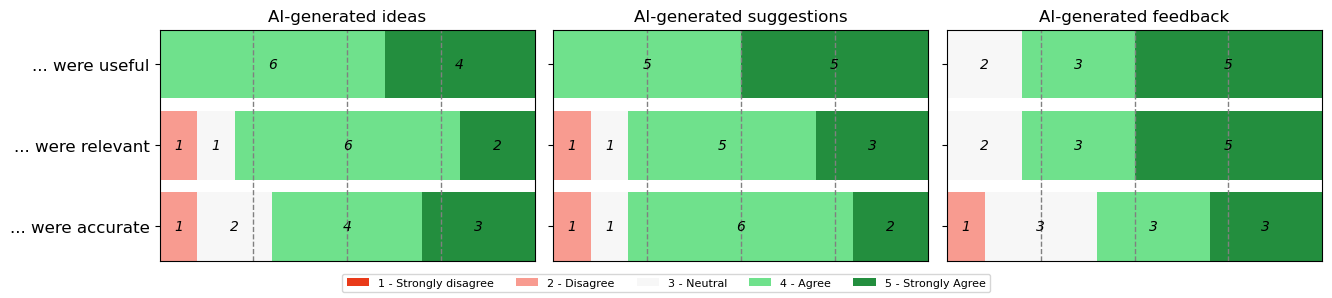}
        \caption{Perceived quality of AI-generated ideas, suggestions, and feedback}
        \label{fig:likert-artifacts}
    \end{subfigure}
    \caption{General perception of the usefulness of \sys{} features and the quality of AI-generated ideas, suggestions, and feedback.}
    \Description[short description]{}
    \label{fig:likert-data}
\end{figure*}

\begin{figure*}
    \centering
    \includegraphics[alt={Screenshot of participant 7's design workspace which includes cards for personas, problem statements, solutions and storyboards to explore the design space: "Gamification of Learning."}, width=\linewidth]{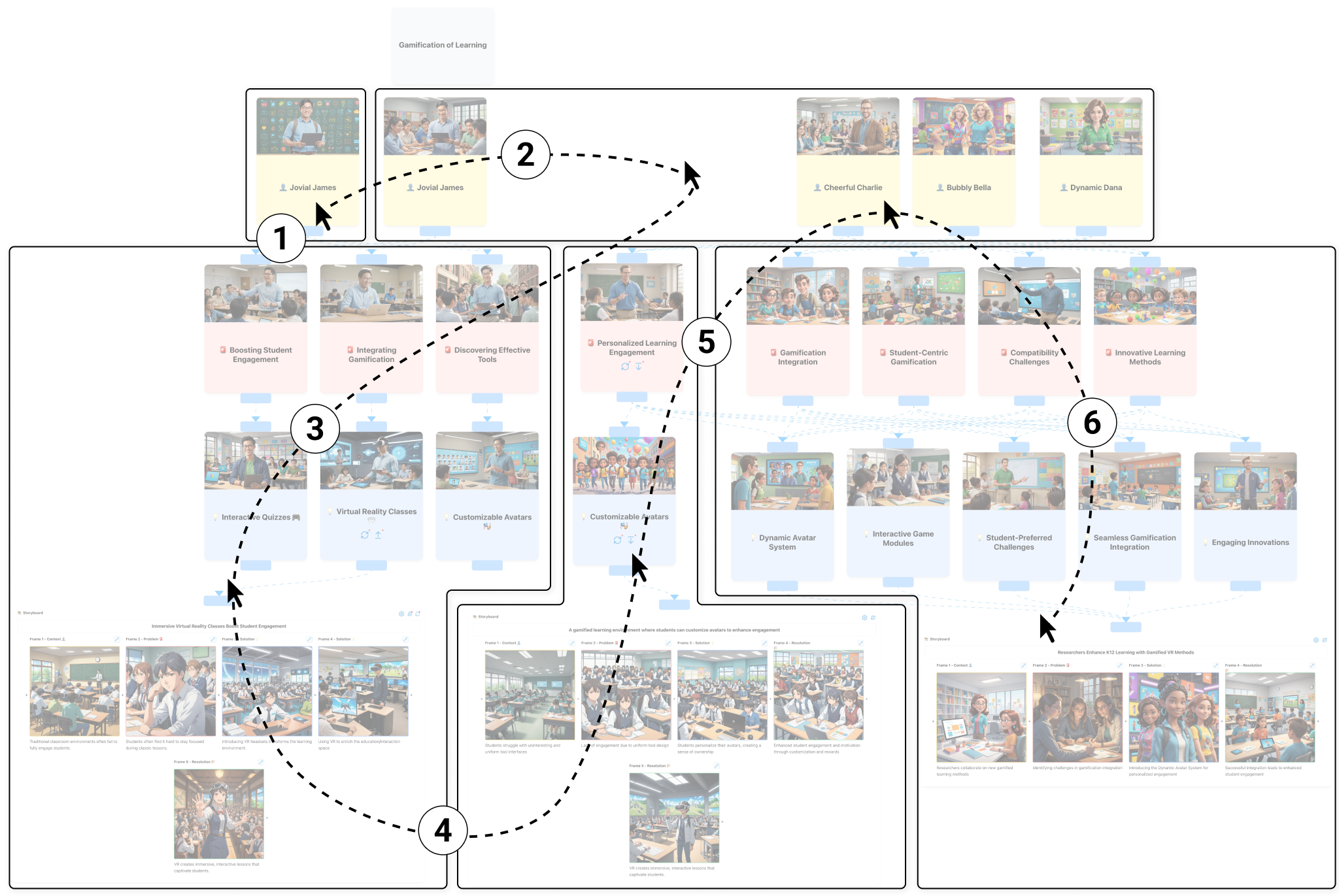}
    \caption{The workspace of P7\textsubscript{stud} after completing the task. P7 followed a hybrid approach that combined manual authoring with AI assistance. (1) P7 manually created an initial persona, then (2) used \textsc{Generate More} to expand the persona space. (3) These personas were used to generate corresponding problems, solutions, and an initial storyboard. (4–5) P7 then used backpropagation from the storyboard to generate and revise earlier artifacts—including persona nodes—and (6) generated up to a storyboard. This process illustrates how StoryEnsemble enables dynamic exploration and iteration.}
    \label{fig:p7-screenshot}
\end{figure*}

\subsection{How Does StoryEnsemble Support Exploration and Iteration?}

\subsubsection{Exploration.}
Participants generally found exploration of ideas easy in \sys{} (`\textsf{it was easy to explore ideas}': 5 Strongly Agree, 3 Agree, 1 Neutral, 1 Disagree). On average, participants created 
$16.\overline{3}$ ideas (PERSONA: $4.\overline{4}$; PROBLEM: $5.\overline{2}$; SOLUTION: $5.\overline{1}$; STORYBOARD: $1.\overline{5}$). Specifically, they all agreed that `\textsf{being able to rapidly generate ideas using AI was useful for exploring ideas}' (7 Strongly Agree, 3 Agree). P6\textsubscript{prac} said: ``I think it's extremely useful, because it generates ideas quickly and it's also easy to edit. The Design Thinking process is currently very manual and in need of being digitised, so what you're doing is great.'' 

Participants also found the AI-generated ideas from \textsc{Start Brainstorming} for the persona, problem, solution, and storyboard nodes very useful (4 Strongly Agree, 6 Agree) and mostly relevant (2 Strongly Agree, 6 Agree, 1 Neutral, 1 Disagree). P1\textsubscript{prac} appreciated how AI provided perspectives outside his experience. He shared that while working on a wedding app, he struggled to see things from a couple’s point of view, as he was not married. \sys{} helped him empathize with personas and better understand their needs. P4\textsubscript{stud} recognized that exploring diverse ideas could help combat ``design fixation,'' a common issue in traditional design processes. 

As shown in Fig.~\ref{fig:likert-data}, they also found AI-generated suggestions extremely useful (5 Strongly Agree, 5 Agree) and for exploring ideas (8 Strongly Agree, 2 Agree). On average, participants generated $22.\overline{6}$ suggestions during their exploration. Most participants found creating additional nodes using the \textsc{Generate More} feature extremely useful for exploration (8 Strongly Agree, 1 Agree, 1 Disagree). Four participants started with a single persona and then expanded their persona set using \textsc{Generate More} instead of generating the full set of ideas from the start. P4\textsubscript{stud}, who selected \textsf{Disagree}, expressed concern about over-reliance on AI for generating synthetic user data. She preferred creating the initial persona set with real data and then using AI to fill in gaps, validating these insights with target users through further research.

\subsubsection{Iteration}

Many participants generally found it easy to iterate on ideas (4 Strongly Agree, 4 Agree, 1 Neutral, 1 Disagree). On average, participants iterated on each node at least twice for all nodes except storyboard (PERSONA: $4.\overline{2}$ ; PROBLEM: $2.\overline{4}$; SOLUTION: $2.\overline{1}$; STORYBOARD: $0.\overline{5}$) by manually editing ($2.\overline{1}$), revising with AI ($2.\overline{6}$), and using feedback to trigger update of the node ($0.\overline{3}$). 

Participants appreciated how the AI-generated feedback (shown in Fig.~\ref{fig:revise-using-ai}) 
nudged them to iterate. P2\textsubscript{prac} said: ``I liked that it made me incorporate feedback and revise.'' Overall, they found AI-generated feedback very useful in general (5 Strongly Agree, 3 Agree, 2 Neutral) and for iterating on ideas (4 Strongly Agree, 5 Agree, 1 Neutral). In addition to the quality of the feedback, the ease with which they could update the node by incorporating the feedback from the list and trigger update of the node seemed to help them effortlessly transition into iteration. P1\textsubscript{prac} said:  ``from feedback, it makes me go, `Oh yeah, I forgot about this.' And then I can just iterate based on that feedback.''

Overall, participants liked the support \sys{} provides to enable flexible iteration, with many participants (P2\textsubscript{prac}, P7-8\textsubscript{stud,inst}, P10\textsubscript{inst}) quoting both the forward propagation and backward propagation as their favorite features. As shown in Fig.~\ref{fig:likert-features}, participants found both propagation mechanisms (Fig.~\ref{fig:propagation-mechanisms}) highly useful for iteration (Forward Prop: 7 Strongly Agree, 3 Agree; Back Prop: 6 Strongly Agree, 2 Agree, 2 Neutral). Most participants (7/9) used `propagate changes' at least twice (Forward Prop: M=$3.\overline{2}$; Backward Prop: M=$1.\overline{2}$). P2 specifically emphasized the value of being able to iterate acr oss different stages:

\begin{displayquote}
    ``I totally like the top down approach of iteration and also bottom up approach where we propagate from one level to another. 
    ... \textbf{I could iterate top, down, bottom up to any levels}. So it felt better than design thinking in a way because, \textbf{I felt like I could iterate on all different levels}.''
\end{displayquote}

\subsection{How Can StoryEnsemble Be Used in Design Process?}

\begin{figure}
    \centering
    \includegraphics[alt={Screenshot of participant 8's design workspace which includes cards for personas, problem statements, solutions and storyboards to explore the design space: `Personalized News Consumption.'}, width=\linewidth]{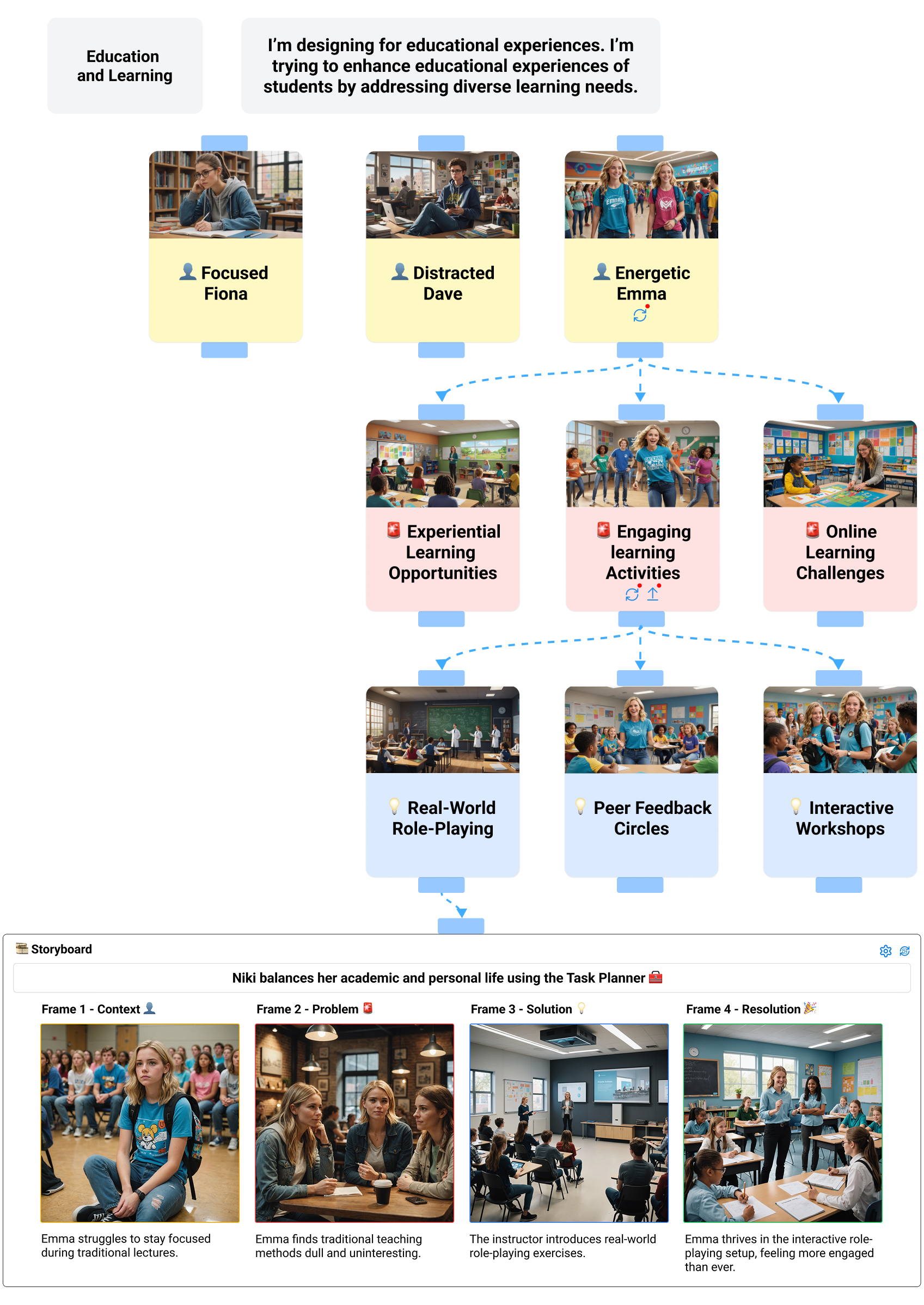}
    \caption{The workspace of P8\textsubscript{inst} after completing the task.}
    \label{fig:p8-screenshot}
\end{figure}

We present three types of workflows (\textbf{W}s) observed in our analysis that reflect a spectrum of design approaches, ranging from mostly user-driven to heavily AI-assisted.

\textit{W1. Bottom-Up (User-driven) Workflow (Fig.~\ref{fig:p1-screenshot}).} P1\textsubscript{prac} followed a bottom-up strategy, generating and iterating within each stage before moving on to the next. In the Empathize Stage, they manually created an initial persona (Persona 1), then used \sys's AI support to generate four additional personas (Persona 2-5). In the Define Stage, P1 generated two problem nodes (Problem 1, 2) connected to Persona 2 and Persona 3. Next, P1 generated 3 solutions (Solution 1-3) based on Problem 1 to explore the Ideate Stage. P1 then created a storyboard based on Solution 1, Problem 1, Persona 2, and Persona 3. Finally, P1 used AI-driven revisions to edit the problem framing within the storyboard and regenerated the problem image to better reflect the design intent in the Prototype Stage. This workflow illustrates how users can actively guide the process while using AI to expand and refine ideas. 

\textit{W2. Top-Down (AI-driven) Workflow (Fig.~\ref{fig:p2-screenshot}).} P2\textsubscript{prac} adopted a more AI-driven, top-down approach, using AI to generate a comprehensive design direction upfront. After manually drafting Persona 1, P2 used AI to revise it and then prompted \sys{} to generate the remaining elements across all design stages in one go: 4 problems (Problem 1-4), 4 solutions (Solution 1-4) and a storyboard linked to Solution 2, Problem 3, and Persona 1 in a single generation from Persona 1. Then P2 iterated on Solution 2 using \sys's ''backward propagation`` feature, which cascaded updates to the linked Problem 2 and Persona 1. This workflow demonstrates \sys{}'s ability to support nonlinear iteration and rapid exploration from a high-level concept.

\textit{W3. Hybrid Workflow (Fig.~\ref{fig:p8-screenshot}).} 
P8\textsubscript{inst} used a hybrid approach, incorporating both designer input and AI suggestions. In the Empathize Stage, P8 generated 3 personas (Persona 1-3) and revised Persona 2 and 3 using designer instructions and AI-recommended prompts. P8 also engaged with AI-generated feedback to further refine Persona 3. In the Define Stage, P8 generated 3 problems (Problem 1-3) based on Persona 3, revising Problem 1 and 2 using AI-recommended prompts before proceeding to brainstorm solutions for Problem 2. This workflow reflects the \sys's flexibility in supporting co-creation and dynamic iteration between user input and AI contributions during design processes.

These workflows illustrate StoryEnsemble's key design principle of supporting flexible, interconnected design exploration, where users can seamlessly move between manual input and AI-assisted generation across different design stages.

\subsection{What Are Potential Applications, Opportunities, and Challenges?}

In addition to its role in supporting exploration and iteration, \sys{} opens up several potential applications and addresses opportunities and challenges in the broader use of iterative design process. Through our study, we identified key use cases in teaching, learning, and professional design, while also uncovering concerns regarding AI's role in the process. Most participants rated StoryEnsemble as extremely useful for \textit{teaching} and \textit{using} design thinking (Teaching: 5 Extremely Useful, 3 Very Useful, 2 Moderately; Using: 7 Extremely Useful, 1 Very Useful, 2 Moderately). In contrast, the ratings for \textit{learning} design thinking were more evenly distributed, ranging from moderate to highly useful (Learning: 3 Extremely Useful, 1 Very Useful, 2 Moderately). This suggests that the tool is particularly valuable for practitioners and instructors, while its benefits for learners are still significant but more varied. Below, we explore the main themes that emerged from the survey and interviews.

\subsubsection{Leveraging \sys{} for Teaching and Demonstrating Design Thinking}

Many participants emphasized the value of \sys{} for teaching design thinking in academic settings. The system’s ability to quickly generate artifacts like personas and solutions, combined with its iterative features, was seen as an effective way to guide students through the design thinking process in a structured, interactive manner.

One of the most prominent themes that emerged from our study was \sys{}'s potential as a teaching tool. Several participants, especially instructors, noted that \sys{} could significantly enhance the way design thinking is taught in classrooms by making the process more interactive and accessible. For instance, P2\textsubscript{prac} said: ``this could help people practice design thinking in the way it's intended to be. It makes more sense to do it this way.''

P8\textsubscript{inst}, an experienced instructor, emphasized how \sys{} could serve as a practical demonstration tool, allowing students to see an entire design thinking process in action. He described how it would help students develop a mental model of what a complete project looks like, giving them a clear understanding of how each stage---persona creation, problem definition, solution ideation, and prototyping---fits together:

\begin{displayquote}
    ``If I had this tool, it would be really neat to demo a toy project... I could \textbf{use the AI to show them how to iterate} through a persona, a problem statement, and a solution.''
\end{displayquote}

Specifically, he was excited about \sys{}'s propagation mechanisms and suggested that the tool could be perfect for demonstrating it:

\begin{displayquote}
    ``\textbf{My favorite feature is the propagate downward and propagate upward} [features], because \textbf{this is what we try to teach to students in the design thinking process}... but you already have a mechanism to allow that by propagating upwards or downwards. So I really like that.''
\end{displayquote}

Another important advantage for teaching is the ability to use real-time feedback and AI-generated suggestions to encourage students to reflect on their decisions. P2\textsubscript{prac} noted how this feature could deepen students’ engagement:

\begin{displayquote}
    ``The fact that you can go back and forth and revise things... it would make them think more about the problem and the users.''
\end{displayquote}

However, some instructors expressed \textbf{concerns about students potentially over-relying on AI} and skipping critical steps in the design process. P4\textsubscript{stud} emphasized the importance of hands-on experience, saying that ``students should have the experience of actually doing it instead of just generating everything randomly.'' Instructors, who identified similar potential issues, suggested controlling when and how students use AI in the classroom. P4\textsubscript{stud} and P10\textsubscript{inst} proposed introducing AI later in the learning process, after students have had a chance to work through initial tasks manually. This could help ensure that students develop essential design skills before using AI to iterate or refine their work. Additionally, P5\textsubscript{prac} suggested that future versions of \sys{} could feature an admin panel, allowing instructors to tailor AI’s involvement to specific tasks and providing them with the flexibility to manage how AI is integrated into different phases of design thinking.

Overall, participants saw \sys{} as a powerful teaching tool, with the potential to make design thinking more accessible, interactive, and engaging for students, while acknowledging the need to maintain a balance between AI assistance and hands-on practice.

\subsubsection{Streamlining Professional Workflows with AI Assistance}

In addition to its potential for education, \sys{} was also seen as a valuable tool for improving design workflows in professional environments. Several participants, particularly those with experience in design practice, noted that \sys{} could help streamline the ideation and iteration process, making it more efficient for small teams or projects with tight deadlines.

P6\textsubscript{prac} reflected on how \sys{} could speed up the design process in situations where resources are limited, saying ``it would \textbf{speed up the process} massively, especially when the design team is small.'' P5\textsubscript{prac} added that \sys{} could drastically reduce the amount of time spent on brainstorming and iteration, allowing teams to focus on more strategic aspects of design. Reflecting on the past design workshop with 8 members of his team, he said that \sys{} could have easily given ``50\% [time] reduction'' for each person, essentially saving ``40 hours'' and allowing that time to be ``used for something else.'' 

In a similar vein, P2\textsubscript{prac}, who learned design thinking through a self-paced, 1-week online course, said that given the low attention span of people, this tool would be amazing for students like her, saying ``if I can learn the same thing in 90 minutes what I learnt in a week, I'll choose the 90 minutes.'' She went onto add that corporations that invest money and time to teach their employees about design thinking would be thrilled to purchase this tool for training purposes.

\subsubsection{Balancing AI Assistance and Hands-On Design Practice}

While participants praised \sys{}'s ability to streamline both teaching and professional workflows, its potential to help students ``get unstuck'' (P4\textsubscript{stud}) and access useful ``feedback'' (P8\textsubscript{inst}), they also raised important concerns about the potential for over-reliance on AI, particularly in educational contexts. Several instructors worried that students might become too dependent on AI-generated outputs, bypassing essential learning experiences such as manual iteration and critical thinking.

P4, a student with prior experience as both an instructor and design practitioner, stressed the importance of hands-on learning, saying ``students should have the experience of actually doing it instead of just generating everything randomly.'' P8\textsubscript{inst} also voiced concern over students relying too heavily on AI-generated assumptions. He said: ``Designing on assumptions is not a good thing... I would tell students to be very cautious with the assumptions the AI engine makes.'' 

To address these concerns, participants suggested ways to control AI’s role and the extent of it in the design process. For instance, P10\textsubscript{inst} recommended introducing AI only after students have completed certain tasks manually, allowing them to first engage with the design process before using AI to support their iterations. P5\textsubscript{prac} proposed an interesting idea, suggesting that there could be a panel that allows instructors to precisely control the level of AI engagement by specifying which node types can use AI capabilities. He said: 

\begin{displayquote}
    ``\textbf{A toggle button to control AI engagement could be useful}... depending on the workshop, AI could be used only for certain tasks.''
\end{displayquote}

He explained that instructors could grant permission to use AI once students complete specific tasks in earlier stages. For example, after conducting interviews and manually creating initial personas, students could then use AI to generate additional personas grounded in the data from their interviews—maintaining a balance between hands-on practice and AI-assisted exploration.

Additionally, participants suggested incorporating validation mechanisms into the system. P7\textsubscript{stud} proposed that after generating AI-driven ideas, \sys{} could prompt users to accept or reject them, encouraging designers to critically assess the outputs rather than blindly accepting them: ``The system could nudge users to accept or reject AI-generated ideas, forcing them to check and validate the data.'' 

\section{Discussion}
\label{sec:discussion}

\subsection{Summary}
In this work, we explored how generative AI can support iterative design processes by developing StoryEnsemble, a system that enables users to easily explore and iterate across multiple stages of design thinking. While prior systems such as DynEx~\cite{ma2025dynex} and IdeaSynth~\cite{pu2025ideasynth} also support divergent and convergent workflows through structured UI generation or research idea synthesis, they focused primarily on generation within specific stages. In contrast, our goal was to address a broader challenge of fragmentation across design stages by introducing mechanisms such as dependency tracking, cascading updates, and backpropagation.

Our evaluation showed that StoryEnsemble effectively supports dynamic exploration and iteration through a combination of AI-generated suggestions, revision prompts, and propagation mechanisms. Iteration was supported through direct editing, feedback-triggered updates, and AI-assisted revisions, with participants frequently using forward and backward propagation to cascade changes across connected nodes. These features enabled participants to revise earlier ideas based on later-stage insights and vice versa, helping them break out of linear flows and engage in multi-level iteration. They also adopted a range of workflows—bottom-up, top-down, and hybrid—demonstrating the flexibility of the system in supporting both structured and improvisational design strategies.

These findings highlight the importance of supporting non-linear, iterative processes. Rather than enforcing a fixed, stage-by-stage progression, design systems should allow users to revisit any stage, propagate changes efficiently, and explore multiple paths with minimal friction. As demonstrated by StoryEnsemble, supporting tightly linked representations, lightweight revisions, and multiple entry points offer a promising model for supporting dynamic exploration and iteration in multi-stage creative processes.

\subsection{Design Implications}

Traditional design processes, such as design thinking, often struggle with complexity and the inherent messiness of interdependent tasks, where changes in one stage affect others, making it difficult to manage under tight deadlines or resource constraints. \sys{} addresses these challenges by visualizing and managing these interdependencies through a node-link interface and allowing bi-directional change propagation to observe the design changes in real-time. As users are freed from manually updating every step, \sys{} facilitates effortless iteration, encouraging a more dynamic and integrated approach to the design process. 

Beyond the system, our work offers comprehensive perspectives from students, instructors, and practitioners regarding the potential of incorporating generative AI in the design process. P9\textsubscript{inst} shared how she has already accepted the fact that she cannot prevent students from using AI and explained that she started asking students to simply describe how they have used it for their work. 
If co-creation becomes the norm in the near future, our work will contribute rich insights on the system design opportunities and tensions for this paradigm shift in the design process.

The mechanisms we developed to manage interconnected design artifacts—such as dependency tracking, cascading updates, and bidirectional propagation—may generalize to other domains with similarly interdependent structures. For example, in software development, updating one module may require adjustments across related components; in content creation, narrative changes in a script may cascade to visuals, audio, or pacing; in education, learning objectives often shape assessments and instructional materials in a co-evolving process; and in video authoring, changes to a scene or storyline may affect timing, transitions, and supporting media assets. Supporting such interdependencies requires tools that can maintain coherence while allowing flexible iteration—a challenge our approach addresses and one that future systems across these domains may benefit from.

\subsection{Ethical and Pedagogical Implications}

One important consideration raised in this study relates to the ethical and pedagogical concerns regarding AI over-reliance, particularly in educational contexts. Several participants highlighted the risks of students relying too heavily on AI-generated content, potentially bypassing critical design skills that can only be developed through hands-on experience and critical reflection.

As noted by P4\textsubscript{stud} and P8\textsubscript{inst}, a balance must be maintained between leveraging AI for efficiency and ensuring students engage with foundational design principles. Without appropriate guidance, students may mistakenly equate the use of AI-generated suggestions with mastering the design process, which may lead to a superficial understanding of key concepts.

To address these concerns, participants recommended diverse ideas such as introducing AI later in the design process, providing a control to regulate AI’s availability, and requiring users to actively reflect on AI-generated outputs via validation prompts. These insights can be useful as we design future systems and tools powered by AI and as AI becomes more integrated into educational practices.

\subsection{Limitations and Future Work}

\textit{Study Design.} While our exploratory study provided broad insights into the potential of \sys{}, more controlled, comparative studies are needed to answer specific questions such as whether AI-enhanced design leads to more creative or diverse solutions than traditional methods. These studies could also explore the role of \sys{} in supporting various design stages in greater depth.

\textit{Long-term Deployment Study.} A long-term deployment study could offer further insights into how generative AI can support sustained design projects and team-based workflows. This would help to evaluate \sys{}'s effectiveness over time and assess its role in supporting co-creation with AI in different contexts.

\textit{Extending System: Adding More Tools.} Future development of \sys{} could expand its utility by incorporating additional tools. As Hehn et al enumerated in their work, there are diverse techniques and methods in design~\cite{hehn2018design}. For example, for need-finding, designers can choose from several methods such as customer journey, empathy map, and 5 whys. For prototyping, in addition to storyboard, designers can choose paper prototype, digital wireframe, as well as high fidelity prototype. In fact, participants such as P8 suggested adding empathy maps or journey maps, which are integral to many design workflows. There is also potential for supporting collaboration and enhancing system's feedback mechanisms by synthesizing user feedback across stages, providing users with a more comprehensive view of insights that emerge from design research or user studies.

\textit{Extending System: Projecting Possibilities through Generative Simulation.} StoryEnsemble's generative capabilities open up exciting possibilities for extending the system into simulation-based research (akin to~\cite{park2022social, nguyen2024simulating}). Future iterations could leverage the tool's persona and scenario generation to help researchers explore potential user impacts across diverse contexts. For instance, policy researchers could use the system to simulate how proposed programs might affect different demographic groups, generating synthetic user scenarios to identify previously overlooked perspectives or potential implementation challenges. Similarly, design teams could use the tool to proactively explore user experiences across varied socioeconomic, cultural, or accessibility contexts, expanding traditional research methodologies by rapidly generating and iterating through multiple potential user scenarios.
This approach would transform StoryEnsemble from a design exploration tool into a powerful platform for scenario planning and social impact assessment, enabling more comprehensive and imaginative approaches to understanding complex user ecosystems.

\textit{Preventing Overreliance on AI.} Our study raised concerns about students over-relying on AI-generated content, especially in educational settings. One solution proposed by participants is to regulate AI engagement based on the task or stage of the process. This could be managed by providing clear guidelines on when to use AI and ensuring that users have the opportunity to engage manually before incorporating AI-generated suggestions. We could also explore techniques such as \textit{frictions}, interventions that force users to cognitively engage with the AI-generated artifact~\cite{kazemitabaar2025exploring}. Our intention is not to replace these critical aspects of the design process but to view AI as a complementary tool, addressing limitations in current design thinking processes such as the costly nature of getting feedback and generating artifacts such as storyboard to communicate ideas~\cite{jung2023toward}.

\textit{Identifying Scenarios Where Generative AI is Most Effective.} While highlighting the potential of AI in HCI, Schmidt et al. also emphasized the need to define when AI should be used~\cite{schmidt2024simulating}. In our evaluation study, we explored, to some extent, how \sys{} and generative AI could be useful or preferred. For example, participants generally felt AI would be helpful for teaching and using design thinking, while they were mixed about using it for learning design thinking. Within the context of learning design thinking, instructors had clear views on when AI would be beneficial. For instance, they saw value in using AI to demonstrate the full design thinking workflow with a simplified example, but expressed concern about novice students relying on AI independently, potentially leading to misunderstandings of core principles—such as the importance of engaging real users in the design process. Further research is needed to understand the nuances and reveal when AI is acceptable or even preferred. For example, Salminen et al. found that using AI to generate more personas can improve the representation of demographically diverse populations and benefit media-based projects where having demographically diverse representation is essential~\cite{salminen2022creating}.

\section{Conclusion}
\label{sec:conclusion}
In this work, we explored how generative AI can address key challenges in the design thinking process, particularly in supporting exploration, iteration, and feedback collection. While design thinking emphasizes iterative, user-centered problem-solving, practical constraints---such as time constraints and difficulty of engaging users for feedback---often hinder its full application. We introduced \sys{}, a system that integrates generative AI to streamline the design thinking workflow, enabling rapid exploration of ideas, flexible iteration across interconnected stages, and efficient feedback collection. Our formative study identified key challenges faced by students, instructors, and practitioners, and our evaluation demonstrated that \sys{} supports rapid exploration and flexible iteration of ideas. This work contributes a novel approach to AI-integrated design thinking, offering insights into the opportunities and challenges of incorporating AI into creative workflows, while emphasizing the importance of recognizing AI’s limitations and ethical considerations.

\bibliographystyle{ACM-Reference-Format}
\bibliography{references}


\begin{thebibliography}{84}


\ifx \showCODEN    \undefined \def \showCODEN     #1{\unskip}     \fi
\ifx \showDOI      \undefined \def \showDOI       #1{#1}\fi
\ifx \showISBNx    \undefined \def \showISBNx     #1{\unskip}     \fi
\ifx \showISBNxiii \undefined \def \showISBNxiii  #1{\unskip}     \fi
\ifx \showISSN     \undefined \def \showISSN      #1{\unskip}     \fi
\ifx \showLCCN     \undefined \def \showLCCN      #1{\unskip}     \fi
\ifx \shownote     \undefined \def \shownote      #1{#1}          \fi
\ifx \showarticletitle \undefined \def \showarticletitle #1{#1}   \fi
\ifx \showURL      \undefined \def \showURL       {\relax}        \fi
\providecommand\bibfield[2]{#2}
\providecommand\bibinfo[2]{#2}
\providecommand\natexlab[1]{#1}
\providecommand\showeprint[2][]{arXiv:#2}

\bibitem[Almeda et~al\mbox{.}(2024)]%
        {almeda2024prompting}
\bibfield{author}{\bibinfo{person}{Shm~Garanganao Almeda}, \bibinfo{person}{JD Zamfirescu-Pereira}, \bibinfo{person}{Kyu~Won Kim}, \bibinfo{person}{Pradeep Mani~Rathnam}, {and} \bibinfo{person}{Bjoern Hartmann}.} \bibinfo{year}{2024}\natexlab{}.
\newblock \showarticletitle{Prompting for Discovery: Flexible Sense-Making for AI Art-Making with Dreamsheets}. In \bibinfo{booktitle}{\emph{Proceedings of the CHI Conference on Human Factors in Computing Systems}}. \bibinfo{pages}{1--17}.
\newblock


\bibitem[Anderson et~al\mbox{.}(2024)]%
        {anderson2024homogenization}
\bibfield{author}{\bibinfo{person}{Barrett~R Anderson}, \bibinfo{person}{Jash~Hemant Shah}, {and} \bibinfo{person}{Max Kreminski}.} \bibinfo{year}{2024}\natexlab{}.
\newblock \showarticletitle{Homogenization effects of large language models on human creative ideation}. In \bibinfo{booktitle}{\emph{Proceedings of the 16th Conference on Creativity \& Cognition}}. \bibinfo{pages}{413--425}.
\newblock


\bibitem[Antony and Huang(2024)]%
        {antony2024id}
\bibfield{author}{\bibinfo{person}{Victor~Nikhil Antony} {and} \bibinfo{person}{Chien-Ming Huang}.} \bibinfo{year}{2024}\natexlab{}.
\newblock \showarticletitle{ID. 8: Co-Creating visual stories with Generative AI}.
\newblock \bibinfo{journal}{\emph{ACM Transactions on Interactive Intelligent Systems}} \bibinfo{volume}{14}, \bibinfo{number}{3} (\bibinfo{year}{2024}), \bibinfo{pages}{1--29}.
\newblock


\bibitem[Archer(1979)]%
        {archer1979whatever}
\bibfield{author}{\bibinfo{person}{L~Bruce Archer}.} \bibinfo{year}{1979}\natexlab{}.
\newblock \showarticletitle{Whatever became of design methodology}.
\newblock \bibinfo{journal}{\emph{Design studies}} \bibinfo{volume}{1}, \bibinfo{number}{1} (\bibinfo{year}{1979}), \bibinfo{pages}{17--18}.
\newblock


\bibitem[Arteaga et~al\mbox{.}(2024)]%
        {arteaga2024across}
\bibfield{author}{\bibinfo{person}{Estefania Arteaga}, \bibinfo{person}{Robbert Biesbroek}, \bibinfo{person}{Johanna Nalau}, {and} \bibinfo{person}{Michael Howes}.} \bibinfo{year}{2024}\natexlab{}.
\newblock \showarticletitle{Across the Great Divide: A Systematic Literature Review to Address the Gap Between Theory and Practice}.
\newblock \bibinfo{journal}{\emph{SAGE Open}} \bibinfo{volume}{14}, \bibinfo{number}{1} (\bibinfo{year}{2024}), \bibinfo{pages}{21582440241228019}.
\newblock


\bibitem[Bangor et~al\mbox{.}(2009)]%
        {bangor2009determining}
\bibfield{author}{\bibinfo{person}{Aaron Bangor}, \bibinfo{person}{Philip Kortum}, {and} \bibinfo{person}{James Miller}.} \bibinfo{year}{2009}\natexlab{}.
\newblock \showarticletitle{Determining what individual SUS scores mean: Adding an adjective rating scale}.
\newblock \bibinfo{journal}{\emph{Journal of usability studies}} \bibinfo{volume}{4}, \bibinfo{number}{3} (\bibinfo{year}{2009}), \bibinfo{pages}{114--123}.
\newblock


\bibitem[Beckman and Barry(2007)]%
        {beckman2007innovation}
\bibfield{author}{\bibinfo{person}{Sara~L Beckman} {and} \bibinfo{person}{Michael Barry}.} \bibinfo{year}{2007}\natexlab{}.
\newblock \showarticletitle{Innovation as a learning process: Embedding design thinking}.
\newblock \bibinfo{journal}{\emph{California management review}} \bibinfo{volume}{50}, \bibinfo{number}{1} (\bibinfo{year}{2007}), \bibinfo{pages}{25--56}.
\newblock


\bibitem[Benharrak et~al\mbox{.}(2024)]%
        {benharrak2024writer}
\bibfield{author}{\bibinfo{person}{Karim Benharrak}, \bibinfo{person}{Tim Zindulka}, \bibinfo{person}{Florian Lehmann}, \bibinfo{person}{Hendrik Heuer}, {and} \bibinfo{person}{Daniel Buschek}.} \bibinfo{year}{2024}\natexlab{}.
\newblock \showarticletitle{Writer-defined AI personas for on-demand feedback generation}. In \bibinfo{booktitle}{\emph{Proceedings of the CHI Conference on Human Factors in Computing Systems}}. \bibinfo{pages}{1--18}.
\newblock


\bibitem[Bilgram and Laarmann(2023)]%
        {bilgram2023accelerating}
\bibfield{author}{\bibinfo{person}{Volker Bilgram} {and} \bibinfo{person}{Felix Laarmann}.} \bibinfo{year}{2023}\natexlab{}.
\newblock \showarticletitle{Accelerating innovation with generative AI: AI-augmented digital prototyping and innovation methods}.
\newblock \bibinfo{journal}{\emph{IEEE Engineering Management Review}} \bibinfo{volume}{51}, \bibinfo{number}{2} (\bibinfo{year}{2023}), \bibinfo{pages}{18--25}.
\newblock


\bibitem[Brade et~al\mbox{.}(2023)]%
        {brade2023promptify}
\bibfield{author}{\bibinfo{person}{Stephen Brade}, \bibinfo{person}{Bryan Wang}, \bibinfo{person}{Mauricio Sousa}, \bibinfo{person}{Sageev Oore}, {and} \bibinfo{person}{Tovi Grossman}.} \bibinfo{year}{2023}\natexlab{}.
\newblock \showarticletitle{Promptify: Text-to-image generation through interactive prompt exploration with large language models}. In \bibinfo{booktitle}{\emph{Proceedings of the 36th Annual ACM Symposium on User Interface Software and Technology}}. \bibinfo{pages}{1--14}.
\newblock


\bibitem[Cao et~al\mbox{.}(2025a)]%
        {Cao2025CompositionalSA}
\bibfield{author}{\bibinfo{person}{Yining Cao}, \bibinfo{person}{Yiyi Huang}, \bibinfo{person}{Anh Truong}, \bibinfo{person}{Hijung~Valentina Shin}, {and} \bibinfo{person}{Haijun Xia}.} \bibinfo{year}{2025}\natexlab{a}.
\newblock \showarticletitle{Compositional Structures as Substrates for Human-AI Co-creation Environment: A Design Approach and A Case Study}.
\newblock
\urldef\tempurl%
\url{https://api.semanticscholar.org/CorpusID:276813065}
\showURL{%
\tempurl}


\bibitem[Cao et~al\mbox{.}(2025b)]%
        {cao2025compositional}
\bibfield{author}{\bibinfo{person}{Yining Cao}, \bibinfo{person}{Yiyi Huang}, \bibinfo{person}{Anh Truong}, \bibinfo{person}{Hijung~Valentina Shin}, {and} \bibinfo{person}{Haijun Xia}.} \bibinfo{year}{2025}\natexlab{b}.
\newblock \showarticletitle{Compositional Structures as Substrates for Human-AI Co-creation Environment: A Design Approach and A Case Study}. In \bibinfo{booktitle}{\emph{Proceedings of the 2025 CHI Conference on Human Factors in Computing Systems}}. \bibinfo{pages}{1--25}.
\newblock


\bibitem[Chen et~al\mbox{.}(2025)]%
        {chen2025dancingboard}
\bibfield{author}{\bibinfo{person}{Longfei Chen}, \bibinfo{person}{Shengxin Li}, \bibinfo{person}{Ziang Li}, {and} \bibinfo{person}{Quan Li}.} \bibinfo{year}{2025}\natexlab{}.
\newblock \showarticletitle{DancingBoard: Streamlining the Creation of Motion Comics to Enhance Narratives}. In \bibinfo{booktitle}{\emph{Proceedings of the 30th International Conference on Intelligent User Interfaces}} \emph{(\bibinfo{series}{IUI '25})}. \bibinfo{publisher}{Association for Computing Machinery}, \bibinfo{address}{New York, NY, USA}, \bibinfo{pages}{477–503}.
\newblock
\showISBNx{9798400713064}
\urldef\tempurl%
\url{https://doi.org/10.1145/3708359.3712167}
\showDOI{\tempurl}


\bibitem[Cherry and Latulipe(2014)]%
        {cherry2014quantifying}
\bibfield{author}{\bibinfo{person}{Erin Cherry} {and} \bibinfo{person}{Celine Latulipe}.} \bibinfo{year}{2014}\natexlab{}.
\newblock \showarticletitle{Quantifying the creativity support of digital tools through the creativity support index}.
\newblock \bibinfo{journal}{\emph{ACM Transactions on Computer-Human Interaction (TOCHI)}} \bibinfo{volume}{21}, \bibinfo{number}{4} (\bibinfo{year}{2014}), \bibinfo{pages}{1--25}.
\newblock


\bibitem[Choi et~al\mbox{.}(2024a)]%
        {choi2024}
\bibfield{author}{\bibinfo{person}{DaEun Choi}, \bibinfo{person}{Sumin Hong}, \bibinfo{person}{Jeongeon Park}, \bibinfo{person}{John Joon~Young Chung}, {and} \bibinfo{person}{Juho Kim}.} \bibinfo{year}{2024}\natexlab{a}.
\newblock \showarticletitle{CreativeConnect: Supporting Reference Recombination for Graphic Design Ideation with Generative AI}. In \bibinfo{booktitle}{\emph{Proceedings of the CHI Conference on Human Factors in Computing Systems}} (Honolulu, HI, USA) \emph{(\bibinfo{series}{CHI '24})}. \bibinfo{publisher}{Association for Computing Machinery}, \bibinfo{address}{New York, NY, USA}, Article \bibinfo{articleno}{1055}, \bibinfo{numpages}{25}~pages.
\newblock
\showISBNx{9798400703300}
\urldef\tempurl%
\url{https://doi.org/10.1145/3613904.3642794}
\showDOI{\tempurl}


\bibitem[Choi et~al\mbox{.}(2024b)]%
        {choi2024proxona}
\bibfield{author}{\bibinfo{person}{Yoonseo Choi}, \bibinfo{person}{Eun~Jeong Kang}, \bibinfo{person}{Seulgi Choi}, \bibinfo{person}{Min~Kyung Lee}, {and} \bibinfo{person}{Juho Kim}.} \bibinfo{year}{2024}\natexlab{b}.
\newblock \showarticletitle{Proxona: Leveraging LLM-Driven Personas to Enhance Creators' Understanding of Their Audience}.
\newblock \bibinfo{journal}{\emph{arXiv preprint arXiv:2408.10937}} (\bibinfo{year}{2024}).
\newblock


\bibitem[Chung and Adar(2023)]%
        {chung2023}
\bibfield{author}{\bibinfo{person}{John Joon~Young Chung} {and} \bibinfo{person}{Eytan Adar}.} \bibinfo{year}{2023}\natexlab{}.
\newblock \showarticletitle{PromptPaint: Steering Text-to-Image Generation Through Paint Medium-like Interactions}. In \bibinfo{booktitle}{\emph{Proceedings of the 36th Annual ACM Symposium on User Interface Software and Technology}} (San Francisco, CA, USA) \emph{(\bibinfo{series}{UIST '23})}. \bibinfo{publisher}{Association for Computing Machinery}, \bibinfo{address}{New York, NY, USA}, Article \bibinfo{articleno}{6}, \bibinfo{numpages}{17}~pages.
\newblock
\showISBNx{9798400701320}
\urldef\tempurl%
\url{https://doi.org/10.1145/3586183.3606777}
\showDOI{\tempurl}


\bibitem[Chung et~al\mbox{.}(2022)]%
        {chung2022talebrush}
\bibfield{author}{\bibinfo{person}{John Joon~Young Chung}, \bibinfo{person}{Wooseok Kim}, \bibinfo{person}{Kang~Min Yoo}, \bibinfo{person}{Hwaran Lee}, \bibinfo{person}{Eytan Adar}, {and} \bibinfo{person}{Minsuk Chang}.} \bibinfo{year}{2022}\natexlab{}.
\newblock \showarticletitle{TaleBrush: Sketching stories with generative pretrained language models}. In \bibinfo{booktitle}{\emph{Proceedings of the 2022 CHI Conference on Human Factors in Computing Systems}}. \bibinfo{pages}{1--19}.
\newblock


\bibitem[Council(2023)]%
        {doublediamond2023}
\bibfield{author}{\bibinfo{person}{British~Design Council}.} \bibinfo{year}{2023}\natexlab{}.
\newblock \bibinfo{title}{The Double Diamond}.
\newblock \bibinfo{howpublished}{\url{https://www.designcouncil.org.uk/our-resources/the-double-diamond/}}.
\newblock
\newblock
\shownote{Retrieved 11 November 2024}.


\bibitem[Dam(2020)]%
        {ideo}
\bibfield{author}{\bibinfo{person}{Rikke~Friis Dam}.} \bibinfo{year}{2020}\natexlab{}.
\newblock \bibinfo{title}{5 Stages in the design thinking process}.
\newblock \bibinfo{howpublished}{\url{https://www.interaction-design.org/literature/article/5-stages-in-the-design-thinking-process?ref=weekly.ui-patterns.com}}.
\newblock


\bibitem[Do and Gross(2001)]%
        {do2001thinking}
\bibfield{author}{\bibinfo{person}{Ellen Yi-Luen Do} {and} \bibinfo{person}{Mark~D Gross}.} \bibinfo{year}{2001}\natexlab{}.
\newblock \showarticletitle{Thinking with diagrams in architectural design}.
\newblock \bibinfo{journal}{\emph{Artificial intelligence review}}  \bibinfo{volume}{15} (\bibinfo{year}{2001}), \bibinfo{pages}{135--149}.
\newblock


\bibitem[D{\"o}rner(1999)]%
        {dorner1999approaching}
\bibfield{author}{\bibinfo{person}{Dietrich D{\"o}rner}.} \bibinfo{year}{1999}\natexlab{}.
\newblock \showarticletitle{Approaching design thinking research}.
\newblock \bibinfo{journal}{\emph{Design Studies}} \bibinfo{volume}{20}, \bibinfo{number}{5} (\bibinfo{year}{1999}), \bibinfo{pages}{407--415}.
\newblock


\bibitem[Dow et~al\mbox{.}(2009)]%
        {dow2009efficacy}
\bibfield{author}{\bibinfo{person}{Steven~P Dow}, \bibinfo{person}{Kate Heddleston}, {and} \bibinfo{person}{Scott~R Klemmer}.} \bibinfo{year}{2009}\natexlab{}.
\newblock \showarticletitle{The efficacy of prototyping under time constraints}. In \bibinfo{booktitle}{\emph{Proceedings of the seventh ACM conference on Creativity and cognition}}. \bibinfo{pages}{165--174}.
\newblock


\bibitem[Duan et~al\mbox{.}(2024)]%
        {duan2024generating}
\bibfield{author}{\bibinfo{person}{Peitong Duan}, \bibinfo{person}{Jeremy Warner}, \bibinfo{person}{Yang Li}, {and} \bibinfo{person}{Bjoern Hartmann}.} \bibinfo{year}{2024}\natexlab{}.
\newblock \showarticletitle{Generating Automatic Feedback on UI Mockups with Large Language Models}. In \bibinfo{booktitle}{\emph{Proceedings of the CHI Conference on Human Factors in Computing Systems}}. \bibinfo{pages}{1--20}.
\newblock


\bibitem[E et~al\mbox{.}(2024)]%
        {jane2024}
\bibfield{author}{\bibinfo{person}{Jane~L. E}, \bibinfo{person}{Yu-Chun~Grace Yen}, \bibinfo{person}{Isabelle~Yan Pan}, \bibinfo{person}{Grace Lin}, \bibinfo{person}{Mingyi Li}, \bibinfo{person}{Hyoungwook Jin}, \bibinfo{person}{Mengyi Chen}, \bibinfo{person}{Haijun Xia}, {and} \bibinfo{person}{Steven~P. Dow}.} \bibinfo{year}{2024}\natexlab{}.
\newblock \showarticletitle{When to Give Feedback: Exploring Tradeoffs in the Timing of Design Feedback}. In \bibinfo{booktitle}{\emph{Proceedings of the 16th Conference on Creativity \& Cognition}} (Chicago, IL, USA) \emph{(\bibinfo{series}{C\&C '24})}. \bibinfo{publisher}{Association for Computing Machinery}, \bibinfo{address}{New York, NY, USA}, \bibinfo{pages}{292–310}.
\newblock
\showISBNx{9798400704857}
\urldef\tempurl%
\url{https://doi.org/10.1145/3635636.3656183}
\showDOI{\tempurl}


\bibitem[Gama et~al\mbox{.}(2023)]%
        {gama2023developers}
\bibfield{author}{\bibinfo{person}{Kiev Gama}, \bibinfo{person}{George Valen{\c{c}}a}, \bibinfo{person}{Pedro Alessio}, \bibinfo{person}{Rafael Formiga}, \bibinfo{person}{Andr{\'e} Neves}, {and} \bibinfo{person}{Nycolas Lacerda}.} \bibinfo{year}{2023}\natexlab{}.
\newblock \showarticletitle{The developers’ design thinking toolbox in hackathons: a study on the recurring design methods in software development marathons}.
\newblock \bibinfo{journal}{\emph{International Journal of Human--Computer Interaction}} \bibinfo{volume}{39}, \bibinfo{number}{12} (\bibinfo{year}{2023}), \bibinfo{pages}{2269--2291}.
\newblock


\bibitem[Goodman et~al\mbox{.}(2011)]%
        {goodman2011understanding}
\bibfield{author}{\bibinfo{person}{Elizabeth Goodman}, \bibinfo{person}{Erik Stolterman}, {and} \bibinfo{person}{Ron Wakkary}.} \bibinfo{year}{2011}\natexlab{}.
\newblock \showarticletitle{Understanding interaction design practices}. In \bibinfo{booktitle}{\emph{Proceedings of the SIGCHI conference on human factors in computing systems}}. \bibinfo{pages}{1061--1070}.
\newblock


\bibitem[Greenwood et~al\mbox{.}(2019)]%
        {greenwood2019dissensus}
\bibfield{author}{\bibinfo{person}{April Greenwood}, \bibinfo{person}{Benjamin Lauren}, \bibinfo{person}{Jessica Knott}, {and} \bibinfo{person}{D{\`a}nielle~Nicole DeVoss}.} \bibinfo{year}{2019}\natexlab{}.
\newblock \showarticletitle{Dissensus, resistance, and ideology: Design thinking as a rhetorical methodology}.
\newblock \bibinfo{journal}{\emph{Journal of Business and Technical Communication}} \bibinfo{volume}{33}, \bibinfo{number}{4} (\bibinfo{year}{2019}), \bibinfo{pages}{400--424}.
\newblock
\urldef\tempurl%
\url{https://doi.org/10.1177/1050651919854063}
\showDOI{\tempurl}


\bibitem[Greever(2015)]%
        {greever2015articulating}
\bibfield{author}{\bibinfo{person}{Tom Greever}.} \bibinfo{year}{2015}\natexlab{}.
\newblock \bibinfo{booktitle}{\emph{Articulating design decisions: Communicate with stakeholders, keep your sanity, and deliver the best user experience}}.
\newblock \bibinfo{publisher}{" O'Reilly Media, Inc."}.
\newblock


\bibitem[H{\"a}m{\"a}l{\"a}inen et~al\mbox{.}(2023)]%
        {hamalainen2023evaluating}
\bibfield{author}{\bibinfo{person}{Perttu H{\"a}m{\"a}l{\"a}inen}, \bibinfo{person}{Mikke Tavast}, {and} \bibinfo{person}{Anton Kunnari}.} \bibinfo{year}{2023}\natexlab{}.
\newblock \showarticletitle{Evaluating large language models in generating synthetic hci research data: a case study}. In \bibinfo{booktitle}{\emph{Proceedings of the 2023 CHI Conference on Human Factors in Computing Systems}}. \bibinfo{pages}{1--19}.
\newblock


\bibitem[Hartmann et~al\mbox{.}(2008)]%
        {hartmann2008design}
\bibfield{author}{\bibinfo{person}{Bj{\"o}rn Hartmann}, \bibinfo{person}{Loren Yu}, \bibinfo{person}{Abel Allison}, \bibinfo{person}{Yeonsoo Yang}, {and} \bibinfo{person}{Scott~R Klemmer}.} \bibinfo{year}{2008}\natexlab{}.
\newblock \showarticletitle{Design as exploration: creating interface alternatives through parallel authoring and runtime tuning}. In \bibinfo{booktitle}{\emph{Proceedings of the 21st annual ACM symposium on User interface software and technology}}. \bibinfo{pages}{91--100}.
\newblock


\bibitem[He et~al\mbox{.}(2024)]%
        {he2024interactive}
\bibfield{author}{\bibinfo{person}{Rui He}, \bibinfo{person}{Huaxin Wei}, {and} \bibinfo{person}{Ying Cao}.} \bibinfo{year}{2024}\natexlab{}.
\newblock \showarticletitle{An Interactive System for Supporting Creative Exploration of Cinematic Composition Designs}. In \bibinfo{booktitle}{\emph{Proceedings of the 37th Annual ACM Symposium on User Interface Software and Technology}}. \bibinfo{pages}{1--15}.
\newblock


\bibitem[Hehn et~al\mbox{.}(2018)]%
        {hehn2018design}
\bibfield{author}{\bibinfo{person}{Jennifer Hehn}, \bibinfo{person}{Falk Uebernickel}, {and} \bibinfo{person}{Matthias Herterich}.} \bibinfo{year}{2018}\natexlab{}.
\newblock \showarticletitle{Design thinking methods for service innovation-A delphi study}.
\newblock  (\bibinfo{year}{2018}).
\newblock
\urldef\tempurl%
\url{https://aisel.aisnet.org/pacis2018/126/}
\showURL{%
\tempurl}


\bibitem[Hung et~al\mbox{.}(2024)]%
        {Hung2024SimTubeGS}
\bibfield{author}{\bibinfo{person}{Yu-Kai Hung}, \bibinfo{person}{Yun-Chien Huang}, \bibinfo{person}{Ting-Yu Su}, \bibinfo{person}{Yen-Ting Lin}, \bibinfo{person}{Lung-Pan Cheng}, \bibinfo{person}{Bryan Wang}, {and} \bibinfo{person}{Shao-Hua Sun}.} \bibinfo{year}{2024}\natexlab{}.
\newblock \showarticletitle{SimTube: Generating Simulated Video Comments through Multimodal AI and User Personas}.
\newblock \bibinfo{journal}{\emph{ArXiv}}  \bibinfo{volume}{abs/2411.09577} (\bibinfo{year}{2024}).
\newblock
\urldef\tempurl%
\url{https://api.semanticscholar.org/CorpusID:274023042}
\showURL{%
\tempurl}


\bibitem[Jiang et~al\mbox{.}(2023)]%
        {jiang2023graphologue}
\bibfield{author}{\bibinfo{person}{Peiling Jiang}, \bibinfo{person}{Jude Rayan}, \bibinfo{person}{Steven~P Dow}, {and} \bibinfo{person}{Haijun Xia}.} \bibinfo{year}{2023}\natexlab{}.
\newblock \showarticletitle{Graphologue: Exploring large language model responses with interactive diagrams}. In \bibinfo{booktitle}{\emph{Proceedings of the 36th annual ACM symposium on user interface software and technology}}. \bibinfo{pages}{1--20}.
\newblock


\bibitem[Jung et~al\mbox{.}(2023)]%
        {jung2023toward}
\bibfield{author}{\bibinfo{person}{Hyunggu Jung}, \bibinfo{person}{Woosuk Seo}, \bibinfo{person}{Seokwoo Song}, {and} \bibinfo{person}{Sungmin Na}.} \bibinfo{year}{2023}\natexlab{}.
\newblock \showarticletitle{Toward Value Scenario Generation Through Large Language Models}. In \bibinfo{booktitle}{\emph{Companion Publication of the 2023 Conference on Computer Supported Cooperative Work and Social Computing}}. \bibinfo{pages}{212--220}.
\newblock


\bibitem[Kazemitabaar et~al\mbox{.}(2025)]%
        {kazemitabaar2025exploring}
\bibfield{author}{\bibinfo{person}{Majeed Kazemitabaar}, \bibinfo{person}{Oliver Huang}, \bibinfo{person}{Sangho Suh}, \bibinfo{person}{Austin~Z Henley}, {and} \bibinfo{person}{Tovi Grossman}.} \bibinfo{year}{2025}\natexlab{}.
\newblock \showarticletitle{Exploring the design space of cognitive engagement techniques with ai-generated code for enhanced learning}. In \bibinfo{booktitle}{\emph{Proceedings of the 30th International Conference on Intelligent User Interfaces}}. \bibinfo{pages}{695--714}.
\newblock


\bibitem[Kim et~al\mbox{.}(2023)]%
        {kim2023metaphorian}
\bibfield{author}{\bibinfo{person}{Jeongyeon Kim}, \bibinfo{person}{Sangho Suh}, \bibinfo{person}{Lydia~B Chilton}, {and} \bibinfo{person}{Haijun Xia}.} \bibinfo{year}{2023}\natexlab{}.
\newblock \showarticletitle{Metaphorian: leveraging large language models to support extended metaphor creation for science writing}. In \bibinfo{booktitle}{\emph{Proceedings of the 2023 ACM Designing Interactive Systems Conference}}. \bibinfo{pages}{115--135}.
\newblock


\bibitem[Kirsh(2010)]%
        {kirsh2010thinking}
\bibfield{author}{\bibinfo{person}{David Kirsh}.} \bibinfo{year}{2010}\natexlab{}.
\newblock \showarticletitle{Thinking with external representations}.
\newblock \bibinfo{journal}{\emph{AI \& society}}  \bibinfo{volume}{25} (\bibinfo{year}{2010}), \bibinfo{pages}{441--454}.
\newblock


\bibitem[Kleinsmann et~al\mbox{.}(2017)]%
        {kleinsmann2017capturing}
\bibfield{author}{\bibinfo{person}{Maaike Kleinsmann}, \bibinfo{person}{Rianne Valkenburg}, {and} \bibinfo{person}{Janneke Sluijs}.} \bibinfo{year}{2017}\natexlab{}.
\newblock \showarticletitle{Capturing the value of design thinking in different innovation practices}.
\newblock \bibinfo{journal}{\emph{International Journal of Design}} \bibinfo{volume}{11}, \bibinfo{number}{2} (\bibinfo{year}{2017}), \bibinfo{pages}{25--40}.
\newblock


\bibitem[Lee et~al\mbox{.}(2022)]%
        {lee2022coauthor}
\bibfield{author}{\bibinfo{person}{Mina Lee}, \bibinfo{person}{Percy Liang}, {and} \bibinfo{person}{Qian Yang}.} \bibinfo{year}{2022}\natexlab{}.
\newblock \showarticletitle{Coauthor: Designing a human-ai collaborative writing dataset for exploring language model capabilities}. In \bibinfo{booktitle}{\emph{Proceedings of the 2022 CHI conference on human factors in computing systems}}. \bibinfo{pages}{1--19}.
\newblock


\bibitem[Lin et~al\mbox{.}(2025)]%
        {Lin2025InkspireSD}
\bibfield{author}{\bibinfo{person}{David Chuan-En Lin}, \bibinfo{person}{Hyeonsu~B Kang}, \bibinfo{person}{Nikolas Martelaro}, \bibinfo{person}{Aniket Kittur}, \bibinfo{person}{Yan-Ying Chen}, {and} \bibinfo{person}{Matthew~K. Hong}.} \bibinfo{year}{2025}\natexlab{}.
\newblock \showarticletitle{Inkspire: Supporting Design Exploration with Generative AI through Analogical Sketching}.
\newblock \bibinfo{journal}{\emph{ArXiv}}  \bibinfo{volume}{abs/2501.18588} (\bibinfo{year}{2025}).
\newblock
\urldef\tempurl%
\url{https://api.semanticscholar.org/CorpusID:275993583}
\showURL{%
\tempurl}


\bibitem[Liu et~al\mbox{.}(2023)]%
        {liu20233dall}
\bibfield{author}{\bibinfo{person}{Vivian Liu}, \bibinfo{person}{Jo Vermeulen}, \bibinfo{person}{George Fitzmaurice}, {and} \bibinfo{person}{Justin Matejka}.} \bibinfo{year}{2023}\natexlab{}.
\newblock \showarticletitle{3DALL-E: Integrating text-to-image AI in 3D design workflows}. In \bibinfo{booktitle}{\emph{Proceedings of the 2023 ACM designing interactive systems conference}}. \bibinfo{pages}{1955--1977}.
\newblock


\bibitem[Ma et~al\mbox{.}(2025)]%
        {ma2025dynex}
\bibfield{author}{\bibinfo{person}{Jenny~GuangZhen Ma}, \bibinfo{person}{Karthik Sreedhar}, \bibinfo{person}{Vivian Liu}, \bibinfo{person}{Pedro~A Perez}, \bibinfo{person}{Sitong Wang}, \bibinfo{person}{Riya Sahni}, {and} \bibinfo{person}{Lydia~B Chilton}.} \bibinfo{year}{2025}\natexlab{}.
\newblock \showarticletitle{Dynex: Dynamic code synthesis with structured design exploration for accelerated exploratory programming}. In \bibinfo{booktitle}{\emph{Proceedings of the 2025 CHI Conference on Human Factors in Computing Systems}}. \bibinfo{pages}{1--27}.
\newblock


\bibitem[Maceli et~al\mbox{.}(2024)]%
        {maceli2024incorporating}
\bibfield{author}{\bibinfo{person}{Monica Maceli}, \bibinfo{person}{Nancy Smith}, {and} \bibinfo{person}{Gatha Bhakta}.} \bibinfo{year}{2024}\natexlab{}.
\newblock \showarticletitle{Incorporating Unanticipated Uses of Generative AI into HCI Education}. In \bibinfo{booktitle}{\emph{Proceedings of the 6th Annual Symposium on HCI Education}}. \bibinfo{pages}{1--7}.
\newblock


\bibitem[Masson et~al\mbox{.}(2025)]%
        {masson2025textoshop}
\bibfield{author}{\bibinfo{person}{Damien Masson}, \bibinfo{person}{Young-Ho Kim}, {and} \bibinfo{person}{Fanny Chevalier}.} \bibinfo{year}{2025}\natexlab{}.
\newblock \showarticletitle{Textoshop: Interactions Inspired by Drawing Software to Facilitate Text Editing}. In \bibinfo{booktitle}{\emph{Proceedings of the 2025 CHI Conference on Human Factors in Computing Systems}}. \bibinfo{pages}{1--14}.
\newblock


\bibitem[Masson et~al\mbox{.}(2024)]%
        {masson2024directgpt}
\bibfield{author}{\bibinfo{person}{Damien Masson}, \bibinfo{person}{Sylvain Malacria}, \bibinfo{person}{G{\'e}ry Casiez}, {and} \bibinfo{person}{Daniel Vogel}.} \bibinfo{year}{2024}\natexlab{}.
\newblock \showarticletitle{Directgpt: A direct manipulation interface to interact with large language models}. In \bibinfo{booktitle}{\emph{Proceedings of the 2024 CHI Conference on Human Factors in Computing Systems}}. \bibinfo{pages}{1--16}.
\newblock


\bibitem[Micheli et~al\mbox{.}(2019)]%
        {micheli2019doing}
\bibfield{author}{\bibinfo{person}{Pietro Micheli}, \bibinfo{person}{Sarah~JS Wilner}, \bibinfo{person}{Sabeen~Hussain Bhatti}, \bibinfo{person}{Matteo Mura}, {and} \bibinfo{person}{Michael~B Beverland}.} \bibinfo{year}{2019}\natexlab{}.
\newblock \showarticletitle{Doing design thinking: Conceptual review, synthesis, and research agenda}.
\newblock \bibinfo{journal}{\emph{Journal of Product innovation management}} \bibinfo{volume}{36}, \bibinfo{number}{2} (\bibinfo{year}{2019}), \bibinfo{pages}{124--148}.
\newblock


\bibitem[Nguyen et~al\mbox{.}(2024)]%
        {nguyen2024simulating}
\bibfield{author}{\bibinfo{person}{Ha Nguyen}, \bibinfo{person}{Victoria Nguyen}, \bibinfo{person}{Sar{\'\i}ah L{\'o}pez-Fierro}, \bibinfo{person}{Sara Ludovise}, {and} \bibinfo{person}{Rossella Santagata}.} \bibinfo{year}{2024}\natexlab{}.
\newblock \showarticletitle{Simulating Climate Change Discussion with Large Language Models: Considerations for Science Communication at Scale}. In \bibinfo{booktitle}{\emph{Proceedings of the Eleventh ACM Conference on Learning@ Scale}}. \bibinfo{pages}{28--38}.
\newblock


\bibitem[Park et~al\mbox{.}(2022)]%
        {park2022social}
\bibfield{author}{\bibinfo{person}{Joon~Sung Park}, \bibinfo{person}{Lindsay Popowski}, \bibinfo{person}{Carrie Cai}, \bibinfo{person}{Meredith~Ringel Morris}, \bibinfo{person}{Percy Liang}, {and} \bibinfo{person}{Michael~S Bernstein}.} \bibinfo{year}{2022}\natexlab{}.
\newblock \showarticletitle{Social simulacra: Creating populated prototypes for social computing systems}. In \bibinfo{booktitle}{\emph{Proceedings of the 35th Annual ACM Symposium on User Interface Software and Technology}}. \bibinfo{pages}{1--18}.
\newblock


\bibitem[Pruitt and Grudin(2003)]%
        {pruitt2003personas}
\bibfield{author}{\bibinfo{person}{John Pruitt} {and} \bibinfo{person}{Jonathan Grudin}.} \bibinfo{year}{2003}\natexlab{}.
\newblock \showarticletitle{Personas: practice and theory}. In \bibinfo{booktitle}{\emph{Proceedings of the 2003 conference on Designing for user experiences}}. \bibinfo{pages}{1--15}.
\newblock


\bibitem[Pu et~al\mbox{.}(2025)]%
        {pu2025ideasynth}
\bibfield{author}{\bibinfo{person}{Kevin Pu}, \bibinfo{person}{KJ~Kevin Feng}, \bibinfo{person}{Tovi Grossman}, \bibinfo{person}{Tom Hope}, \bibinfo{person}{Bhavana Dalvi~Mishra}, \bibinfo{person}{Matt Latzke}, \bibinfo{person}{Jonathan Bragg}, \bibinfo{person}{Joseph~Chee Chang}, {and} \bibinfo{person}{Pao Siangliulue}.} \bibinfo{year}{2025}\natexlab{}.
\newblock \showarticletitle{Ideasynth: Iterative research idea development through evolving and composing idea facets with literature-grounded feedback}. In \bibinfo{booktitle}{\emph{Proceedings of the 2025 CHI Conference on Human Factors in Computing Systems}}. \bibinfo{pages}{1--31}.
\newblock


\bibitem[Quesenbery and Brooks(2010)]%
        {quesenbery2010storytelling}
\bibfield{author}{\bibinfo{person}{Whitney Quesenbery} {and} \bibinfo{person}{Kevin Brooks}.} \bibinfo{year}{2010}\natexlab{}.
\newblock \bibinfo{booktitle}{\emph{Storytelling for user experience: Crafting stories for better design}}.
\newblock \bibinfo{publisher}{Rosenfeld Media}.
\newblock


\bibitem[Razzouk and Shute(2012)]%
        {razzouk2012design}
\bibfield{author}{\bibinfo{person}{Rim Razzouk} {and} \bibinfo{person}{Valerie Shute}.} \bibinfo{year}{2012}\natexlab{}.
\newblock \showarticletitle{What is design thinking and why is it important?}
\newblock \bibinfo{journal}{\emph{Review of educational research}} \bibinfo{volume}{82}, \bibinfo{number}{3} (\bibinfo{year}{2012}), \bibinfo{pages}{330--348}.
\newblock


\bibitem[Rogers(2004)]%
        {rogers2004new}
\bibfield{author}{\bibinfo{person}{Yvonne Rogers}.} \bibinfo{year}{2004}\natexlab{}.
\newblock \showarticletitle{New theoretical approaches for HCI}.
\newblock \bibinfo{journal}{\emph{Annual review of information science and technology}} \bibinfo{volume}{38}, \bibinfo{number}{1} (\bibinfo{year}{2004}), \bibinfo{pages}{87--143}.
\newblock


\bibitem[Rose and Reimer(2022)]%
        {rose2022}
\bibfield{author}{\bibinfo{person}{Emma~J Rose} {and} \bibinfo{person}{Cody Reimer}.} \bibinfo{year}{2022}\natexlab{}.
\newblock \bibinfo{booktitle}{\emph{6. Iteration}}.
\newblock \bibinfo{publisher}{WAC Clearinghouse}.
\newblock
\urldef\tempurl%
\url{https://doi.org/10.37514/TPC-B.2022.1725.2.06}
\showDOI{\tempurl}


\bibitem[Rosson and Carroll(2007)]%
        {rosson2007scenario}
\bibfield{author}{\bibinfo{person}{Mary~Beth Rosson} {and} \bibinfo{person}{John~M Carroll}.} \bibinfo{year}{2007}\natexlab{}.
\newblock \showarticletitle{Scenario-based design}.
\newblock In \bibinfo{booktitle}{\emph{The human-computer interaction handbook}}. \bibinfo{publisher}{CRC Press}, \bibinfo{pages}{1067--1086}.
\newblock


\bibitem[Salminen et~al\mbox{.}(2022)]%
        {salminen2022creating}
\bibfield{author}{\bibinfo{person}{Joni Salminen}, \bibinfo{person}{Soon-Gyo Jung}, \bibinfo{person}{Lene Nielsen}, {and} \bibinfo{person}{Bernard Jansen}.} \bibinfo{year}{2022}\natexlab{}.
\newblock \showarticletitle{Creating More Personas Improves Representation of Demographically Diverse Populations: Implications Towards Interactive Persona Systems}. In \bibinfo{booktitle}{\emph{Nordic Human-Computer Interaction Conference}}. \bibinfo{pages}{1--11}.
\newblock


\bibitem[Salminen et~al\mbox{.}(2024)]%
        {salminen2024deus}
\bibfield{author}{\bibinfo{person}{Joni Salminen}, \bibinfo{person}{Chang Liu}, \bibinfo{person}{Wenjing Pian}, \bibinfo{person}{Jianxing Chi}, \bibinfo{person}{Essi H{\"a}yh{\"a}nen}, {and} \bibinfo{person}{Bernard~J Jansen}.} \bibinfo{year}{2024}\natexlab{}.
\newblock \showarticletitle{Deus Ex Machina and Personas from Large Language Models: Investigating the Composition of AI-Generated Persona Descriptions}. In \bibinfo{booktitle}{\emph{Proceedings of the CHI Conference on Human Factors in Computing Systems}}. \bibinfo{pages}{1--20}.
\newblock


\bibitem[Schmidt et~al\mbox{.}(2024)]%
        {schmidt2024simulating}
\bibfield{author}{\bibinfo{person}{Albrecht Schmidt}, \bibinfo{person}{Passant Elagroudy}, \bibinfo{person}{Fiona Draxler}, \bibinfo{person}{Frauke Kreuter}, {and} \bibinfo{person}{Robin Welsch}.} \bibinfo{year}{2024}\natexlab{}.
\newblock \showarticletitle{Simulating the human in HCD with ChatGPT: Redesigning interaction design with AI}.
\newblock \bibinfo{journal}{\emph{Interactions}} \bibinfo{volume}{31}, \bibinfo{number}{1} (\bibinfo{year}{2024}), \bibinfo{pages}{24--31}.
\newblock


\bibitem[Schrage(1999)]%
        {schrage1999serious}
\bibfield{author}{\bibinfo{person}{Michael Schrage}.} \bibinfo{year}{1999}\natexlab{}.
\newblock \bibinfo{booktitle}{\emph{Serious play: How the world's best companies simulate to innovate}}.
\newblock \bibinfo{publisher}{Harvard Business Press}.
\newblock


\bibitem[Schuller et~al\mbox{.}(2024)]%
        {schuller2024generating}
\bibfield{author}{\bibinfo{person}{Andreas Schuller}, \bibinfo{person}{Doris Janssen}, \bibinfo{person}{Julian Blumenr{\"o}ther}, \bibinfo{person}{Theresa~Maria Probst}, \bibinfo{person}{Michael Schmidt}, {and} \bibinfo{person}{Chandan Kumar}.} \bibinfo{year}{2024}\natexlab{}.
\newblock \showarticletitle{Generating personas using LLMs and assessing their viability}. In \bibinfo{booktitle}{\emph{Extended Abstracts of the CHI Conference on Human Factors in Computing Systems}}. \bibinfo{pages}{1--7}.
\newblock


\bibitem[Shi et~al\mbox{.}(2025)]%
        {Shi2025BrickifyEE}
\bibfield{author}{\bibinfo{person}{Xinyu Shi}, \bibinfo{person}{Yinghou Wang}, \bibinfo{person}{Ryan Rossi}, {and} \bibinfo{person}{Jian Zhao}.} \bibinfo{year}{2025}\natexlab{}.
\newblock \showarticletitle{Brickify: Enabling Expressive Design Intent Specification through Direct Manipulation on Design Tokens}.
\newblock \bibinfo{journal}{\emph{ArXiv}}  \bibinfo{volume}{abs/2502.21219} (\bibinfo{year}{2025}).
\newblock
\urldef\tempurl%
\url{https://api.semanticscholar.org/CorpusID:276649308}
\showURL{%
\tempurl}


\bibitem[Shin et~al\mbox{.}(2024)]%
        {shin2024understanding}
\bibfield{author}{\bibinfo{person}{Joongi Shin}, \bibinfo{person}{Michael~A Hedderich}, \bibinfo{person}{Bart{\l}omiej~Jakub Rey}, \bibinfo{person}{Andr{\'e}s Lucero}, {and} \bibinfo{person}{Antti Oulasvirta}.} \bibinfo{year}{2024}\natexlab{}.
\newblock \showarticletitle{Understanding Human-AI Workflows for Generating Personas}. In \bibinfo{booktitle}{\emph{Proceedings of the 2024 ACM Designing Interactive Systems Conference}}. \bibinfo{pages}{757--781}.
\newblock


\bibitem[Shokrizadeh et~al\mbox{.}(2025)]%
        {Shokrizadeh2025DancingWC}
\bibfield{author}{\bibinfo{person}{Atefeh Shokrizadeh}, \bibinfo{person}{Boniface~Bahati Tadjuidje}, \bibinfo{person}{Shivam Kumar}, \bibinfo{person}{Sohan Kamble}, {and} \bibinfo{person}{Jinghui Cheng}.} \bibinfo{year}{2025}\natexlab{}.
\newblock \showarticletitle{Dancing With Chains: Ideating Under Constraints With UIDEC in UI/UX Design}.
\newblock \bibinfo{journal}{\emph{ArXiv}}  \bibinfo{volume}{abs/2501.18748} (\bibinfo{year}{2025}).
\newblock
\urldef\tempurl%
\url{https://api.semanticscholar.org/CorpusID:276079770}
\showURL{%
\tempurl}


\bibitem[Son et~al\mbox{.}(2024)]%
        {son2024}
\bibfield{author}{\bibinfo{person}{Kihoon Son}, \bibinfo{person}{DaEun Choi}, \bibinfo{person}{Tae~Soo Kim}, \bibinfo{person}{Young-Ho Kim}, {and} \bibinfo{person}{Juho Kim}.} \bibinfo{year}{2024}\natexlab{}.
\newblock \showarticletitle{GenQuery: Supporting Expressive Visual Search with Generative Models}. In \bibinfo{booktitle}{\emph{Proceedings of the CHI Conference on Human Factors in Computing Systems}} (Honolulu, HI, USA) \emph{(\bibinfo{series}{CHI '24})}. \bibinfo{publisher}{Association for Computing Machinery}, \bibinfo{address}{New York, NY, USA}, Article \bibinfo{articleno}{180}, \bibinfo{numpages}{19}~pages.
\newblock
\showISBNx{9798400703300}
\urldef\tempurl%
\url{https://doi.org/10.1145/3613904.3642847}
\showDOI{\tempurl}


\bibitem[Stolterman(2008)]%
        {stolterman2008nature}
\bibfield{author}{\bibinfo{person}{Erik Stolterman}.} \bibinfo{year}{2008}\natexlab{}.
\newblock \showarticletitle{The nature of design practice and implications for interaction design research}.
\newblock \bibinfo{journal}{\emph{International Journal of Design}} \bibinfo{volume}{2}, \bibinfo{number}{1} (\bibinfo{year}{2008}).
\newblock


\bibitem[Suh et~al\mbox{.}(2024)]%
        {suh2024luminate}
\bibfield{author}{\bibinfo{person}{Sangho Suh}, \bibinfo{person}{Meng Chen}, \bibinfo{person}{Bryan Min}, \bibinfo{person}{Toby Jia-Jun Li}, {and} \bibinfo{person}{Haijun Xia}.} \bibinfo{year}{2024}\natexlab{}.
\newblock \showarticletitle{Luminate: Structured Generation and Exploration of Design Space with Large Language Models for Human-AI Co-Creation}. In \bibinfo{booktitle}{\emph{Proceedings of the CHI Conference on Human Factors in Computing Systems}}. \bibinfo{pages}{1--26}.
\newblock


\bibitem[Suh et~al\mbox{.}(2023)]%
        {suh2023sensecape}
\bibfield{author}{\bibinfo{person}{Sangho Suh}, \bibinfo{person}{Bryan Min}, \bibinfo{person}{Srishti Palani}, {and} \bibinfo{person}{Haijun Xia}.} \bibinfo{year}{2023}\natexlab{}.
\newblock \showarticletitle{Sensecape: Enabling multilevel exploration and sensemaking with large language models}. In \bibinfo{booktitle}{\emph{Proceedings of the 36th Annual ACM Symposium on User Interface Software and Technology}}. \bibinfo{pages}{1--18}.
\newblock


\bibitem[Tantiyaswasdikul(2020)]%
        {tantiyaswasdikul2020design}
\bibfield{author}{\bibinfo{person}{Kallaya Tantiyaswasdikul}.} \bibinfo{year}{2020}\natexlab{}.
\newblock \showarticletitle{How design thinking can foster environmental sustainability: Integrating design thinking into circular design guide}. In \bibinfo{booktitle}{\emph{Proceedings of the 6th International Conference on Industrial and Business Engineering}}. \bibinfo{pages}{157--162}.
\newblock


\bibitem[Tham(2022)]%
        {tham2022keywords}
\bibfield{author}{\bibinfo{person}{Jason Chew~Kit Tham}.} \bibinfo{year}{2022}\natexlab{}.
\newblock \bibinfo{booktitle}{\emph{Keywords in Design Thinking: A Lexical Primer for Technical Communicators \& Designers}}.
\newblock \bibinfo{publisher}{WAC Clearinghouse}.
\newblock


\bibitem[Truong et~al\mbox{.}(2006)]%
        {truong2006storyboarding}
\bibfield{author}{\bibinfo{person}{Khai~N Truong}, \bibinfo{person}{Gillian~R Hayes}, {and} \bibinfo{person}{Gregory~D Abowd}.} \bibinfo{year}{2006}\natexlab{}.
\newblock \showarticletitle{Storyboarding: an empirical determination of best practices and effective guidelines}. In \bibinfo{booktitle}{\emph{Proceedings of the 6th conference on Designing Interactive systems}}. \bibinfo{pages}{12--21}.
\newblock


\bibitem[Urgo and Arguello(2022)]%
        {urgo2022learning}
\bibfield{author}{\bibinfo{person}{Kelsey Urgo} {and} \bibinfo{person}{Jaime Arguello}.} \bibinfo{year}{2022}\natexlab{}.
\newblock \showarticletitle{Learning assessments in search-as-learning: A survey of prior work and opportunities for future research}.
\newblock \bibinfo{journal}{\emph{Information Processing \& Management}} \bibinfo{volume}{59}, \bibinfo{number}{2} (\bibinfo{year}{2022}), \bibinfo{pages}{102821}.
\newblock


\bibitem[Wadinambiarachchi et~al\mbox{.}(2024)]%
        {wadinambiarachchi2024effects}
\bibfield{author}{\bibinfo{person}{Samangi Wadinambiarachchi}, \bibinfo{person}{Ryan~M Kelly}, \bibinfo{person}{Saumya Pareek}, \bibinfo{person}{Qiushi Zhou}, {and} \bibinfo{person}{Eduardo Velloso}.} \bibinfo{year}{2024}\natexlab{}.
\newblock \showarticletitle{The Effects of Generative AI on Design Fixation and Divergent Thinking}. In \bibinfo{booktitle}{\emph{Proceedings of the CHI Conference on Human Factors in Computing Systems}}. \bibinfo{pages}{1--18}.
\newblock


\bibitem[Wang et~al\mbox{.}(2024a)]%
        {wang2024lave}
\bibfield{author}{\bibinfo{person}{Bryan Wang}, \bibinfo{person}{Yuliang Li}, \bibinfo{person}{Zhaoyang Lv}, \bibinfo{person}{Haijun Xia}, \bibinfo{person}{Yan Xu}, {and} \bibinfo{person}{Raj Sodhi}.} \bibinfo{year}{2024}\natexlab{a}.
\newblock \showarticletitle{LAVE: LLM-Powered Agent Assistance and Language Augmentation for Video Editing}. In \bibinfo{booktitle}{\emph{Proceedings of the 29th International Conference on Intelligent User Interfaces}} (Greenville, SC, USA) \emph{(\bibinfo{series}{IUI '24})}. \bibinfo{publisher}{Association for Computing Machinery}, \bibinfo{address}{New York, NY, USA}, \bibinfo{pages}{699–714}.
\newblock
\showISBNx{9798400705083}
\urldef\tempurl%
\url{https://doi.org/10.1145/3640543.3645143}
\showDOI{\tempurl}


\bibitem[Wang et~al\mbox{.}(2024b)]%
        {wang2024roomdreaming}
\bibfield{author}{\bibinfo{person}{Shun-Yu Wang}, \bibinfo{person}{Wei-Chung Su}, \bibinfo{person}{Serena Chen}, \bibinfo{person}{Ching-Yi Tsai}, \bibinfo{person}{Marta Misztal}, \bibinfo{person}{Katherine~M Cheng}, \bibinfo{person}{Alwena Lin}, \bibinfo{person}{Yu Chen}, {and} \bibinfo{person}{Mike~Y Chen}.} \bibinfo{year}{2024}\natexlab{b}.
\newblock \showarticletitle{RoomDreaming: Generative-AI Approach to Facilitating Iterative, Preliminary Interior Design Exploration}. In \bibinfo{booktitle}{\emph{Proceedings of the CHI Conference on Human Factors in Computing Systems}}. \bibinfo{pages}{1--20}.
\newblock


\bibitem[Wang and Xing(2024)]%
        {wang2024research}
\bibfield{author}{\bibinfo{person}{Wenjing Wang} {and} \bibinfo{person}{Baixi Xing}.} \bibinfo{year}{2024}\natexlab{}.
\newblock \showarticletitle{Research on the Integration and Application of Design Thinking and Large Language Models in the Innovation Design of Fintech Products}.
\newblock In \bibinfo{booktitle}{\emph{Design Studies and Intelligence Engineering}}. \bibinfo{publisher}{IOS Press}, \bibinfo{pages}{673--682}.
\newblock


\bibitem[Wang et~al\mbox{.}(2025)]%
        {Wang2025AIdeationDA}
\bibfield{author}{\bibinfo{person}{Wen-Fan Wang}, \bibinfo{person}{Chien-Ting Lu}, \bibinfo{person}{Nil~Ponsa Campanya}, \bibinfo{person}{Bing-Yu Chen}, {and} \bibinfo{person}{Mike~Y. Chen}.} \bibinfo{year}{2025}\natexlab{}.
\newblock \showarticletitle{AIdeation: Designing a Human-AI Collaborative Ideation System for Concept Designers}.
\newblock \bibinfo{journal}{\emph{ArXiv}}  \bibinfo{volume}{abs/2502.14747} (\bibinfo{year}{2025}).
\newblock
\urldef\tempurl%
\url{https://api.semanticscholar.org/CorpusID:276482514}
\showURL{%
\tempurl}


\bibitem[Xiang et~al\mbox{.}(2024)]%
        {xiang2024simuser}
\bibfield{author}{\bibinfo{person}{Wei Xiang}, \bibinfo{person}{Hanfei Zhu}, \bibinfo{person}{Suqi Lou}, \bibinfo{person}{Xinli Chen}, \bibinfo{person}{Zhenghua Pan}, \bibinfo{person}{Yuping Jin}, \bibinfo{person}{Shi Chen}, {and} \bibinfo{person}{Lingyun Sun}.} \bibinfo{year}{2024}\natexlab{}.
\newblock \showarticletitle{SimUser: Generating Usability Feedback by Simulating Various Users Interacting with Mobile Applications}. In \bibinfo{booktitle}{\emph{Proceedings of the CHI Conference on Human Factors in Computing Systems}}. \bibinfo{pages}{1--17}.
\newblock


\bibitem[Xu et~al\mbox{.}(2024)]%
        {xu2024jamplate}
\bibfield{author}{\bibinfo{person}{Xiaotong Xu}, \bibinfo{person}{Jiayu Yin}, \bibinfo{person}{Catherine Gu}, \bibinfo{person}{Jenny Mar}, \bibinfo{person}{Sydney Zhang}, \bibinfo{person}{Jane~L E}, {and} \bibinfo{person}{Steven~P Dow}.} \bibinfo{year}{2024}\natexlab{}.
\newblock \showarticletitle{Jamplate: Exploring LLM-Enhanced Templates for Idea Reflection}. In \bibinfo{booktitle}{\emph{Proceedings of the 29th International Conference on Intelligent User Interfaces}}. \bibinfo{pages}{907--921}.
\newblock


\bibitem[Yen et~al\mbox{.}(2024)]%
        {yen2024processgallery}
\bibfield{author}{\bibinfo{person}{Yu-Chun~Grace Yen}, \bibinfo{person}{Jane~L E}, \bibinfo{person}{Hyoungwook Jin}, \bibinfo{person}{Mingyi Li}, \bibinfo{person}{Grace Lin}, \bibinfo{person}{Isabelle~Yan Pan}, {and} \bibinfo{person}{Steven~P Dow}.} \bibinfo{year}{2024}\natexlab{}.
\newblock \showarticletitle{ProcessGallery: Contrasting Early and Late Iterations for Design Principle Learning}.
\newblock \bibinfo{journal}{\emph{Proceedings of the ACM on Human-Computer Interaction}} \bibinfo{volume}{8}, \bibinfo{number}{CSCW1} (\bibinfo{year}{2024}), \bibinfo{pages}{1--35}.
\newblock


\bibitem[Zhang et~al\mbox{.}(2024)]%
        {zhang2024auto}
\bibfield{author}{\bibinfo{person}{Xishuo Zhang}, \bibinfo{person}{Lin Liu}, \bibinfo{person}{Yi Wang}, \bibinfo{person}{Xiao Liu}, \bibinfo{person}{Hailong Wang}, \bibinfo{person}{Chetan Arora}, \bibinfo{person}{Haichao Liu}, \bibinfo{person}{Weijia Wang}, {and} \bibinfo{person}{Thuong Hoang}.} \bibinfo{year}{2024}\natexlab{}.
\newblock \showarticletitle{Auto-Generated Personas: Enhancing User-centered Design Practices among University Students}. In \bibinfo{booktitle}{\emph{Extended Abstracts of the CHI Conference on Human Factors in Computing Systems}}. \bibinfo{pages}{1--7}.
\newblock


\bibitem[Zhang et~al\mbox{.}(2023)]%
        {zhang2023personagen}
\bibfield{author}{\bibinfo{person}{Xishuo Zhang}, \bibinfo{person}{Lin Liu}, \bibinfo{person}{Yi Wang}, \bibinfo{person}{Xiao Liu}, \bibinfo{person}{Hailong Wang}, \bibinfo{person}{Anqi Ren}, {and} \bibinfo{person}{Chetan Arora}.} \bibinfo{year}{2023}\natexlab{}.
\newblock \showarticletitle{Personagen: A tool for generating personas from user feedback}. In \bibinfo{booktitle}{\emph{2023 IEEE 31st International Requirements Engineering Conference (RE)}}. IEEE, \bibinfo{pages}{353--354}.
\newblock


\bibitem[Zhang and Liu(2020)]%
        {zhang2020users}
\bibfield{author}{\bibinfo{person}{Yao Zhang} {and} \bibinfo{person}{Chang Liu}.} \bibinfo{year}{2020}\natexlab{}.
\newblock \showarticletitle{Users' Knowledge Use and Change during Information Searching Process: A Perspective of Vocabulary Usage}. In \bibinfo{booktitle}{\emph{Proceedings of the ACM/IEEE joint conference on digital libraries in 2020}}. \bibinfo{pages}{47--56}.
\newblock


\end{thebibliography}

\newpage

\onecolumn
\appendix
\section{Appendix}

\subsection{User Study}
\label{sec:user-study}

\begin{figure}[htb!]
    \centering
    \includegraphics[alt={Screenshot of participant 1's design workspace which includes cards for personas, problem statements, solutions and storyboards to explore a design space about encouraging social interaction.}, width=\linewidth]{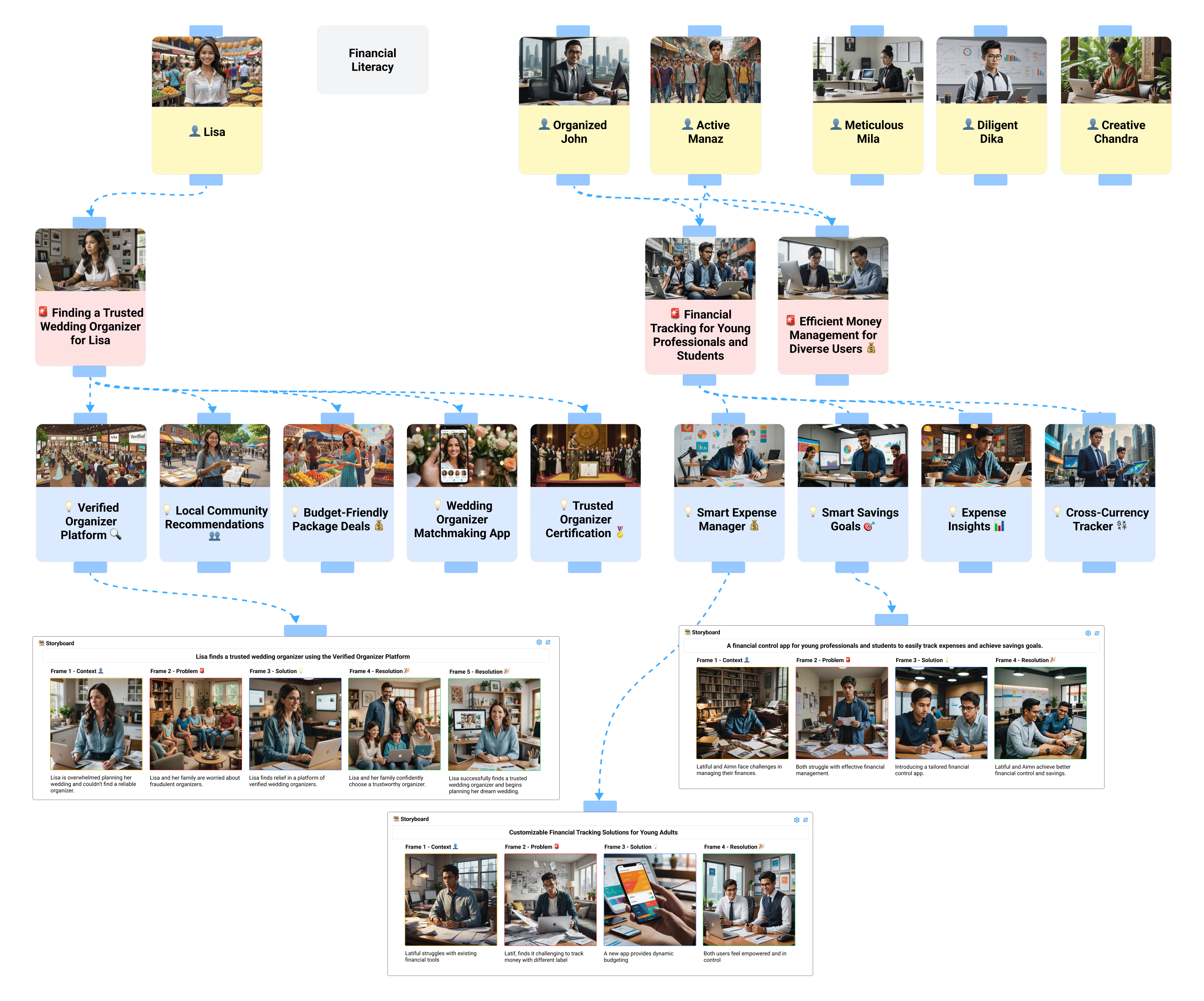}
    \caption{The workspace of P1\textsubscript{prac} after completing the user study task}
    \label{fig:p1-screenshot}
\end{figure}

\begin{figure}[htb!]
    \centering
    \includegraphics[alt={Screenshot of participant 2's design workspace which includes cards for personas, problem statements, solutions and storyboards to explore the design space: `Personalized News Consumption.'}, width=0.75\linewidth]{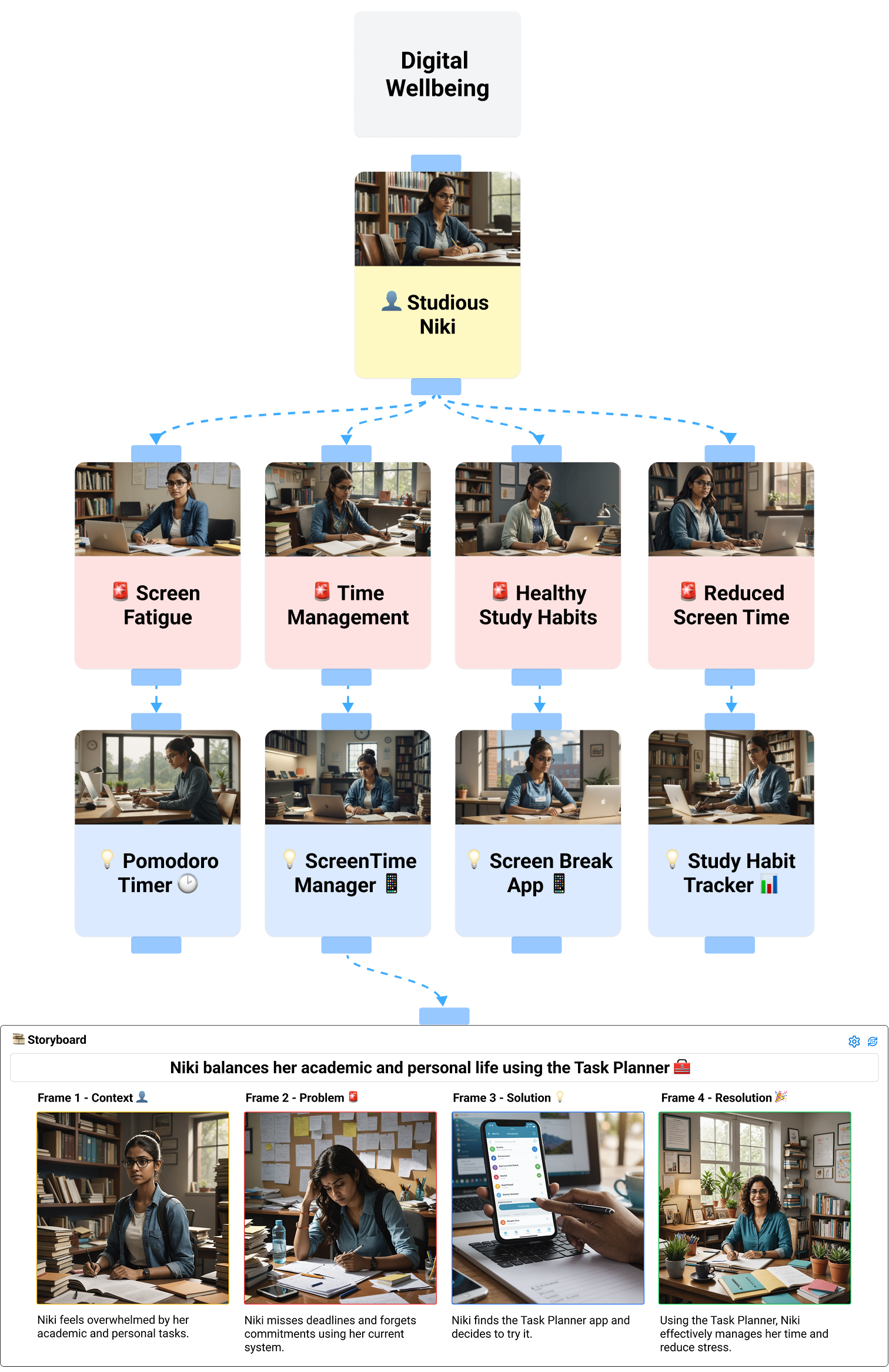}
    \caption{The workspace of P2\textsubscript{prac} after completing the user study task}
    \label{fig:p2-screenshot}
\end{figure}



\begin{table*}[h]
    \caption{Some of the topics provided to user study participants as a reference to choose from. They were also free to choose a topic outside this list. The bold indicates topics participants chose.}
    \label{table:task-topics}
    \centering
    \resizebox{\textwidth}{!}{
    \begin{tabular}{ p{4cm}  p{9cm} p{5cm} }
        \toprule
\textbf{Topic}      
& \textbf{Background}   
& \textbf{Task} \\\midrule
\textbf{Financial Literacy} (P1) & Many individuals struggle with managing their finances due to a lack of accessible and engaging educational resources. & Help users better understand and manage their finances. \\\hline
\textbf{Health and Wellness Tracking} (P5) & Tracking health and wellness metrics can help individuals lead healthier lives, but existing tools often fail to engage users consistently. & Design a tool that helps users engage with their health and wellness. \\\hline
\textbf{Education and Learning} (P8) & Educational experiences vary widely in effectiveness, often depending on how well they cater to individual learning needs and styles. & Enhance educational experiences by addressing diverse learning needs. \\\hline
\textbf{Personalized News Consumption} (P4) & The rise of digital news has led to concerns about echo chambers and misinformation. Users need tools that allow them to consume news that is both personalized and balanced. & Enable personalized and balanced news consumption. \\\hline
\textbf{Gamification of Learning} (P7) & Gamification has been shown to increase engagement and motivation in learning, but its application in education is still evolving. Effective design is key to balancing fun and educational value. & Make learning more engaging through innovative approaches. \\\hline
\textbf{Sustainable Living} (P3) & Environmental sustainability is becoming increasingly important, but many people struggle to incorporate sustainable practices into their daily lives. & Create a way to encourage and support sustainable living practices. \\\hline
\textbf{Digital Wellbeing} (P2) & The pervasive use of digital devices can lead to challenges in maintaining a balanced lifestyle, impacting both mental and physical health. & Develop a solution that promotes healthy digital habits.\\\hline
Accessibility in Public Spaces & Public spaces are essential for community interaction, but they are often not fully accessible to all individuals. This can create barriers to participation and inclusion. & Design a solution that enhances accessibility in public spaces for diverse users. \\\hline
Remote Work Collaboration & The rise of remote work has transformed how teams collaborate, bringing new challenges in communication, coordination, and maintaining a sense of connection. & Improve the experience of remote team collaboration. \\\hline
Smart Home Interfaces & As smart home technology becomes more prevalent, users face challenges in managing multiple devices with varying levels of complexity. & Simplify and improve the interaction with smart home technology. \\\hline
Transportation and Mobility & Urban mobility is a critical issue, with challenges such as traffic congestion, inefficient public transport, and accessibility for people with disabilities. & Improve urban mobility and the user experience of transportation. \\\hline
        \bottomrule
    \end{tabular}
}
\end{table*}

\newpage

\subsection{Pipeline \& Prompts}
\label{sec:pipeline-n-prompts}

\begin{table}[h!]
    \caption{System-wide prompts used with other prompts to enforce response format.}
    \label{table:prompts-system}
    \centering
    \begin{tabular}{ p{4cm}  p{10cm} }
        \toprule
        \textbf{Prompt Type} & \textbf{Prompt} \\\midrule
        Enforce response JSON schema & Output a JSON object that fits the schema based on the user message. \newline\newline
        \texttt{JSON SCHEMA:} \texttt{"""}\newline
        \textcolor{ACMOrange}{\texttt{\{jsonSchema\}}}\newline
        \texttt{"""}
        \\\bottomrule
    \end{tabular}
\end{table}

\begin{longtable}{ p{4cm}  p{10cm} }
    \caption{Prompts for generating and regenerating user personas.}
    \label{table:prompts-persona} \\
    \toprule
    \textbf{Prompt Type} & \textbf{Prompt} \\
    \midrule
    \endfirsthead

    \caption*{(Table \thetable. continued)} \\
    \toprule
    \textbf{Prompt Type} & \textbf{Prompt} \\
    \midrule
    \endhead

    \bottomrule
    \endlastfoot

    Generate personas & User personas are detailed descriptions of fictional characters that represent different user types. They help designers gain empathy for their users and understand their needs, goals, and behaviors. \newline\newline
    Keep the values short and to the point and use emojis where possible. \newline
    Avoid repeating the same information in different values. \newline\newline
    Use an alliterative adjective and first name to create a memorable name for the persona. \newline
    Create \textcolor{ACMOrange}{\{numberOfVariations\}} personas based on the given context. \newline\newline
    Context: \texttt{"""}\newline
    \textcolor{ACMOrange}{\texttt{\{context\}}} \newline
    \texttt{"""} \\\midrule
    
    Regenerate personas & User personas are detailed descriptions of fictional characters that represent different user types. They help designers gain empathy for their users and understand their needs, goals, and behaviors. \newline\newline
    Keep the values short and to the point and use emojis where possible. \newline
    Avoid repeating the same information in different values. \newline\newline
    Use an alliterative adjective and first name to create a memorable name for the persona. \newline\newline
    Update the given personas based on the new context/instructions, keeping the essence of the persona intact, but making necessary changes to ensure cohesiveness. \newline\newline
    When the context requests regenerating personas based on updated problems: \newline
    - Rewrite the personas to directly address the problems. \newline
    - Remove any information that isn't directly related to the new problems. \newline\newline
    Personas: \texttt{"""}\newline
    \textcolor{ACMOrange}{\texttt{\{personas\}}} \newline
    \texttt{"""} \newline\newline
    Context: \texttt{"""}\newline
    \textcolor{ACMOrange}{\texttt{\{context\}}} \newline
    \texttt{"""}
    \\
\end{longtable}

\begin{longtable}{ p{4cm}  p{10cm} }
    \caption{Prompts for generating and regenerating problem statements.}
    \label{table:prompts-problem} \\
    \toprule
    \textbf{Prompt Type} & \textbf{Prompt} \\
    \midrule
    \endfirsthead

    \caption*{(Table \thetable. continued)} \\
    \toprule
    \textbf{Prompt Type} & \textbf{Prompt} \\
    \midrule
    \endhead

    \bottomrule
    \endlastfoot

    Generate problem statements & A problem statement is a description of an issue designers are trying to solve. \newline
    A good problem statement should be centered on specific people and their needs. \newline
    It should be narrow enough to be manageable but broad enough to explore a variety of solutions. \newline
    A problem statement should be a statement and not a question. \newline
    A problem statement should not be a hypothetical solution or strongly suggest a single solution. \newline\newline
    Keep values short and to the point and use emojis where possible. \newline
    Avoid repeating the same information in different values. \newline
    Create \textcolor{ACMOrange}{\{numberOfVariations\}} problem statements based on the given context. \newline\newline
    Context: \texttt{"""}\newline
    \textcolor{ACMOrange}{\texttt{\{context\}}} \newline
    \texttt{"""} \\\midrule
    
    Regenerate problem statements & A problem statement is a description of an issue designers are trying to solve. \newline
    A good problem statement should be centered on specific people and their needs. \newline
    It should be narrow enough to be manageable but broad enough to explore a variety of solutions. \newline
    A problem statement should be a statement and not a question. \newline
    A problem statement should not be a hypothetical solution or strongly suggest a single solution. \newline\newline
    Keep values short and to the point and use emojis where possible. \newline
    Avoid repeating the same information in different values. \newline\newline
    Update the given problem statements based on the new context/instructions. \newline
    Keep the essence of the problem statement intact, but make necessary changes to ensure cohesiveness. \newline\newline
    When the context requests regenerating problems based on updated personas: \newline
    - Rewrite the problem statements to directly address the personas. \newline
    - Remove any information that isn't directly related to the new personas. \newline\newline
    When the context requests regenerating problems based on updated solutions: \newline
    - Rewrite the problem statements to directly address the solutions. \newline
    - Remove any information that isn't directly related to the new solutions. \newline\newline
    Problems: \texttt{"""}\newline
    \textcolor{ACMOrange}{\texttt{\{problems\}}} \newline
    \texttt{"""} \newline\newline
    Context: \texttt{"""}\newline
    \textcolor{ACMOrange}{\texttt{\{context\}}} \newline
    \texttt{"""}
    \\
\end{longtable}

\begin{longtable}{ p{4cm}  p{10cm} }
    \caption{Prompts for generating and regenerating solutions.}
    \label{table:prompts-solutions} \\
    \toprule
    \textbf{Prompt Type} & \textbf{Prompt} \\
    \midrule
    \endfirsthead

    \caption*{(Table \thetable. continued)} \\
    \toprule
    \textbf{Prompt Type} & \textbf{Prompt} \\
    \midrule
    \endhead

    \bottomrule
    \endlastfoot

    Generate solutions & A design solution is a proposal for addressing specific problems or user needs. \newline
    Solutions don't have to be perfect, but instead should be a starting point for further refinement and iteration. \newline\newline
    Keep values short and to the point and use emojis where possible. \newline
    Avoid repeating the same information in different values. \newline
    Create \textcolor{ACMOrange}{\{numberOfVariations\}} solutions based on the given context. \newline\newline
    Context: \texttt{"""}\newline
    \textcolor{ACMOrange}{\texttt{\{context\}}} \newline
    \texttt{"""} \\\midrule
    
    Regenerate solutions & A design solution is a proposal for addressing specific problems or user needs. \newline
    Solutions don't have to be perfect, but instead should be a starting point for further refinement and iteration. \newline\newline
    Keep values short and to the point and use emojis where possible. \newline
    Avoid repeating the same information in different values. \newline\newline
    Update the given solutions based on the new context/instructions. \newline
    Keep the essence of the problem statement intact, but make necessary changes to ensure cohesiveness. \newline\newline
    When the context requests regenerating solutions based on updated problems: \newline
    - Rewrite the solutions to directly address the problems. \newline
    - Remove any information that isn't directly related to the new problems. \newline\newline
    When the context requests regenerating solutions based on updated storyboards: \newline
    - Rewrite the solutions to directly address the storyboards. \newline
    - Remove any information that isn't directly related to the new storyboards. \newline\newline
    Solutions: \texttt{"""}\newline
    \textcolor{ACMOrange}{\texttt{\{solutions\}}} \newline
    \texttt{"""} \newline\newline
    Context: \texttt{"""}\newline
    \textcolor{ACMOrange}{\texttt{\{context\}}} \newline
    \texttt{"""}
    \\
\end{longtable}

\begin{longtable}{ p{4cm}  p{10cm} }
    \caption{Prompts for generating a storyboard outline and image prompts.}
    \label{table:prompts-storyboard} \\
    \toprule
    \textbf{Prompt Type} & \textbf{Prompt} \\
    \midrule
    \endfirsthead

    \caption*{(Table \thetable. continued)} \\
    \toprule
    \textbf{Prompt Type} & \textbf{Prompt} \\
    \midrule
    \endhead

    \bottomrule
    \endlastfoot

    Generate storyboard outline & You are an AI assistant tasked with creating a detailed storyboard outline for design thinking based on specific dimensions and their assigned values. \newline
    Use the given dimensions to generate a coherent storyboard outline to visualize a problem, solution, and the surrounding context. \newline\newline
    The title is a sentence summarizing a story about the user situation, problem, solution, and resolution. \newline
    Example title: "A seed catalog app that lets users watch videos instead of reading plant info." \newline\newline
    Generate an outline with at least 4 frames. \newline\newline
    The frame outline contains a brief description for each frame in the storyboard. \newline
    The description should tell a cohesive story about the user situation, problem, solution, and resolution. \newline
    Each frame description should describe key observations and actions in the frame. \newline\newline
    Each frame outline also contains a caption that will be displayed directly below the visuals as part of the storyboard. \newline
    The caption should add clarity and context to the visual image. \newline
    The caption should be shorter than the frame description since it will be displayed directly below the visual image. \newline\newline
    When the context requests regenerating storyboards based on updated personas, problems, and solutions: \newline
    - Revise the storyboard to address the personas, problems, and solutions. \newline
    - Remove any information that isn't directly related to the new personas, problems, and solutions. \newline\newline
    Outline Example: \newline
    Context: \texttt{"""}\newline
    \textcolor{ACMOrange}{\texttt{\{context\}}} \newline
    \texttt{"""} \\\midrule
    
    Generate storyboard image prompts & Given the outline for a storyboard with \textcolor{ACMOrange}{\{numFrames\}} frames, generate a list of image prompts for each frame. \newline\newline
    Each frame outline contains an image prompt and imageNegativePrompt used to generate visuals for the storyboard frame. \newline
    The image prompt should be based on the frame description but focus on describing the visual elements. \newline
    The image prompt should briefly describe the subject using key adjectives. Additionally, it should describe the setting, lighting, and any other relevant visual details. \newline
    Wrap terms in \texttt{(\#weight)} to adjust the importance of the term in the image. The weight should be a number between 0 and 1. \newline
    Do not set the weight for random terms, only for essential terms required for the subject and action. \newline
    The image prompt within a single outline should be consistent in style and form to ensure cohesive visual narrative. \newline
    Consider using specific details such as colors, textures, and other style modifiers. \newline
    Shot types such as wide shot, medium shot, close-up, etc. can be used to specify the framing of the image. Choose the shot type that best suits the frame description. \newline
    The negative prompt should omit things that are difficult for AI to generate or are not relevant to the frame. \newline
    Include the full name of people to ensure consistent characters across all frames. \newline\newline
    Frames: \texttt{"""}\newline
    \textcolor{ACMOrange}{\texttt{\{frames\}}} \newline
    \texttt{"""} \newline\newline
    Visual Character Descriptions: \texttt{"""}\newline
    \textcolor{ACMOrange}{\texttt{\{visualCharacterDescriptions\}}} \newline
    \texttt{"""}
    \\
\end{longtable}

\begin{longtable}{ p{4cm}  p{10cm} }
    \caption{Prompts for generating image prompts and visual character descriptions.}
    \label{table:prompts-images} \\
    \toprule
    \textbf{Prompt Type} & \textbf{Prompt} \\
    \midrule
    \endfirsthead

    \caption*{(Table \thetable. continued)} \\
    \toprule
    \textbf{Prompt Type} & \textbf{Prompt} \\
    \midrule
    \endhead

    \bottomrule
    \endlastfoot

    Generate illustrative image prompt & Using the following idea, generate an image prompt and image negative prompt to generate an illustrative image that represents the key elements of the idea. \newline
    Describe a scene which is a visual metaphor for the persona, problem, or solution described in the idea. \newline
    Prompts should be short and focus on the visual metaphor and not overly describe the characters. \newline
    Describe the image in literal visual elements of the image, not the message or meaning of the image. \newline\newline
    Idea: \texttt{"""}\newline
    \textcolor{ACMOrange}{\texttt{\{idea\}}} \newline
    \texttt{"""} \\\midrule
    
    Generate problem illustrative image prompt & Generate an image that depicts a problem and helps to build empathy. \newline
    Generate a prompt for the image which describes the scene that depicts the problem. Negative prompts describe what the scene should not include. \newline\newline
    Problem: \texttt{"""}\newline
    \textcolor{ACMOrange}{\texttt{\{problem\}}} \newline
    \texttt{"""} \\\midrule
    
    Generate visual character descriptions & Generate visual character descriptions used to ensure visual consistency between different artists and illustrations. \newline
    Generate a visual character description used to illustrate the following idea. \newline\newline
    Idea: \texttt{"""}\newline
    \textcolor{ACMOrange}{\texttt{\{idea\}}} \newline
    \texttt{"""} \\
\end{longtable}

\begin{longtable}{ p{4cm}  p{10cm} }
    \caption{Prompts for generating feedback on design artifacts.}
    \label{table:prompts-design-feedback} \\
    \toprule
    \textbf{Prompt Type} & \textbf{Prompt} \\
    \midrule
    \endfirsthead

    \caption*{(Table \thetable. continued)} \\
    \toprule
    \textbf{Prompt Type} & \textbf{Prompt} \\
    \midrule
    \endhead

    \bottomrule
    \endlastfoot

    Generate persona feedback & You are an UX designer given a persona. \newline
    Brainstorm questions to evaluate this persona being used for design thinking. \newline\newline
    - Consider any information that is missing, but could be useful. \newline
    - Example: If we know the persona is a salesperson, it may be useful to know what kind of product they sell? \newline
    - Consider the accuracy and consistency of the persona. \newline
    - Example: If a persona is experienced, but only has a few years of experience this may be incorrect. \newline
    - Consider alternative persona values which may be useful. \newline
    - Example: If a persona is a salesperson, could a sales manager be another useful persona to consider. \newline\newline
    Persona: \texttt{"""}\newline
    \textcolor{ACMOrange}{\texttt{\{persona\}}} \newline
    \texttt{"""} \\\midrule
    
    Generate problem feedback & You are an UX designer given a problem statement. \newline
    Brainstorm questions to evaluate this problem statement being used for design thinking. \newline
    Focus on the quality of the problem statement rather than solutions. \newline\newline
    - Consider any information that is missing, but could be useful. \newline
    - Example: If we know a salesperson is struggling to find qualified leads, it may be useful to know what strategies they have tried. \newline
    - Consider the accuracy and consistency of the problem statement. \newline
    - Example: If a salesperson struggles to make a connection, but can find qualified leads, this may be incorrect since qualified leads require connections. \newline
    - Consider alternative or related problem statements which may be useful. \newline
    - Example: If a salesperson struggles to find qualified leads, they may also struggle with finding leads in general. \newline\newline
    Problem: \texttt{"""}\newline
    \textcolor{ACMOrange}{\texttt{\{problem\}}} \newline
    \texttt{"""} \\\midrule
    
    Generate solution feedback & You are an UX designer given a solution. \newline
    Brainstorm questions to evaluate this solution being used for design thinking. \newline\newline
    - Consider if any information that is missing, but could be useful. \newline
    - Example: If we want to create a networking mobile app, what are the key features of the app? \newline
    - Consider the accuracy and consistency of the solution. \newline
    - Example: A high-tech drone solution to deliver food to low-income families may be infeasible as the solution may be too expensive. \newline
    - Consider alternative or related solutions which may be useful. \newline
    - Example: If our solution helps a salesperson find leads, it may be useful to also have a solution for qualifying those leads. \newline\newline
    Solution: \texttt{"""}\newline
    \textcolor{ACMOrange}{\texttt{\{solution\}}} \newline
    \texttt{"""} \\\midrule
    
    Generate storyboard feedback & You are an UX designer given a storyboard. \newline
    Brainstorm questions to evaluate this storyboard being used for design thinking. \newline\newline
    Storyboard: \texttt{"""}\newline
    \textcolor{ACMOrange}{\texttt{\{storyboard\}}} \newline
    \texttt{"""} \\\midrule
    
    Generate group feedback & You are an UX designer given a list of nodes representing personas, problems, solutions, and storyboards. \newline
    Brainstorm questions to evaluate these nodes being used for design thinking. \newline\newline
    Nodes: \texttt{"""}\newline
    \textcolor{ACMOrange}{\texttt{\{nodes\}}} \newline
    \texttt{"""} \\
\end{longtable}

\begin{longtable}{ p{4cm}  p{10cm} }
    \caption{Prompts for generating autocomplete recommendations.}
    \label{table:prompts-autocomplete-recommendations} \\
    \toprule
    \textbf{Prompt Type} & \textbf{Prompt} \\
    \midrule
    \endfirsthead

    \caption*{(Table \thetable. continued)} \\
    \toprule
    \textbf{Prompt Type} & \textbf{Prompt} \\
    \midrule
    \endhead

    \bottomrule
    \endlastfoot

    Generate problems recommendations & Generate autocomplete suggestions for descriptions of a general group of problems based on the provided personas. \newline
    Suggestions should be a couple of words. \newline\newline
    Nodes: \texttt{"""}\newline
    \textcolor{ACMOrange}{\texttt{\{nodes\}}} \newline
    \texttt{"""} \\\midrule
    
    Generate solution recommendations & Generate autocomplete suggestions for descriptions of a general group of solutions based on the provided problems. \newline
    Suggestions should be a couple of words. \newline\newline
    Nodes: \texttt{"""}\newline
    \textcolor{ACMOrange}{\texttt{\{nodes\}}} \newline
    \texttt{"""} \\\midrule
    
    Generate dependent storyboard recommendations & Generate autocomplete suggestions for the title of a general group of storyboards based on the provided personas, problems, and solutions. \newline
    The title should generally describe the plot of the storyboard. \newline\newline
    Nodes: \texttt{"""}\newline
    \textcolor{ACMOrange}{\texttt{\{nodes\}}} \newline
    \texttt{"""} \\\midrule
    
    Generate more personas recommendations & Generate autocomplete suggestions for descriptions of a general group of personas which differ from the provided personas. \newline
    Suggestions should be a couple of words. \newline\newline
    Nodes: \texttt{"""}\newline
    \textcolor{ACMOrange}{\texttt{\{nodes\}}} \newline
    \texttt{"""} \\\midrule
    
    Generate more problems recommendations & Generate autocomplete suggestions for descriptions of a general group of problems which differ from the provided problems. \newline
    Suggestions should be a couple of words. \newline\newline
    Nodes: \texttt{"""}\newline
    \textcolor{ACMOrange}{\texttt{\{nodes\}}} \newline
    \texttt{"""} \\\midrule
    
    Generate more solutions recommendations & Generate autocomplete suggestions for descriptions of a general group of solutions which differ from the provided solutions. \newline
    Suggestions should be a couple of words. \newline\newline
    Nodes: \texttt{"""}\newline
    \textcolor{ACMOrange}{\texttt{\{nodes\}}} \newline
    \texttt{"""} \\\midrule
    
    Generate more storyboard recommendations & Generate autocomplete suggestions for the title of a general group of storyboards which differ from the provided storyboards. \newline
    The title should generally describe the plot of the storyboard. \newline\newline
    Nodes: \texttt{"""}\newline
    \textcolor{ACMOrange}{\texttt{\{nodes\}}} \newline
    \texttt{"""} \\\midrule
    
    Revise node recommendations & Generate suggestions for specific instructions to update/change the following design thinking nodes. \newline
    Nodes will be either personas, problems, solutions, or storyboards. \newline
    Suggestions should focus on variations or edits to existing nodes. \newline
    Constrain suggestions to 100 characters. \newline\newline
    Nodes: \texttt{"""}\newline
    \textcolor{ACMOrange}{\texttt{\{nodes\}}} \newline
    \texttt{"""} \\
\end{longtable}

\begin{figure*}[htb!]
	\centering
        \includegraphics[alt={A flowchart illustrating the process of converting text inputs into a complete visual storyboard. At the top, there are four inputs: "Persona Inputs," "Problem Inputs," "Solution Inputs," and "Visual Character Descriptions." These inputs flow into a central box labeled "Generate Storyboard Outline (OpenAI)." Next, the storyboard outline is passed to another box labeled "Generate Storyboard Image Prompts (OpenAI)." Finally, the image prompts flow into four separate boxes representing the storyboard image generation. Each image is generated using Stability AI.}, width=0.8\textwidth]{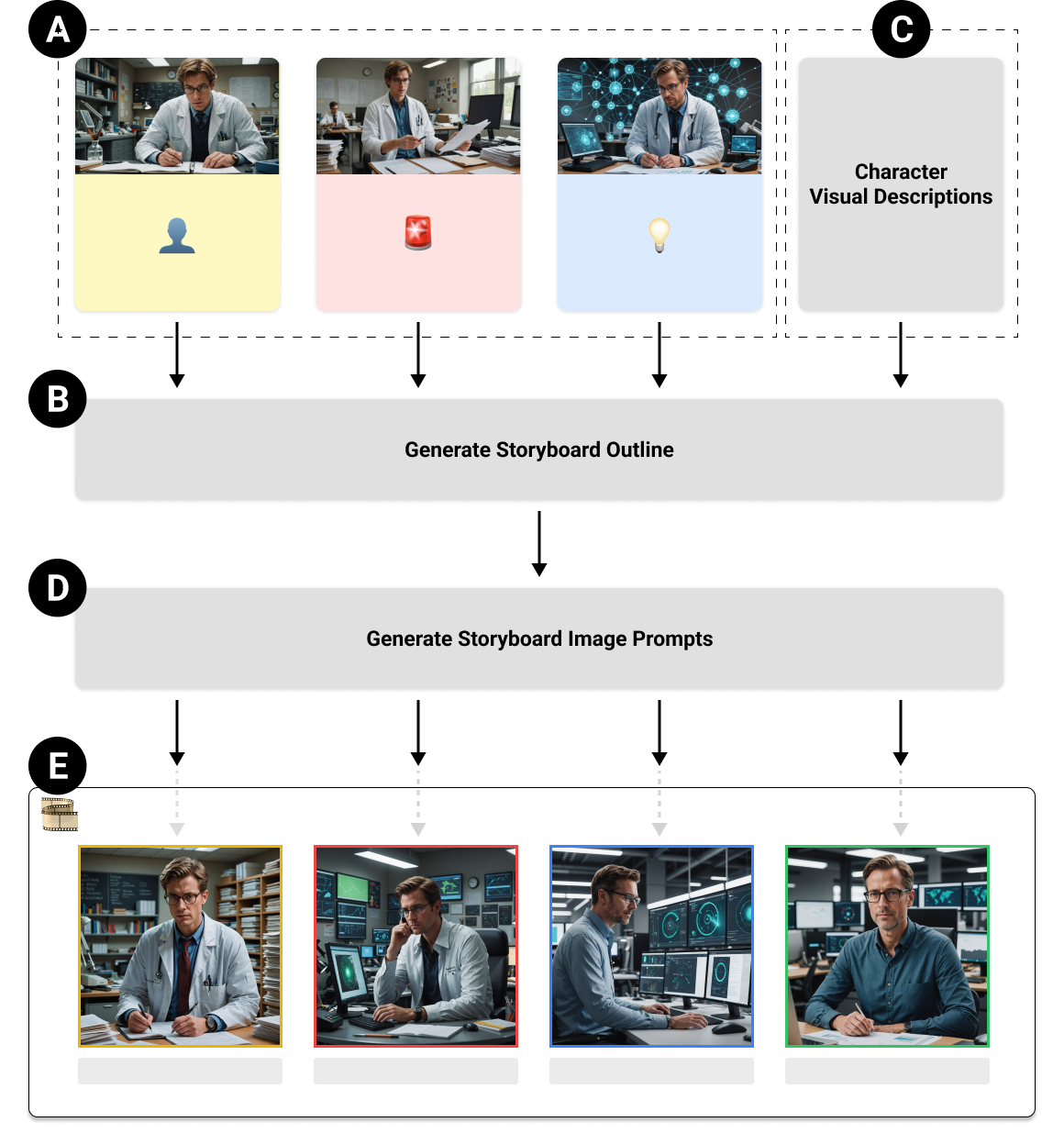}
        \caption{\textsc{Pipeline for building a storyboard}: (A) Persona, problem, and solution inputs are used to generate a (B) text storyboard outline that includes a title and frames consisting of a caption and description. (C) Visual character descriptions help to maintain visual consistency between the design artifacts and the storyboard. The storyboard outline is used to generate (D) image prompts in a single LLM invocation to ensure continuity across frames. (E) Images are generated in parallel using image prompts from the previous step.}
	\label{fig:pipeline-building-storyboard}
\end{figure*}

\end{document}